\documentclass[11pt]{article} 
\usepackage{jheppub}

\usepackage{amsmath,amsfonts,amsbsy,amssymb,array,accents,dsfont}
\usepackage{enumerate}
\usepackage[shortlabels]{enumitem}
\usepackage{graphicx,verbatim} 
\usepackage{subcaption} 
\usepackage{upgreek}
\usepackage{bm}
\usepackage{slashed}
\usepackage{cancel}

\let\includefigures=\iftrue
\let\useblackboard==\iftrue
\newfam\black

\usepackage[dvipsnames]{xcolor}

\definecolor{myblue}{RGB}{85,130,255}
\definecolor{myred}{RGB}{200, 45, 40}

\PassOptionsToPackage{ 
     colorlinks=true,
     linkcolor=myBlue,
     urlcolor=blue,
     filecolor=black,
     citecolor=myRed,
}{hyperref}

\usepackage{xparse}
\usepackage{xspace}

\NewDocumentCommand\eqn{om}{%
  \IfNoValueTF{#1}
     {\[ #2 \]}
     {\begin{equation}\label{#1} #2  \end{equation} \expandafter\newcommand\csname #1\endcsname{\eqref{#1}\xspace}\ignorespaces}
}
\NewDocumentCommand\eqna{om}{%
  \IfNoValueTF{#1}
    {\begin{align*} #2 \end{align*}}
    {\begin{equation}\label{#1}\begin{split} #2  \end{split}\end{equation} \expandafter\def\csname #1\endcsname{\eqref{#1}\xspace}\ignorespaces}
}

\newcommand{\rcite}{\cite}

\def\ttr{{\mathtt r}}
\def\ttt{{\mathtt t}}
\def\tty{{\mathtt y}}

\def\ytil{{\tilde y}}

\newcommand{\Msl}{M}
\newcommand{\Msu}{M'}
\newcommand{\wsl}{w}
\newcommand{\wsu}{w'}

\newcommand{\bMsl}{\bar{M}}
\newcommand{\bMsu}{\bar{M}'}
\newcommand{\bwsl}{w}
\newcommand{\bwsu}{\bar{w}'}
\newcommand{\jsl}{j}
\newcommand{\jsu}{j'}

\def\sl{\text{sl}}
\def\su{\text{su}}

\def\vareps{\varepsilon}

\def\str{{\rm str}}

\def\sltwo{\ensuremath{SL(2,\bR)}}

\def\sutwo{{SU(2)}}

\def\uone{U(1)}

\def\tight#1{\! #1 \!}  

\def\({\left(}
\def\){\right)}
\def\[{\left[}
\def\]{\right]}

\def\ie{{i.e.}}
\def\eg{{e.g.}}

\def\etc{{etc}}

\def\qu{{\rm qu}}

\def\ext{{\rm ext}}

\def\tot{{\rm tot}}

\def\gstr{g_{\textit s}^{\;}}
\def\gstrsq	{g_{\textit s}^{2}}

\def\btz{{\sst \rm BTZ}}

\def\nfive{{n_5}}

\def\none{{n_1}}

\def\sfA{{\mathsf A}}
\def\sfB{{\mathsf B}}

\def\sfD{{\mathsf D}}
\def\sfF{{\mathsf F}}

\def\sfH{{\mathsf H}}

\def\sfR{{\mathsf R}}

\def\sfT{{\mathsf T}}

\def\sfX{{\mathsf X}}

\def\sfZ{{\mathsf Z}}
\def\sfa{{\mathsf a}}
\def\sfb{{\mathsf b}}

\def\sfm{{\mathsf m}}
\def\sfn{{\mathsf n}}

\def\sfp{{\mathsf p}}

\def\sfr{{\mathsf r}}

\def\sft{{\mathsf t}}

\def\sfx{{\mathsf x}}
\def\sfy{{\mathsf y}}
\def\sfz{{\mathsf z}}

\def\mfa{{\mathfrak a}}
\def\mfb{{\mathfrak b}}

\DeclareMathSymbol{\medhatsym}{\mathord}{largesymbols}{"62} 

\DeclareMathSymbol{\medtildesym}{\mathord}{largesymbols}{"65}

\makeatletter
\newcommand*\rel@kern[1]{\kern#1\dimexpr\macc@kerna}
\newcommand*\widebar[1]{%
  \begingroup
  \def\mathaccent##1##2{%
    \rel@kern{0.8}%
    \overline{\rel@kern{-0.8}\macc@nucleus\rel@kern{0.2}}%
    \rel@kern{-0.2}%
  }%
  \macc@depth\@ne
  \let\math@bgroup\@empty \let\math@egroup\macc@set@skewchar
  \mathsurround\z@ \frozen@everymath{\mathgroup\macc@group\relax}%
  \macc@set@skewchar\relax
  \let\mathaccentV\macc@nested@a
  \macc@nested@a\relax111{#1}%
  \endgroup
}
\makeatother

\def\ytil{{\tilde y}}

\def\rhat{r}
\def\runsc{r_\flat}
\def\tunsc{t_\flat}
\def\yunsc{y_\flat}

\def\half{\frac12}
\def\coeff#1#2{{\textstyle \frac{#1}{#2}}}
\def\hf{\coeff12}

\def\One{{\hbox{1\kern-1mm l}}}

\def\barray{\begin{array}}
\def\earray{\end{array}}
\def\be{\begin{equation}}
\def\ee{\end{equation}}
\def\bea{\begin{align}}
\def\eea{\end{align}}
\def\bal{\begin{align}}
\def\eal{\end{align}}
\def\nn{\nonumber}

\newcommand{\bF}{{\mathbb F}}

\newcommand{\bR}{{\mathbb R}}
\newcommand{\bS}{{\mathbb S}}
\newcommand{\bT}{{\mathbb T}}
\newcommand{\bX}{{\mathbb X}}
\newcommand{\bZ}{{\mathbb Z}}

\definecolor{cardinal}{rgb}{0.6,0,0}
\definecolor{darkgreen}{rgb}{0,0.4,0}
\definecolor{green}{rgb}{0,0.4,0}
\definecolor{golden}{rgb}{0.92, 0.7, 0}
\definecolor{midnight}{rgb}{0, 0, 0.5}
\definecolor{darkblue}{rgb}{0, 0, 0.7}

\numberwithin{equation}{section}

\mathchardef\mhyphen="2D

 \def\cB{\mathcal {B}} 
\def\cD{\mathcal {D}} \def\cE{\mathcal {E}} \def\cF{\mathcal {F}}
\def\cG{\mathcal {G}} \def\cH{\mathcal {H}} \def\cI{\mathcal {I}}
\def\cJ{\mathcal {J}}  
\def\cM{\mathcal {M}} \def\cN{\mathcal {N}} 
\def\cP{\mathcal {P}}  
\def\cS{\mathcal {S}} \def\cT{\mathcal {T}}

\def\one{{\hbox{\kern+.5mm 1\kern-.8mm l}}}
\def\zero{{\hbox{0\kern-1.5mm 0}}}

\def\id{\textrm{id}}

\def\id{{1 \kern-.28em {\rm l}}}

\def\journal#1&#2(#3){\unskip, \sl #1\ \bf #2 \rm(19#3) }
\def\andjournal#1&#2(#3){\sl #1~\bf #2 \rm (19#3) }

\def\ie{{\it i.e.}}
\def\eg{{\it e.g.}}

\def\etc{{\it etc}}

\def\sst{\scriptscriptstyle}

\def\coeff#1#2{{\textstyle{\frac{#1}{ #2}}}}
\def\half{\frac12}
\def\hf{{\textstyle\half}}

\def\One{{1\hskip -3pt {\rm l}}}

\catcode`\@=11
\def\slash#1{\mathord{\mathpalette\c@ncel{#1}}}
\def\vareps{\varepsilon}
\def\underrel#1\over#2{\mathrel{\mathop{\kern\z@#1}\limits_{#2}}}

\catcode`\@=12

\def\det{{\rm det}}

\def\det{{\rm det}}
\def\exp{{\rm exp}}

\def\ie{{\it i.e.}}
\def\eg{{\it e.g.}}

\title{
{
BPS Fivebrane Stars and BTZ Black Holes
}}

\author{Emil J. Martinec$^a$, Yoav Zigdon$^{b}$\\}

\affiliation[a]{
The Leinweber Institute for Theoretical Physics,\\ Enrico Fermi Institute, and Department of Physics\\ 
University of Chicago\\ 
5640 S. Ellis Ave.\\
Chicago IL 60637  USA\\ 
}

\affiliation[b]{School of Physics and Astronomy, Tel Aviv University, Ramat Aviv, 69978, Israel}

 \emailAdd{e-martinec@uchicago.edu}
 \emailAdd{yoavzi(at)tauex.tau.ac.il}

\abstract{
We construct a large new class of BPS supergravity solutions parametrized by arbitrary chiral wave profiles on fundamental string and Neveu-Schwarz fivebrane sources.  
The effective action approach we employ describes the regime where the brane sources are slightly separated.    
String probes see these backgrounds as smooth and effectively capped off at the scale of that separation.
Within the space of solutions is a large ensemble of ultracompact objects exhibiting many of the features of extremal BTZ black holes, in particular having a deep $AdS_2$ throat.
We present the details of solutions with circular source profiles, and compare them to known superstratum geometries. 

The separation of the sources makes them structurally more like BPS fivebrane stars than like extremal black hole microstates. 
When evaluated on the supergravity background, the brane effective action governs (chaotic) near-BPS dynamics, and exhibits a collective mode analogous to the Schwarzian mode of $AdS_2$ black holes.
Finally, we discuss the approach to the black hole phase as the sources coalesce, in which the semiclassical approximation breaks down due to the appearance of light D-brane excitations; this bulk realization of the fivebrane deconfinement transition connects directly to the Hawking-Page transition.
}

\begin{document}
\maketitle
\hypersetup{pageanchor=true}
\pagenumbering{arabic}






\section{Introduction} 
\label{sec:intro}

The fuzzball paradigm posits that string theory resolves black hole singularities with stringy matter of horizon scale extent, so that a black hole essentially becomes a compact object composed of the exotic extended objects of the theory.  The puzzles associated to black hole thermodynamics would then be resolved by quantum coherence of this object on the scale of the classical horizon.

One context where this idea has been extensively explored takes the stringy matter to consist of fivebranes compactified on $\bT^4\times\bS^1$, bound to strings wrapping the $\bS^1$ (which we parametrize by a coordinate $y$, with periodicity $2\pi R_y$).  BPS black holes carrying these charges, as well as momentum along $\bS^1_y$, have an entropy~\rcite{Strominger:1996sh}
\be
\label{3chgent}
S_{BH} = 2\pi \sqrt{\vphantom{d}n_5n_1n_p}
\ee
in terms of the integer winding and momentum charge quanta.

In the duality frame where the branes are Neveu-Schwarz fivebranes (NS5) and fundamental strings (F1), a qualitative explanation of this formula posits that in the presence of $n_5$ coincident NS5-branes, a fundamental string fractionates its charge and tension into $n_5$ constituent {\it little strings}~\rcite{Strominger:1996sh,Maldacena:1996ya,Dijkgraaf:1998iz}, so that the entropy $S_{\str}\sim \sqrt{\vphantom{d}w n_p}$ of a string of winding $w$ and carrying $n_p$ units of momentum gains a factor of $n_5$ under the square root because the constituent little string winding is $w=n_5n_1$ rather than $n_1$.

Little strings are the chief actor in non-abelian NS5-brane dynamics, the analogue of W-particles in non-abelian gauge theory.  They are thus made massive by separating the fivebranes onto their Coulomb branch, where they eventually become D2-brane strips stretched between the NS5's, sufficiently heavy that they can be decoupled (for a discussion in the present context, see~\rcite{Martinec:2019wzw}).  On the Coulomb branch, the NS5 dynamics abelianizes, and one can treat that dynamics self-consistently using a low-energy effective action for the branes coupled to bulk supergravity~\rcite{Martinec:2024emf}.

This approach applies even in the fivebrane decoupling limit, which takes the asymptotic value of the string coupling to zero, and commensurately scales the brane locations $\sfF^\flat_\sfm$, $\sfm=1...\nfive$ in their asymptotically flat transverse space~\rcite{Giveon:1999px,Giveon:1999tq}
\be
\label{decoupling}
\gstr\to 0 
~~,~~~~
\hat\sfF_\sfm = \frac{\sfF^\flat_\sfm}{\gstr} 
~~ {\rm fixed}~.
\ee
As we review in section~\ref{sec:review} below, the effective fivebrane throat geometry seen by strings is capped off at the scale of the brane separations $|\hat\sfF_\sfm-\hat\sfF_{\sfm'}|$; if the smallest such scale is sufficiently large, non-abelian little string excitations are decoupled and string perturbation theory is valid.

The separation of the fivebranes effectively resolves the singularities in the classical geometry that they source.  Strings are unable to penetrate the throats of individual isolated fivebranes, as has been discussed in detail in previous work~\rcite{Martinec:2017ztd,Martinec:2018nco,Martinec:2019wzw,Martinec:2020gkv,Martinec:2022okx,Martinec:2024emf,Martinec:2025xoy}.  As a result, the geometry is effectively smeared and capped off at the scale of the fivebrane separation.

In this abelianized regime, the three charged constituents (NS5-F1-P) form BPS bound states that are perhaps better thought of as brane stars rather than black holes.  But when the configuration of these charged constituents is sufficiently compact, such that the radius of the bound state $r_*$ is much smaller than each of the charge radii
\be
r_*^2\ll Q_5,Q_1,Q_p ~~,
\ee
the object lives inside an $AdS_2$ throat~-- it is {\it ultracompact}, in the sense that the area of the surface that surrounds the object is essentially the same as that of the horizon of the corresponding three-charge black hole.  Nevertheless the object is {\it not} a black hole, because the radius $r_*$ at which the geometry caps off is larger than the radial scale $r_{\qu}$ at which little strings become light and enter the dynamics; also, the ``brane star'' entropy is parametrically much less than the black hole entropy~\eqref{3chgent}~\rcite{Shigemori:2019orj}.

The scale $r_\qu$ at which non-abelian effects kick in is the scale at which the system enters the ensemble of black hole states.  As we argue in section~\ref{sec:review}, this non-abelian scale agrees with the quantum scale of BPS $AdS_2$ black holes~\rcite{Preskill:1991tb,Lin:2022zxd,Heydeman:2020hhw,Lin:2022rzw}~-- the $AdS_2$ throat depth at which quantum gravity  fluctuations cannot be ignored, and modify the classical picture of the geometry.  In the process of collapse of the near-BPS brane star, the constituent branes shrink their extent in the transverse space of the fivebranes; the cap in the geometry descends the $AdS_2$ throat until it reaches the quantum scale $r_\qu$, at which point the little string excitations are liberated and the system enters the black hole phase.

\begin{figure}[ht]
\centering
\includegraphics[scale=0.4]{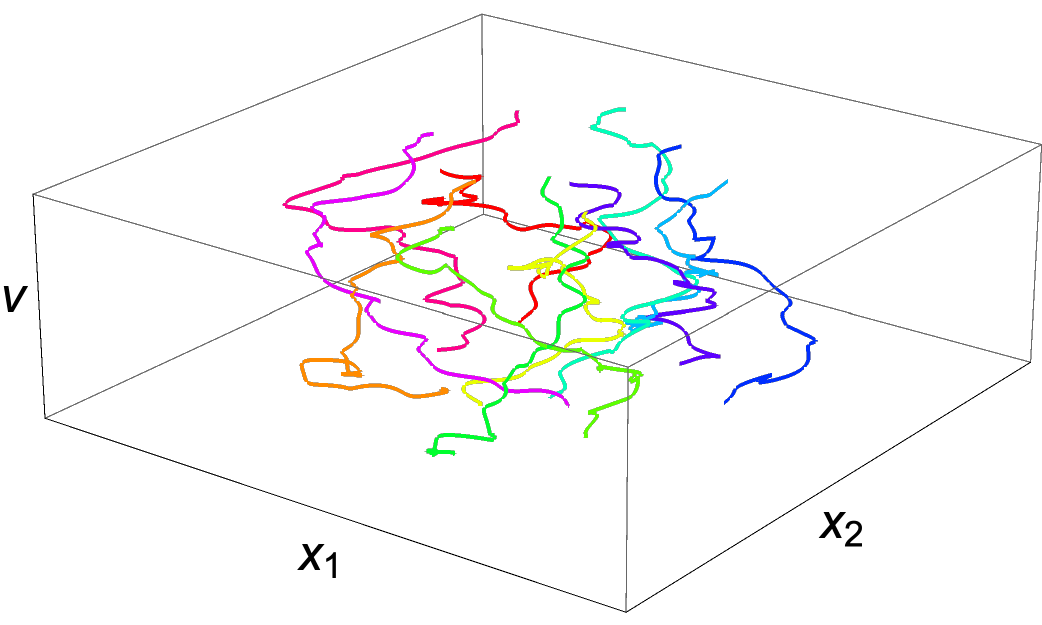}
\caption{\it A highly excited BPS fivebrane multiply wrapping the y-circle ($v=t+y$) executes a random walk in its transverse space. The hues of the fivebrane strands evolve around the color wheel {\rm\small (ROYGBIV)} to indicate their connectivity. }
\label{fig:NS5-P}
\end{figure}

When the fivebranes are compactified on $\bT^4\times\bS^1_y$, they have a large but finite mass, and their transverse locations are quantum variables living in some wavefunction rather than having fixed expectation values.  The validity of string perturbation theory requires that this wavefunction be predominantly supported on the Coulomb branch of NS5 configurations.  This property can be engineered by binding the fivebranes together via a twisted boundary condition on $\bS^1_y$ that wraps them all together into a single fivebrane, and highly exciting the scalars $\sfF$ which specify their transverse location, see figure~\ref{fig:NS5-P}.  The configuration is BPS provided the excitations are purely chiral (say, left-moving).  The higher the level of excitation of the scalars $\sfF$, the more the strands of the winding fivebrane are separated, and little string dynamics is more highly suppressed. 
%
%

In the absence of F1 charge, the geometry sourced by such a fivebrane configuration is called an NS5-P {\it supertube}~\rcite{Lunin:2001fv,Kanitscheider:2007wq}.  These geometries can be derived~\rcite{Martinec:2024emf} from an effective action for the fivebranes~\rcite{Eyras:1998hn} coupled to supergravity.  The typical source configuration (\ie\ drawn at random from the phase space of classical solutions) is a random walk of  size~\rcite{Alday:2006nd,Balasubramanian:2008da,Raju:2018xue,Martinec:2023xvf}
\be
r_*^2 = \mu^2 \Big(\frac{\pi^2 n_5 n_p^\perp}{6}\Big)^{\half}
~~,~~~~
\mu^2 = \frac{g_s^2(\alpha')^3}{V_4} ~,
\ee
where $V_4$ is the volume of $\bT^4$ and $n_p^\perp$ is the total momentum carried by the scalars $\sfF$.%
\footnote{In general, the momentum carried by the fivebrane is partitioned into BPS excitations of the transverse scalars $\sfF$ and internal excitations of the fivebrane gauge multiplet.}
The scale $r_*$ is parametrically larger than the quantum scale $r_\qu$ provided $n_p^\perp$ is sufficiently large.  If these excitations are sufficiently generic, and comprise a finite fraction of the total momentum $n_p$, the typical such NS5-P supertube is a fivebrane star rather than a stringy extremal black hole.

When one adds the third (F1) charge, there are BPS black holes with macroscopic horizon area and associated entropy~\eqref{3chgent}.  In addition, there are three-charge BPS fivebrane stars.  Examples of the latter are {\it superstrata}, in which an NS5-F1 supertube is decorated with a supergraviton wave carrying momentum along $\bS^1_y$.  The T-dual of these backgrounds along the $y$-circle are NS5-P supertubes bound to a condensate of winding fundamental strings.

In this work, we extend our effective action analysis to three-charge BPS geometries by adding F1-P sources to two-charge NS5-P backgrounds (the T-duals of superstrata thus comprise a special case where the winding strings carry no momentum along $\bS^1_y$).  The BPS field equations have the layered structure~\rcite{Bena:2011dd,Niehoff:2012wu} used in the construction of superstrata~\rcite{Bena:2015bea,Bena:2017xbt} (see~\rcite{Shigemori:2020yuo} for a review).  
We augment the supergravity plus fivebrane effective action with the Polyakov action for fundamental strings, and solve the resulting BPS equations of motion in closed form in terms of the left-moving source excitation profiles $\sfF(t\tight+y)$ of the fivebranes and $\sfX(t\tight+y)$ of the strings.

The relevant three-charge (NS5-F1-P) BPS solutions were in fact already largely found in the work of~\rcite{Niehoff:2012wu} (in the S-dual D5-D1-P frame).  Our work completes their analysis by matching the bulk solutions to the particular brane configurations that source them.  The various harmonic forms and functions appearing in the solution all have pole singularities at the sources, but it is known from previous work on two-charge solutions~\rcite{Martinec:2017ztd,Martinec:2018nco,Martinec:2019wzw,Martinec:2020gkv,Martinec:2022okx,Martinec:2025xoy,Martinec:2024emf} that the singularities associated to the fivebranes are not manifested in string perturbation theory on such backgrounds.  The classical singularities associated to fundamental string sources are as usual resolved by string quantization, and the inherent fuzziness which that entails on the length scale $\alpha'$ set by the string tension.  Thus perturbative string theory is valid in this class of backgrounds. 

Among the solutions are configurations that closely approximate extremal BTZ black holes; this feature is in fact generic for fivebrane stars in certain regimes of the BPS charge lattice and moduli of the compactification.  
One can arrange the hierarchy of scales $Q_5\gg Q_1\gg Q_p\gg r_*^2$.  The first inequality results from the limit of small $g_s$, which makes the fivebranes heavy relative to the strings; the second inequality results from taking large $R_y$, which makes the energy cost of winding strings much larger than that of momentum excitations.  The last inequality arises when the NS5-F1-P source is sufficiently compact.  Since $Q_p\propto n_p$ and the size of the fivebrane random walk scales as $(n_{p,5}^\perp)^{1/2}$, the hierarchy $Q_p\gg r_*^2$ arises when the quantum numbers are large.

\begin{figure}[ht]
\centering
\includegraphics[scale=0.5]{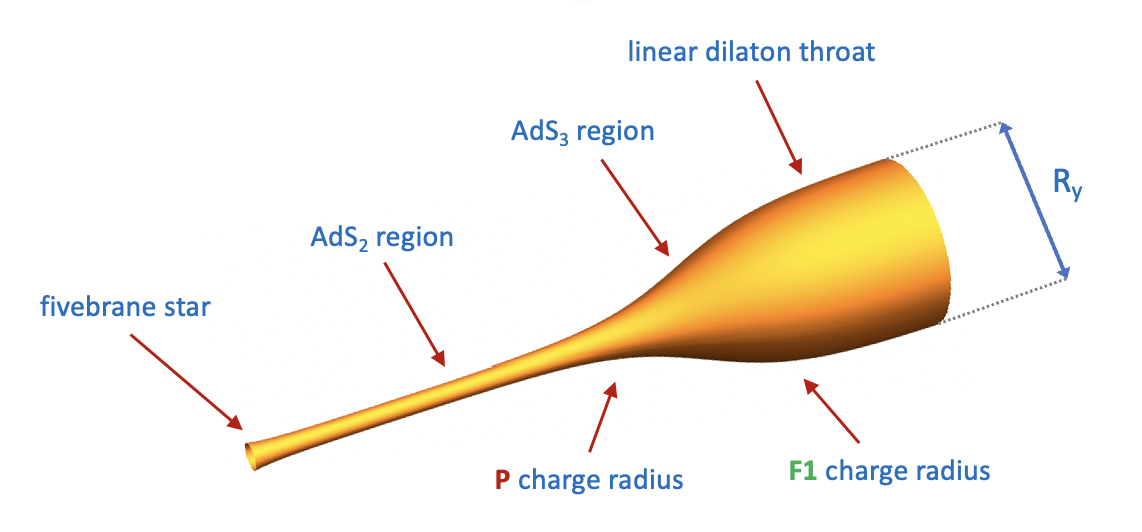}
\caption{\it Proper size of $\bS^1_y$ as a function of radius in the typical three-charge BPS state we construct, in the regime of weak string coupling and large $R_y$. }
\label{fig:3chg}
\end{figure}

The typical geometry with this hierarchy of scales has the linear dilaton throat of $\nfive$ fivebranes at lage radius, with the proper size of the $y$-circle approximately constant at its asymptotic value $R_y$ until one reaches the radial scale set by the string charge.  At this point the geometry rolls over to $AdS_3$, with the $y$-circle forming the $AdS_3$ azimuthal direction; it thus starts shrinking with decreasing radius, until one reaches the radial scale set by the momentum charge (provided the sources are sufficiently compact that their support lies within this momentum charge radius).  At this point, the $y$-circle stops shrinking, being stabilized by the pressure exerted by the momentum of the sources; at smaller scales one encounters a nearly-$AdS_2$ regime with a frozen angular geometry.  The typical geometry is exponentially close to the extremal BTZ solution until one reaches the surface of the fivebrane star composed of the constituent strings and fivebranes.  This typical geometry is depicted in figure~\ref{fig:3chg}.

%

If the fivebrane is sufficiently puffed up that its strands are well separated, \ie\ the fivebrane star is well outside its Schwarzshild radius, then string theory is weakly coupled and a perturbative string description is valid.  When the star approaches its Schwarzschild radius, the perturbative description breaks down, as the little string excitations of nonabelian fivebranes become light and no longer decouple.

This behavior is precisely what we would expect in gauge-gravity duality, where the Hawking-Page transition in gravity, associated to black hole formation, is dual to the deconfinement transition in the gauge theory, where non-abelian degrees of freedom are liberated~\rcite{Witten:1998zw}.  Our work exhibits this gauge theory mechanism on the gravity (more precisely, string theory) side of the duality (see also~\rcite{Martinec:2015pfa,Martinec:2019wzw}).  While we cannot at present follow the subsequent development of the fivebrane source once it collapses to this point, we do see that new, light degrees of freedom emerge that are well beyond those of supergravity, and which must be incorporated in any effective description.  

The appearance of new, light, strongly coupled stringy degrees of freedom at the horizon, that are responsible for (1) eliminating the black hole singularity of general relativity, (2) explaining black hole thermodynamics, and (3) resolving the black hole information paradox, is the heart of the fuzzball paradigm~\rcite{Mathur:2005zp}.  In the present context, our suggestion is that little string theory provides those degrees of freedom~\rcite{Martinec:2015pfa,Martinec:2019wzw,Martinec:2024emf}, and that the appropriate effective bulk description of black holes should involve bulk supergravity (and perturbative strings) coupled to the Hagedorn phase of little string theory at the would-be horizon~-- in other words, that the fivebranes constitute the ``membrane'' of the membrane paradigm~\rcite{Price:1986yy}.  

Now, according to the effective supergravity theory, all of this horizon-scale structure should collapse into a singularity.
There needs to be a mechanism that maintains such a structure supported at the horizon scale in order that black holes radiate information as ordinary bodies do.  
It has been proposed that having the microstates self-consistently available at the horizon scale implies a breakdown of classical gravity at that scale~\rcite{Kraus:2015zda}; perhaps this idea provides a heuristic (if somewhat circular) rationale. 
Perhaps related is the fact that, as discussed in~\rcite{Martinec:2014gka,Martinec:2019wzw,Martinec:2021vpk}, the little string is effectively always at its correspondence point~\rcite{Horowitz:1996nw}~-- its effective tension $(n_5\alpha')^{-1}$ is equal to the $AdS_3$ curvature.  The fundamental string in such a situation (\ie~curvature equals string tension) floats near the $AdS_3$ boundary (its radial wavefunction is rigid~\rcite{Eberhardt:2018ouy}, and correlation functions localize on worldsheets supported out at the $AdS_3$ boundary~\rcite{Gaberdiel:2020ycd,Dei:2022pkr}).  A similar story holds for the throat of a single, isolated fivebrane~\rcite{Dei:2025ilx}.  Perhaps the little string floats as well, out to the radial scale beyond which it becomes confined, \ie\ the ``horizon'' scale.  This mechanism might then explain how the stringy matter that makes up black holes manages to evade the singularity theorems of general relativity~-- the idea is that geometry is not the dominant effect when a closed trapped surface wants to form, but instead the wavefunction of stringy matter has support out to that surface.

What is clear from our analysis here is that, at the moment of near-extremal horizon formation, degrees of freedom beyond supergravity become available, at least in the adiabatic collapse of near-BPS fivebrane stars, and so supergravity alone is not the correct effective theory.  Unitarity demands that these degrees of freedom remain available, and we expect that their incorporation into the effective bulk description would lead to a qualitative if not quantitative picture of black hole microstructure and its unitary evolution, from the bulk perspective.

So what does the black hole consist of in string theory?  
What we know thus far suggests a sort of ``bag model'' of black holes, in analogy to a qualitative picture of hadronic structure developed in the early days of QCD~\rcite{Chodos:1974je,Chodos:1974pn}, in which the exterior of a hadron consists of the confining QCD vacuum, and the interior is a nearly free gas of deconfined constituents (even though QCD is strongly coupled on the scale of the hadron).  Indeed, the fact that the emergence of black holes in the spectrum of $AdS_5\times\bS^5$ is associated to the deconfinement transition in $\cN=4$ SYM~\rcite{Witten:1998zw}, and that supergravitons in the bulk are associated to single-trace operators, supports the idea that the horizon is such a phase boundary, outside of which the nonabelian constituents are effectively confined, and inside which they are liberated and dominate the dynamics.  That the interior constituents are in some respects approximately free is suggested by the interpretation of general $\bT^5$ and $\bT^6$-compactified supergravity black hole solutions in~\rcite{Horowitz:1996ay,Horowitz:1996ac} (see also~\rcite{Mathur:2005ai}), in which the black hole is composed of brane-antibrane constituents whose energetic cost is essentially additive in the number of constituents.

This picture of a black hole interior cannot be solely that of vacuum solutions of general relativity.
The inner horizon, even in general relativity, is expected to be the locus of large tidal forces for an infalling probe~\rcite{Marolf:2011dj,Horowitz:2022leb}, and the point where the geometry caps off.
For this reason, it is natural to suspect that in a complete theory, there is structure at the quantum scale of an extremal horizon, where inner and outer horizons coincide, which probes go ``splat'' upon.
It is less clear whether the outer ``horizon'' (the surface of the ultracompact object) is a hard surface away from extremality, that probes go ``splat'' upon.  Non-extremality separates the inner and outer horizon (to the extent a geometrical picture of the black hole interior exists).
If the surface of the non-extremal, collapsed brane star is simply a crossover between confined and deconfined phases of the little string, then pursuing the bag model analogy, a highly blueshifted infalling probe may simply make a ``jet'' of fractionated constituents as it enters the star.  Due to that fractionation, the probe is completely trapped once it passes the ``horizon'' (or more generally passes inside a ``trapped surface'' which extends out beyond the star's surface to meet it).  It necessarily thermalizes into the non-abelian soup in the interior, perhaps suddenly upon reaching the inner horizon, to be re-emitted as Hawking radiation.  The region between inner and outer ``horizons'', if it exists, would simply be the ``thermal atmosphere'' of the little strings (see sections 2 and 6.2 of~\rcite{Martinec:2014gka}).  In accordance with the fuzzball proposal, Hawking radiation would be the process of assembling singlet fundamental strings from their fractionated constituents in this atmosphere, and emitting them from the outer surface of the ultracompact object.   

If little strings behave like fundamental strings, they might not impart large transverse momentum to an infalling probe when it hits this atmosphere, the evolution consisting of a large number of (little) string-scale momentum transfers rather than a small number of large momentum transfers (the little string scale is the ambient curvature scale in $AdS_3\times \bS^3$, and so the natural scale of tidal forces on a probe).  The infall would turn a singlet excitation into a collimated jet of fractionated constituents once it crosses a trapped surface, which for a time could maintain something approximating the coherence of a single particle.   
This scenario provides a concrete realization of the idea of ``fuzzball complementarity''~\rcite{Mathur:2012jk,Mathur:2013gua}.  Central to this idea is that Hawking radiation does {\it not} result from a vacuum pair creation process in an effective field theory containing only supergravity~-- because that's the effective field theory of the confined phase, that ignores and omits the entropic degrees of freedom of the deconfined phase.  A necessary property of the putative interior here needs to be that gravity is emergent from little string theory but not the main driver of little string dynamics, otherwise the consequences of trapped surfaces are the singularity theorems of general relativity and the information paradox.  If on the other hand, horizons are simply the places where little string dynamics takes over, and the little string ``floats'' because it is at its correspondence point, then the information paradox is resolved by string theory and its extended objects. 

Of course, it is also possible that there is no such interior, and geometry effectively caps off at the outer horizon, in which case the fivebrane fuzzball provides some realization of a ``firewall'' structure.  The geometric features of a smooth interior would then be simply the result of analytically continuing a solution of the effective field theory of the confined phase past a phase boundary, where supergravity is not the correct effective field theory, and none of its features survive in any approximation.

The layout of the paper is as follows.  In section~\ref{sec:review}, we review the results of~\rcite{Martinec:2024emf}, which derives the two-charge NS5-P supertube solutions using an effective action formalism in which the bulk type II supergravity action is coupled to the worldvolume action of NS5-branes~\rcite{Eyras:1998hn}.  We also review the black fivebrane and BTZ geometries that the fivebrane star approximates well away from the source locus, and exhibit the scaling limits that isolate (1) the fivebrane throat, decoupling the asymptotically flat region of spacetime; (2) the $AdS_3$ attractor geometry of strings and fivebranes; and (3) the $AdS_2$ attractor geometry of the very near-horizon region.
We also discuss the various quantum scales associated to black holes and brane stars.

Then, in section~\ref{sec:STsolns}, we add the fundamental string action to the mix, and derive classical BPS solutions having both NS5-P and F1-P sources.  The effective action, evaluated on the supergravity solution sourced by the branes, yields in section~\ref{sec:joint} a joint effective action for the strings and fivebranes through which they interact with one another.  This effective action
reproduces and extends a result of~\rcite{Papadopoulos:2000hb} for unexcited strings and fivebranes (\ie~having only center-of-mass motion).  It has a form familiar from other contexts, where a sigma model on the moduli space of BPS configurations governs the leading near-BPS dynamics (see for example~\rcite{Manton:1981mp,Douglas:1996yp}).
In section~\ref{sec:decoupling}, we take the fivebrane decoupling limit~\eqref{decoupling} of these results.

In section~\ref{sec:superstrata}, we consider the simple example of concentric, circular string and fivebrane sources in which only a single momentum mode is excited on each, and compare the resulting geometry with the WKB limit of round superstrata recently analyzed in~\rcite{Bena:2025uyg}.

Section~\ref{sec:discussion} concludes with a discussion of the near-BPS regime.  Among the topics are the collective breathing modes of the star, which play a role similar to the Schwarzian modes of near-extremal black holes; 
the chaotic dynamics governed by the joint effective action; absorption and emission of radiation from the effective string sources; the fate of the star when perturbed~-- specifically, a possible mechanism of collapse toward the black hole regime; and the onset of the deconfined phase of little strings.  We also preview forthcoming work on ensemble averages of our solutions.

Several Appendices contain supplementary material.  Appendix~\ref{app:details} reviews several duality relations used in our analysis.  Appendix~\ref{app:soln sum} collects in one place all the various coefficients in the supergravity fields of our solution.  Appendix~\ref{app:GWZW} contains an analysis of the exactly solvable worldsheet theory for strings in a round NS5-P supertube background, which links the helical source solutions of section~\ref{sec:helices} and the effective single-mode superstrata
discussed in section~\ref{sec:singlemode},  Appendix~\ref{app:om-F aves} contains details of the averaging procedure for metric coefficients of the circular source solution.


\section{Review of previous results} 
\label{sec:review}

To begin, let us review section 3.1 of paper III~\rcite{Martinec:2024emf}, where an effective action approach was used to derive the Lunin-Mathur solutions in Type IIA supergravity compactified on $T^4 \times S^1 _{\tilde{y}}$, coupled to explicit fivebrane sources. We restrict the discussion to the four bosonic, transverse excitations of the fivebranes, turning off all fermionic modes, as well as the R-R internal scalar and the chiral two-form of the IIA NS5-brane. 

A point of notation: In the following, we denote worldsheet fields of strings and fivebranes with sans-serif letters ($\sfF,\sfX$), while spacetime fields are largely denoted by math italic letters (\eg\ $G,B,Z$, \etc).  We also denote the four coordinates parametrizing the space transverse to the fivebranes by $\sfx$.

\subsection{Two-charge BPS solutions: NS5-P}
\label{sec:NS5-P}

The effective action approach has the advantage that it determines the residues of the harmonic forms that parametrize the supersymmetric ansatz that preserves eight supercharges:%
\footnote{Here we use the notation in common use to describe three-charge BPS backgrounds.  The notation used in our previous work is related by
$$
Z_2\to \sfH_5
~~,~~~~
\cF\to -\sfH_p
~~,~~~~
\omega\to -\sfA
~~,~~~~
a_1-\omega \to \sfB^{(1)}
~~,~~~~
\gamma_2\to \sfb_2 ~
$$
}
\begin{align}
\label{NS5P Ansatz}
 ds^2  &= -2du\,dv-2\omega_i\, d\sfx^i dv  -\cF\, dv^2 + Z_2\, d\sfx ^2 + ds^2(\bT^4)
\nonumber\\[.1cm]
 e^{2\Phi} &= g_s^2\,Z_2  
\\[.1cm]
 B^{(2)} &= \big(a_1-\omega\big) \wedge dv + \gamma_2 ~,
\nn
\end{align}
where $\sfx^i$, $i=1,...,4$ parametrize the warped $\bR^4$ transverse to the fivebranes, and the fields satisfy the duality relations%
\footnote{The duality relation for $\omega$ given here corrects the expression in~\rcite{Martinec:2024emf}, Eq.~(3.36).  That expression holds when the sources are smeared over $v$, as they are in NS5-F1 frame solutions.  For unsmeared sources, the additional term $\partial_v\gamma_2$ appears on the R.H.S.}
\begin{align}
d\omega+*_4d\omega =  da_1+ \partial_v\gamma_2
 ~~,~~~~ 
 d\gamma_2 = *_4dZ_2~.
\end{align}
The Hodge star $*_4$~is in these relations is taken with respect to a flat four-dimensional transverse space, and
\be
\label{uvdef}
u = \frac{1}{\sqrt{2}}(t-y) 
~~,~~~~
v = \frac{1}{\sqrt{2}}(t+y)   ~.
\ee
We parametrize $\bT^4$ by coordinates $\sfz^I$, $I=1,...,4$.

We define 
\be
\label{kappa0}
\kappa_{0}^2 = 2^6 \pi^7 g_s ^2 (\alpha')^4 ~,
\ee
where $g_s$ is the asymptotic string coupling.  Also, the six-dimensional gravitational coupling  
\be
\label{kappa6}
\kappa_{6}^2 = \frac{\kappa_{0}^2}{(2\pi)^4 V_4} = \frac{4\pi^3g_s^2(\alpha')^4}{V_4}
\ee
is sometimes useful, where $(2\pi)^4V_4$ is the volume of the four-torus compactification.

The bosonic part of the action of Type II supergravity reduced along $\bT^4$, in the absence of R-R fields and in the string frame, is given by
\begin{align}
\label{SUGRA}
\cI =~& \frac{1}{2\kappa_6^2} \int d^6 x \sqrt{-G}\, e^{-2\Phi}\left( R+ 4 G^{\mu \nu} \partial_{\mu} \Phi \partial_{\nu} \Phi  -\frac{1}{12} H_{\alpha \beta \gamma} H^{\alpha \beta \gamma} \right)~.
\end{align}
The NS5-brane worldvolume effective action contains a bosonic part, which consists of
\begin{equation}
\label{OriginalNS5Action}
 \cS_{\text{source}} = -T_{5} \int d^6 \xi \Big( e^{-2\Phi} \sqrt{-G} + \widetilde{B} \Big)~, 
\end{equation}
where the tension of the fivebrane is given by
\begin{equation}
 T_{5} = \frac{2\pi^2 \alpha'}{\kappa_{0}^2}~,
\end{equation}
and the bulk fields $G,\widetilde B$, and $\Phi$ are pulled back to the worldvolume.
The potential $\widetilde B$ is the magnetic dual of $B$,
\begin{equation}
\label{Hodge}
	*d\widetilde{B}^{(6)} = e^{-2\Phi} dB^{(2)}~.
\end{equation}
Assuming that no fields depend on the torus directions, one can also dimensionally reduce the source action~(\ref{OriginalNS5Action}) on~$\bT^4$. Further introducing an intrinsic metric brings about an effective string action of Polyakov form for a magnetic string in a 6d target space:
\begin{equation}
\label{magstring}
 \cS_{\text{source}} = -\frac{\tau_{\sst\rm NS5}}{2}\int d^2 \sigma \sqrt{-\gamma} \Big( \gamma^{ab} e^{-2\Phi} G_{\mu \nu} \partial_{a} \sfF^{\mu} \partial_b \sfF^{\nu} +\epsilon^{ab} \widetilde{B}_{\mu \nu}\partial_{a} \sfF^{\mu} \partial_b \sfF^{\nu}\Big)~.  
\end{equation}
Here
\begin{equation}
\label{tauNS5}
 \tau_{\sst\rm NS5} = \frac{2\pi^2 \alpha'}{\kappa_{6}^2} =\frac{1}{2\pi \mu^2} 
~~,~~~~~
\mu^2\equiv 
\frac{g_s^2 (\alpha')^3}{V_4}
\end{equation}
is the tension of the effective 6d magnetic string obtained by compactifying the fivebrane on $\bT^4$.
Note that $\mu$ is the natural scale of quantum ``fuzziness'' of the fivebranes sourcing such a geometry, just as $\ell_\str=\sqrt{\alpha'}$ is the natural scale of fuzziness of the quantized fundamental string.

For the purpose of deriving the bulk supergravity equations of motion with sources, this NS5-brane effective action can be rewritten as a six-dimensional bulk integral involving the four dimensional transverse space, the time direction, and the dimension along the circle $S^1 _{\tilde{y}}$: 
\begin{equation}
\label{SourceAction}
 \cS_{\text{source}} = -\frac{\tau_{\sst\rm NS5}}{2}\int d^2\sigma \sqrt{-\gamma} \Big( \gamma^{ab} e^{-2\Phi} G_{\mu \nu} \partial_{a} \sfF^{\mu} \partial_b \sfF^{\nu} +\epsilon^{ab} \widetilde{B}_{\mu \nu}\partial_{a} \sfF^{\mu} \partial_b \sfF^{\nu}\Big) \delta^4 (\sfx -\sfF)~.  
\end{equation}
The equations of motion for BPS transverse excitations $\sfF^{i}$ are solved by any purely left-moving $\sfF^i (v)$. The component $\sfF^v$ is a free field; it is convenient to choose a gauge where it is given by $\sfF^v=v=\sigma^+$. The field $\sfF^u$ is a sum of left- and right-moving components; the right-moving part can be set to $u=\sigma^-$ by the remaining reparametrization invariance, while the left-moving piece is determined by the Virasoro constraints.  The $n_5$ fivebranes are taken to be bound together into a single fivebrane wrapping $n_5$ times around $\bS^1_\ytil$.

The next step is to plug the ansatz of Eqs.~(\ref{NS5P Ansatz}) into the equations of motion derived from the sum of the brane and supergravity effective actions (\ref{SUGRA})+(\ref{SourceAction}).
The determination of the harmonic functions and harmonic one-form can be organized into layers. 
The first layer is the dilaton equation, which implies that
\begin{equation}
\label{H5soln}
  Z_2 (v,\sfx) =  1+ \alpha'\sum_{\sfm =1} ^{\nfive} \frac{1}{|\sfx-\sfF_\sfm (v)|^2}~.
\end{equation}
The second layer uses the previous result from the dilaton equation; the $v$-$i$ metric equations then imply that
\begin{equation}
\label{Asoln}
 \omega_i (v,\sfx)=  \alpha'\sum_{\sfm =1} ^{\nfive} \frac{\partial_v \sfF_{\sfm i}}{|\sfx-\sfF_\sfm (v)|^2}~.
\end{equation}
Finally, the third layer uses the these equations as well as the $v$-$v$ metric equation of motion, and the Virasoro constraint from varying the $\gamma_{vv}$ intrinsic metric component. These equations determine the momentum harmonic function $\cF$:
\begin{equation}
\label{Hpsoln}
  \cF (v,\sfx)=  -\alpha'\sum_{\sfm =1} ^{\nfive} \frac{|\partial_v \sfF_{\sfm }|^2}{|\sfx-\sfF_\sfm (v)|^2}~.
\end{equation}

In the typical solution, the highly excited effective magnetic string profile $\sfF(v)$ is a random walk of size
\be
\label{walksize}
|\delta\sfx| \sim \mu\, \big(n_p^\perp \nfive\big)^{1/4}
~,
\ee
where $n_p^\perp\nfive$ is the total excitation level of the scalars $\sfF$, and the magnetic string length $\mu$ is defined in~\eqref{tauNS5}. 
This scale can be understood as follows: The number of steps in the random walk is the square root of the total excitation level,%
\footnote{Treating the left-moving excitations as a 1d relativistic gas in a box of size $n_5R_y$, whose temperature is $T_L$, one has $P=\frac{n_5n_p^\perp}{n_5R_y}=c(n_5R_y)T_L^2$ (where $c$ is the number of species); the thermal wavelength is thus 
$\lambda_L=n_5R_y[c/(n_p^\perp n_5)]^{1/2}$.  There are thus of order $\sqrt{n_5n_p^\perp }$ thermal wavelengths in the box of size $n_5R_y$.  The scalars are coherent over a thermal wavelength, and so there are of order this many steps in their random walk.}
and the size of the walk is the square root of the number of steps; the dimensionful scale $\mu$ is the string length set by the magnetic string tension.

These solutions are self-consistently weakly coupled if the wiggly fivebrane source is well-separated from itself.  Consider a D-string stretched between strands of the fivebrane.  In the DBI action, the warp factor $Z_2$ in the metric and dilaton cancel one another, and the factors of the asymptotic string coupling $g_s$ cancel between the explicit factor in the dilaton~\eqref{NS5P Ansatz} and a corresponding factor in the string scale $\mu$ appearing in the size~\eqref{walksize} of the bound state.  As we show below, the end result is that the mass of such a D-string stretching between strands $i,j$ of the fivebrane is simply $M\approx|\hat\sfF_i-\hat\sfF_j|/(2\pi \alpha')$ in terms of the appropriately scaled fivebrane source locations~\eqref{decoupling}.  

String perturbation theory is valid when such excitations are parametrically much heavier than those of fundamental strings, \ie\ when the separation of the fivebranes is much larger than the magnetic string scale $\mu$. 
The probability distribution $\cP(d)$ of strand separation $d$ was computed in~\rcite{Martinec:2023xvf}.  It has the approximate form
\be
\cP(d) \approx \Big[\frac{3}{2\pi^2 N}\Big]^{\frac12} x^2 e^{-x^2}
~~,~~~~
x=\Big[\frac{3}{2\pi^2N}\Big]^{\!\frac14} \frac{d}{\mu}  
~~,~~~~
N=n_pn_5
~.
\ee
Then the likelihood to find two strands within a distance $d\approx\mu$ in a typical two-charge BPS state scales like $1/N$, and can be self-consistently ignored in the large $N$ limit.

\subsection{Three-charge BPS solutions}
\label{sec:blackness1}

In this work we are interested in generalizing the above solutions by the addition of sources carrying a third BPS charge.  BPS solutions in Type II supergravity carrying NS5-F1-P charges, and having vanishing R-R fields, can be written in the form
\begin{align}
\label{AnsatzIIB}
	ds_{10} ^{2} &= -\frac{2}{Z_1}(dv+\beta)\left[du + \omega + \frac{\mathcal{F}}{2}(dv+\beta)\right]+Z_2 \,ds^2 (\mathcal{B})+ds^2 (\mathcal{M})~,\nn \\
	B^{(2)}  &=du\wedge dv -\frac{1}{Z_1}(du+\omega) \wedge (dv+\beta) +a_1 \wedge (dv+\beta)+\gamma_2 ~,\nn\\
	e^{2\Phi} &= g_s^2\frac{Z_2 }{Z_1} ~, \\
	\widetilde{B}^{(6)} &= \frac{1}{g_s^2}\,\text{vol}(\mathcal{M}) \wedge \left[-\frac{1}{Z_2}(du+\omega) \wedge (dv+\beta) +a_2 \wedge (dv+\beta)+\gamma_1\right]~. \nn
\end{align}
Here ${\rm vol}(\cM)$ is the volume form on the compactification $\cM=\bT^4$, and
we have introduced a new harmonic function $Z_1$ that encodes the string winding charge, as well as an associated two-form $\gamma_{1}$ and one-form $a_2$ in $\widetilde{B}$.
We take the transverse space $\cB$ to be $\bR^4$.%
\footnote{This form of the solution follows from an S-duality transformation on the more traditional D5-D1-P frame ansatz used in the literature, as we show in Appendix~\ref{app:details}.}  
With only NS-NS fields present, the same ansatz applies for both IIA and IIB duality frames, with the map that we write in Eq.~(\ref{Tduality map}).%
\footnote{In~\rcite{Martinec:2024emf}, R-R excitations (but not R-R charges) were also considered.  We ignore them here, as all the basic physics results can be achieved without them, and the calculational overhead is much reduced. }

\subsection{BPS black fivebranes and extremal BTZ: Scales and rescalings}
\label{sec:blackness2}

As an illustration, let us review the structure of the three-charge BPS black hole solution, which includes near horizon $AdS_3$ and $AdS_2$ regions~\cite{Strominger:1998yg}.  This solution exemplifies the scaling limit that decouples the asymptotically flat region from the near-source geometry of the fivebranes and strings.  Moreover, we will want to compare the horizonless geometries we construct below to the corresponding black hole carrying the same charges.  In the process, we can use this exercise to set some conventions.

In the F1-NS5 duality frame, the asymptotically flat BPS black fivebrane geometry, wrapped on $\bT^4\times \bS^1_y$ and carrying F1 and P charges, is given by
\begin{align}
\label{BTZ metric}
ds^2 &= \frac{1}{Z^\flat_1} \Big( -d\tunsc^2+d\yunsc^2 -\frac{1}{2} \cF^\flat\big(d\tunsc+d\yunsc\big)^2\Big) + Z^\flat_2 \, d\sfx^2 + ds^2_{\bT^4} 
\nn\\[.2cm]
&= \frac{\runsc^2}{\runsc^2+Q_1} \Big( -d\tunsc^2+d\yunsc^2 + \frac{Q_p}{\runsc^2}\big(d\tunsc+d\yunsc\big)^2\Big) + \Big(1+\frac{Q_5}{\runsc^2}\Big)\big(d\runsc^2 +\runsc^2d\Omega_3^{\,2}\big) + ds^2_{\bT^4}
\end{align}
where the subscript $\flat$ denotes coordinates in asymptotically flat spacetime, and
\be
\label{charges}
Q_5 = \nfive \alpha'
\quad,\qquad
Q_1 = \frac{\gstrsq \none (\alpha')^3}{V_4}
\quad,\qquad
Q_p = \frac{\gstrsq n_p (\alpha')^4}{V_4 R_y^2}~.
\ee
Here the $y$-circle has periodicity $2\pi R_y$ asymptotically, and the $\bT^4$ has volume $(2\pi)^4 V_4$.  The $B$-field and dilaton are given by 
\begin{align}
\label{Bfield}
B^{(2)} &= \bigg(1 - \frac{\runsc^2}{\runsc^2+Q_1}\bigg) \, dt_\flat\wedge dy_\flat + \gamma_2
\nn\\[.2cm]
e^{2\Phi} &= g_s^2\, \frac{\runsc^2+Q_5}{\runsc^2+Q_1}  ~,
\end{align}
where $d\gamma_2= -Q_5{\rm vol}(\Omega_3)$. 
To simplify the analysis, we focus on solutions carrying no angular momentum on $\bS^3$, and thus have set $\beta=\omega=0$ in~\eqref{AnsatzIIB}.

\paragraph{The fivebrane decoupling limit:}
To decouple the fivebrane throat from the asymptotically flat region of spacetime, we define the rescaled coordinates
\be
\label{rhat defn}
\tunsc = R_y \,t
~~,~~~~
\yunsc = R_y \,y  
~~,~~~~
\runsc = a\,\rhat
~~,~~~~
a^2 =\frac{g_s^2 (\alpha')^4}{R_y^2 V_4}\,\nfive\none  = \frac{Q_1Q_5}{R_y^2}
\ee
(in particular, $y\sim y+2\pi$).
The fivebrane decoupling limit then takes $\gstr\to 0$ at fixed $\rhat$.  In this limit, the dilaton is approximately linear in the region $\rhat\gg \frac{R_y}{\sqrt{\nfive \alpha'}}$, and approximately constant at smaller radius.  The metric~\eqref{BTZ metric} becomes
\begin{align}
\label{NS5 3chg}
ds^2_{\sst\rm NS5} = \frac{{\nfive \alpha'}R_y^2\,\rhat^2}{{\nfive \alpha'}\rhat^2+ {R_y^2}}\bigg[ -dt^2+dy^2+\frac{n_p}{\none\nfive}\frac{1}{\rhat^2}\big(dt+dy\big)^2\bigg]+\nfive \alpha' \bigg[\Big(\frac{d\rhat}{\rhat}\Big)^2+d\Omega_3^2\bigg]+ds^2_{\bT^4} ~,
\end{align}
while the $B$-field and dilaton become
\begin{align}
B^{(2)} &= R_y^2 \bigg( 1 -\frac{\nfive \alpha' r^2}{  \nfive \alpha' r^2+R_y^2} \bigg) dt\wedge dy +\gamma_2
\nn\\[.2cm]
e^{2\Phi} &= \frac{\nfive V_4R_y^2}{\none (\alpha')^2} \,\frac{1}{{\nfive \alpha'}\rhat^2+{R_y^2}} ~. 
\end{align}

The $y$-circle has the constant proper size $R_y$ in the linear dilaton regime at large $r$, and the angular $\bS^3$ has radius $\sqrt{\nfive \alpha'}$.
When we get to the radial scale $\rhat = \frac{R_y}{\sqrt{\nfive \alpha'}}\equiv \rhat_{\sst\rm AdS3}$, the geometry rolls over to $AdS_3$, and the dilaton stops running.

Note that in this decoupling limit, all factors of $g_s$ have disappeared.  From the perspective of someone sitting at fixed $r$, the place where the throat opens out onto asymptotically flat spacetime is pushed off to infinite $r$, and the asymptotic throat geometry is a cylinder $\bR\times\bS^3$, with an angular three-sphere of fixed size $\nfive\alpha'$, and a dilaton linear in the logarithmic radial coordinate $\rho=\log r$.    

\paragraph{The $\bf AdS_3$ decoupling limit:}
One can exhibit the $AdS_3$ geometry by taking the further decoupling limit $R_y \to \infty$ with fixed $t,y$ and~$r$. In this limit, the line element reads
\begin{align}
\label{AdS3}
ds^2_{\sst\rm AdS3} = \nfive \alpha' \Big( \rhat^2\big(-dt^2+dy^2\big) +\frac{d\rhat ^2}{\rhat^2 }+d\Omega_3^2\Big)+\frac{n_p }{\none}\alpha'\big(dt+dy\big)^2+ds^2_{\bT^4} ~.
\end{align}
The dilaton is fixed at the value
\begin{align}
 e^{2\Phi} = \frac{ \nfive V_4 }{\none (\alpha')^2}~.
\end{align}
However, for the most part we will keep $R_y$ finite; then the $AdS_3$ geometry only extends out to $\nfive\alpha' r^2\sim R_y^2$, beyond which it rolls over to the linear dilaton asymptotics.

\paragraph{The $\bf AdS_2$ decoupling limit:}
Below the scale $\rhat\approx \rhat_{\sst\rm AdS3}$, the $y$-circle is the $AdS_3$ aximuthal direction, and so starts shrinking with decreasing radius, until it stabilizes at the momentum charge radius 
\be
\label{rads2}
\rhat \sim \Big(\frac{n_p}{\none\nfive}\Big)^{1/2} \equiv \rhat_{\!\sst\rm AdS2} ~,
\ee
where its proper size is of order 
\be
\label{Ryhor}
\frac{R_{y,{\it hor}}}{\sqrt{\alpha'}} = \sqrt{\frac{n_p}{\none}} ~.
\ee
Let us suppose that $n_p \gg \none$, so that the size of the $y$-circle at small $r$ is larger than the string scale; later, we will consider the opposite regime $n_p \ll \none$.
At this radial scale the geometry~\eqref{AdS3} rolls over to a constant size $y$-circle, fibered over (nearly) $AdS_2$.
A canonical parametrization of the $AdS_2$ line element arises upon defining new coordinates (denoted by sans-serif font) 
\begin{equation}
\label{AdS2coords}
\sfr \equiv 2\sqrt{\frac{\nfive \none}{n_p}}\, r^2
~~,~~~~
\sft = t
~~,
\end{equation} 
so that the geometry asymptotes to
\be
\label{AdS2one}
ds^2_{\sst\rm AdS2} \sim \frac{\nfive \alpha'}{ 4} \Big( -\sfr^2 d\sft^2 + \frac{d\sfr^2}{\sfr^2} \Big) + \frac{n_p}{\none}\alpha'(dy+d\sft)^2 +\nfive \alpha' d\Omega_3 ^2 + ds_{\bT^4} ^2
\ee
for $\sfr \ll \rhat_{\!\sst\rm AdS2}$.  
The geometry looks like figure~\ref{fig:3chg}, except instead of capping off at some finite depth, the $AdS_2$ throat extends to infinite depth.  There is a finite size extremal horizon at $\rhat=0$, where the $y$-circle has proper radius~\eqref{Ryhor} and the three-sphere has radius $\sqrt{\nfive \alpha'}$.  The horizon area in Planck units yields the extremal BTZ entropy $S_{\it BH} = 2\pi\sqrt{\nfive \none n_p\vphantom{l}}$.

Now consider the possibility that $n_p \ll \none$.  We see from~\eqref{Ryhor} that at some point before we reach the $AdS_2$ throat, we need to switch to a T-dual description of the $y$-circle.
Indeed, in the geometry~\eqref{AdS3}, the size of this circle becomes comparable to the string length at the radial scale
\begin{equation}
\label{rd}
 r_d \equiv \frac{1}{\sqrt{\nfive}}~,
\end{equation}
and when $r< r_d$, the use of effective field theory requires working with the T-dual background, as discussed in~\cite{Martinec:2023xvf}.  
The resulting T-dual metric
can be written (in the NS5 decoupling limit)
\begin{align}
\label{Tdual 3chg}
d\tilde s^2_{\sst\rm NS5} &= -\frac{{ R_\ytil^2\,}\nfive^2 r^4\, d{\tilde t}^{2}}{\big(\nfive r^2\tight+\frac{n_p}{\none}\big)\big(\nfive r^2\tight+\frac{R_y^2}{\alpha'}\big)} 
+ { R_\ytil^2}\frac{\big(\nfive r^2\tight+\frac{R_y^2}{\alpha'}\big)}{\big(\nfive r^2\tight+\frac{n_p}{\none}\big)}\bigg(\!d\ytil + \frac{ R_y ^2 d\tilde t}{\nfive \alpha' r^2\tight+R_y^2}\bigg)^{\!2}
\!\!+\nfive \alpha' \bigg(\frac{dr^2}{r^2} \!+ d\Omega_3 ^2 \bigg) \!+ ds_{\bT^4}^2
~,
\end{align}
where $R_\ytil=\alpha'/R_y$, and $R_\ytil d\tilde t = R_y dt$.
At large $r$ this is a flat cylinder with the $\ytil$ circle having proper size $R_\ytil$, though of course the metric is only appropriate in the regime $r<r_d$, where the proper size of the $\ytil$-circle is growing with decreasing radius, until one reaches the scale $r\approx r_{\sst\rm AdS2}$.  For $r\ll r_{\sst\rm AdS2}$, one has the $\ytil$ circle fibered over $AdS_2$ similar to~\eqref{AdS2one}
\be
\label{AdS2two}
d\tilde s^2_{\sst\rm AdS2} \sim \frac{\nfive \alpha'}4 \Big( -\tilde\sfr^2 d\tilde\sft^2 + \frac{d\tilde\sfr^2}{\tilde\sfr^2} \Big) + \frac{\none}{n_p}\alpha'(d\tilde y+d\tilde\sft)^2 +\nfive \alpha' d\Omega_3 ^2 + ds_{\bT^4} ^2 ~.
\ee
in terms of the coordinates
\begin{equation}
\label{AdS2tilcoords}
\tilde\sfr \equiv \frac{2R_\ytil^2}{\alpha'}\sqrt{\frac{\nfive \none}{n_p}}\, r^2
~~,~~~~
\tilde\sft = \tilde t
~~.
\end{equation}
The proper size of the $\ytil$ circle in this regime is
\be
\label{Ryhor2}
\frac{R_{\ytil,{\it hor}}}{\sqrt{\alpha'}} = \sqrt{\frac{\none}{n_p}} ~.
\ee

The string coupling in this T-dualized geometry also grows with decreasing radius
\begin{equation}
 e^{2\Phi} = \frac{V_4}{(\alpha')^2 \none \big(r^2 +\frac{n_p}{\none \nfive}\big)}~;
\end{equation}
the dilaton saturates in the IR at
\begin{align}
 e^{2\Phi} \approx \frac{\nfive V_4}{n_p (\alpha')^2}~,
\end{align}
which we assume to be small.

\paragraph{The quantum scale:}
There is one more scale that arises in the quantum theory~-- the scale that characterizes the breakdown of black hole thermodynamics due to quantum effects~\rcite{Preskill:1991tb,Iliesiu:2020qvm}.  As the temperature is lowered to zero, the black hole's quantum fluctuations start to compete with thermal fluctuations when the energy $T$ of a typical thermal (Hawking) quantum is of order the available energy above extremality $M-M_{\rm ext}$.  We can regard the horizon radius associated to this scale as a kind of ``stretched horizon'', the scale at which quantum fluctuations in the (near) extremal geometry make the classical solution unreliable.  In particular, the infinite throat of the classical extremal solution, \eg~\eqref{AdS3}, should be cut off at this scale.   

In canonical Schwarzschild/Kerr-like coordinates (denoted here by typewriter font), the BTZ solution reads 
\be
ds^2_\btz = \ell_{\sst\rm AdS}^2\bigg[-f(\ttr) d\ttt^2+\frac{d\ttr^2}{f(\ttr)} + \ttr^2\Big(d\tty-\frac{\ttr_+\ttr_-}{\ttr^2} d\ttt\Big)^2\bigg]
~~,~~~~
f(\ttr) = \frac{(\ttr^2-\ttr_+^2)(\ttr^2-\ttr_-^2)}{\ttr^2}
\ee
The mass, $\tty$-momentum and temperature of the black hole are given (in $AdS_3$ units) by
\be
M = \frac N2\big(\ttr_+^2+\ttr_-^2\big)
~~,~~~~
J = n_p = N\ttr_+\ttr_-
~~,~~~~
T = \frac{\ttr_+^2-\ttr_-^2}{2\pi \ttr_+}
\ee
where $N=\none\nfive$.  We also note that the extremal mass for fixed $J$ is given by $M_{\rm ext} = J \equiv N\ttr_{\rm ext}^2$, or $\ttr_{\rm ext}^2 = \ttr_+\ttr_-$.  Then one has the quantum scale
\be
\label{rqbtz}
M-M_{\rm ext} = T
~~\Longrightarrow~~
\ttr_+ = \ttr_{\rm ext} + \frac{1}{\pi N} \equiv \ttr_{\qu}  ~.
\ee

Now we note that the relation between the radial coordinate $r$ of~\eqref{AdS3} and the BTZ radial coordinate $\ttr$ is (see for instance the review~\rcite{Peet:2000hn})
\be
r^2 = \ttr^2-\ttr_-^2 ~,
\ee
and so
\be
r^2_{\qu} = 4\ttr_\ext (\ttr_{\qu}-\ttr_\ext) = \frac{4}{\pi}\frac{n_p^{1/2}}{N^{3/2}}  ~.
\ee 
As a result, the top of the $AdS_2$ throat at $r^2=n_p/N$, Eq.~\eqref{rads2}, is at $\ttr=\sqrt2\,\ttr_\ext = \sqrt{2n_p/N}$.

The proper distance from the quantum scale $r_\qu$ to the top $r_{\sst\rm AdS2}$ of the $AdS_2$ throat is then (in units of the $AdS_2$ radius)
\be
\label{depth}
\frac{L_{\rm throat}}{R_{\rm AdS2}} = 2\int_{r_\qu}^{r_{\sst\rm AdS2}}\! \frac{dr}{r} \approx 2\bigg[\log\Big( \frac{n_p^{1/2}}{N^{1/2}} \Big) - \log\Big( \frac{n_p^{1/4}}{N^{3/4}} \Big)\bigg] = \log\big(\sqrt{\nfive \none n_p\vphantom{l}}\,\big) \approx \log\big(S_0\big) ~.
\ee

Another route to this result begins with the effective 2d Jackiw–Teitelboim (JT) gravity theory that captures the gravitational dynamics of a nearly-$AdS_2$ black hole throat.  As an $AdS_2$ black hole nears extremality, quantum fluctuations of JT gravity's ``Schwarzian modes'' start to dominate.  An analysis of the near-BPS dynamics as well as the wavefunction of the BPS ground state of JT supergravity~\rcite{Lin:2022rzw} corroborates the estimate~\eqref{depth} of the length of the extremal throat in the quantum theory.

\subsection{Scales in superstrata}
\label{sec:strata scales}

Horizonless 3-charge {\it superstratum} geometries~\rcite{Bena:2015bea} can have long $AdS_2$ throats~\rcite{Bena:2017xbt}, and comprise a large class of examples of fivebrane stars, as we will explore in section~\ref{sec:superstrata}.

\begin{figure}[ht]
\centering
\includegraphics[scale=0.6]{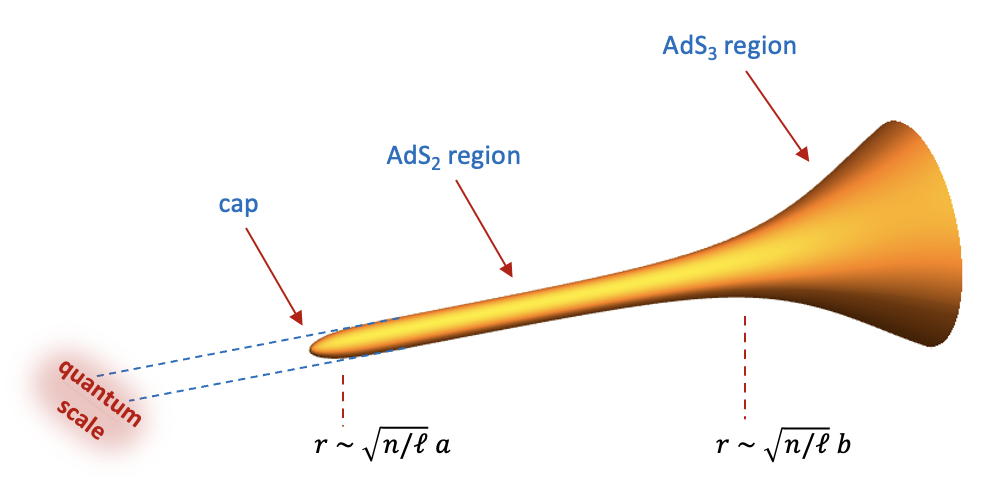}
\caption{\it Sketch of the $(\ell,0,n)$ superstratum geometry.  
The geometry has an $AdS_2$ throat beginning at $r\sim \sqrt{n/\ell}\,b$, supported by a momentum wave at $r\sim \sqrt{n/\ell}\,a$; inside the momentum wave, the geometry is approximately $AdS_3$ again down to $r=0$, where it caps off smoothly. }
\label{fig:k0n stratum}
\end{figure}

The superstratum geometries most studied in the literature have the characteristic structure depicted in figure~\ref{fig:k0n stratum}.  While they are highly coherent states having only a handful of underlying brane oscillation modes excited, and thus are far from typical, they do exhibit some of the basic features of the extremal BTZ geometry of the previous subsection.
%
%

The simplest ``$(\ell,0,n)$'' superstrata have a single coherent momentum mode excited on a background NS5-F1 supertube, characterized by two radial scales $a$ and $b$ in the coordinates $r$ of~\eqref{AdS3} (see for instance the discussion in~\rcite{Martinec:2020cml} around figure~1).  The $\bS^3$ angular momentum in the transverse space is given by $a^2\approx \frac{\mu^2\alpha'}{R_y^2}2J_{R}$, which sets the radius of gyration of the fivebrane supertube; the $\bS^1_y$ momentum charge is determined by $\frac{n}{\ell}\frac{b^2}{2}=Q_p$; also one has the constraint $a^2+\half b^2=\frac{\mu^2\alpha'}{R_y^2}N$.  The regime of long $AdS_2$ throats is $b\gg a$, in which one has 
\be
\frac{b^2}{a^2} \sim \frac{N}{J_R} ~.
\ee
The proper length of the $AdS_2$ region in units of the $AdS_2$ length scale is given approximately by (recalling the relation~\eqref{AdS2coords} between the canonical $AdS_3$ and $AdS_2$ radial coordinates) 
\be
\label{AdS2 len}
\frac{L_{ba}}{R_{\sst\rm AdS2}} = \log\Big(\frac{b^2}{a^2}\Big) \approx \log\Big(\frac{N}{J_R}\Big) ~.
\ee
As one passes inside the momentum wave at $r\sim a\sqrt{n/\ell}$, one passes from a three-charge geometry to that of the underlying two-charge NS5-F1 supertube, which is approximately $AdS_3$.
Thus the proper distance from the transition scale $r\sim a\sqrt{n/\ell}$ to the tip of the cap at $r=0$ is crudely approximated using the metric of global $AdS_3$, which at large $n,\ell$ yields 
\be
\frac{L_{\rm cap}}{R_{\sst\rm AdS2}}
\approx \int_0^{a\sqrt{n/\ell}}\! \frac{dr}{\sqrt{r^2+a^2}}
\approx \log \sqrt{n/\ell} ~.
\ee
All told, the proper length of the throat is
\be
\label{Lstratum}
\frac{L_\tot}{R_{\sst\rm AdS2}} = \frac{L_{ba}+L_{\rm cap}}{R_{\sst\rm AdS2}} \approx \log\Big(\frac{N\sqrt{n/\ell}}{J_R}\Big) = \log\Big(\frac{\sqrt{\none \nfive n_p\vphantom{d}}}{J_R}\Big) ~.
\ee
We see that as we dial down the radius of gyration $r_5$ of the fivebranes by decreasing their angular momentum $J_R$, the throat lengthens, and approaches that of the extremal quantum black hole throat~\eqref{depth} when $r_5^2=a^2\sim 1/N$.  This scale corresponds to $J_R\sim O(1)$, where indeed the quantum fluctuations of the transverse size of the fivebrane state are order one.%
\footnote{Supergravity solutions are coherent states that superpose states of fixed $J_R$, weighted by $a^{2J_R}$~\rcite{Bena:2017xbt}.  When $J_R\sim O(1)$, the fluctuations in $a$ are order one.}
Indeed, the radius of gyration of the fivebranes is the scale $\mu^2$ set by the magnetic string tension of the fivebranes, the natural width of their quantum wavefunction.

\subsection{Scales in supertubes}
\label{sec:ST scales}

We should also note a few other quantum scales, that arise in the analysis of two-charge states.  Suppose we partition the fivebrane excitation budget into the transverse excitations of the scalars $\sfF$ that described where the fivebrane sits in its transverse space, and the internal excitations of the fields $\mfa^{\sst AB}$ which are the BPS excitations of the internal gauge multiplet on the fivebranes, with total excitation levels $n_p^\perp$ and $n_p^\parallel$ respectively.
The classical analysis of the two-charge states using the magnetic effective string action~\eqref{magstring} is valid when the nonabelian excitations of the fivebrane are heavy and thus can be consistently decoupled.  But how heavy are they?  

One can estimate the mass of a D-string stretching between NS5-branes in type IIB string theory using the DBI effective action
\be
\cS_{\rm DBI} = \tau_{D1}\int\! e^{-\Phi}\sqrt{\det\big[G\tight+(B\tight-F)\big]} + \tau_{D1} \int\! C\wedge e^{B-F} ~.
\ee
The tension of the D1-brane is given by
\begin{equation}
   \tau_{D1} = \frac{1}{2\pi \alpha' g_s} \equiv \frac{1}{\ell_{na}^2} ~,
\end{equation}
and sets the scale for non-abelian fivebrane excitations.
For static fivebranes on their Coulomb branch, the geometry is given by~\eqref{NS5P Ansatz}, \eqref{H5soln}-\eqref{Hpsoln} with $v$-independent source locations $\sfF$.  In the DBI action, the warping of the transverse metric cancels against the dilaton, with the result that the mass is simply given by the flat space result
\be
\label{MD1}
M_{\rm D1} = \frac{\big| \sfF^\flat_\sfm(v)-\sfF^\flat_{\sfm'}(v)\big|}{\ell_{na}^2}  ~.
\ee

Note that in the decoupling limit~\eqref{decoupling}, the factors of $\gstr$ cancel between the brane separation and the D1 inverse tension, and the mass is simply the coordinate separation of the NS5-branes in string units.
Similarly, the tension of a stretched D2-brane in type IIA in the decoupling limit is 
\be
\label{tauD2}
\tau^{~}_{\rm D2} = \frac{\big| \hat\sfF_\sfm(v)-\hat\sfF_{\sfm'}(v)\big|}{4\pi^2 (\alpha')^{3/2}}  ~.
\ee
These stretched D2-branes are in fact the non-abelian little strings, made massive by the separation of the fivebranes in their transverse space.
When the fivebrane separation becomes of order one in string units, little string excitations are as light as fundamental string excitations; string perturbation theory breaks down, and one enters the regime of non-abelian (\ie~black) fivebrane dynamics.

These results continue to hold for general wiggly profiles $\sfF(v)$~\rcite{Martinec:2019wzw} sourcing the general two-charge NS5-P background.  It is useful to consider the more general 6d metric~\eqref{AnsatzIIB}, setting $\beta=0$, which is the form of the three-charge NS5-F1-P geometries we derive below; the two-charge case is recovered by setting $Z_1=1$.   The metric can be reorganized as
\be
ds^2 = -\frac2{Z_1} du\, dv + Z_2 \Big( d\sfx^i - \frac{\omega^i}{Z_1Z_2} dv\Big)^2 - \Big(\frac{\cF}{Z_1} + \frac{\omega_i \omega^i}{Z_1^2Z_2}\Big)dv^2  ~.
\ee
Passing to co-moving coordinates
\be
\label{comoving}
du' = du + \half\Big(\cF + \frac{\omega_i \omega^i}{Z_1Z_2}\Big)dv
~~,~~~~
dv'= dv
~~,~~~~
d\sfx'_i = d\sfx_i - \frac{\omega_i}{Z_1Z_2} dv
\ee
the DBI action (with $e^{2\Phi} = g_s^2{Z_2}/{Z_1}$) has the same form as in the static Coulomb branch for $Z_1=1$,%
\footnote{The pullback of the $B$-field vanishes for comoving D-strings.}
leading once again to the results~\eqref{MD1}, \eqref{tauD2}.  For $Z_1\ne 1$, there is a cancellation between the redshift factor $Z_1$ in the metric and the factor of $Z_1$ in the dilaton. 

Thus the non-abelian scale always corresponds to fivebrane coordinate separation $\delta\hat\sfF\sim \sqrt{\alpha'}$, independent of the presence of string or momentum charges.

In typical two-charge states, the separation of successive windings of the NS5 is governed by the thermal correlator of the left-moving excitations.
As discussed around Eq.~\eqref{walksize}, the typical configuration is a random walk of $(n_5n_p^\perp)^{1/2}$ steps.  Successive windings are $(n_p^\perp/n_5)^{1/2}$ steps apart, and so are separated by an amount of order $\mu (n_p^\perp/n_5)^{1/4}$.
We thus want $n_p^\perp\gg n_5$ in order that the fivebranes are separated by more than a magnetic string length after going once around the $y$-circle, so that successive strands of the winding fivebrane don't overlap and require consideration of nonabelian effects.

There is an important point to bear in mind in this discussion.  The scale $\mu$ set by the fivebrane tension differs from the non-abelian scale at which non-abelian effects arise, by a factor $[V_4/(\alpha')^2]^{1/4}$.  The non-abelian scale, where stretched D-branes become as light as fundamental strings is $|\delta\hat\sfF|^2=O(\alpha')$, while the intrinsic width of the fivebrane center-of-mass wavefunction is~$|\delta\hat\sfF|^2=O\big({\alpha'}/(V_4)^{\half}\big)$.  Increasing $V_4$ makes the fivebranes heavier, and the size of the NS5-P supertube smaller.  On the other hand, the energy cost of stretched D1- or D2-branes is independent of $V_4$.  For $V_4$ larger than the string scale, non-abelian effects between a pair of fivebranes kick in somewhat {\it before} their center-of-mass wavefunctions overlap.  In particular, the superstratum throat length~\eqref{Lstratum} at which quantum effects take over is not where $J_R\sim O(1)$, but rather $J_R\sim O\big((V_4)^\half/{\alpha'}\big)$.  With this caveat, we will treat these two scales as roughly interchangeable, keeping in mind that if the torus is significantly larger than string scale, quantum effects kick in somewhat earlier than fivebrane separations of order~$\mu$.

Another scale is that associated to the onset of large quantum fluctuations of a two-charge BPS fivebrane star~-- the size of the region where the quantum fluctuations of the metric coefficients become large.
In~\rcite{Martinec:2023gte}, the quantum fluctuations of the harmonic functions $Z_2,\cF$ (there called $H_5,H_p$) were calculated for the particular case of the round supertube.  Not surprisingly, these fluctuations become large at the scale of the magnetic string length $\mu$.

The fivebrane star geometries we discuss below are extremely close to the BTZ geometry~\eqref{BTZ metric} (and its linear dilaton extension) outside the star radius, but instead of descending to an extremal horizon at infinite throat depth and infinite redshift, one encounters fivebrane and string sources at the star's surface and the geometry begins to cap off.  
The fivebrane star radius $r_5$ is once again the radius of gyration of the fivebrane blob.  One can arrange the hierarchy of scales $r_5\ll \rhat_{\!\sst\rm AdS2}$.  As the size of the star approaches the non-abelian scale, the description of it as a geometry sourced by a classical brane configuration breaks down, as non-abelian fluctuations of the brane become important.  Since the deconfinement transition is associated to black hole formation, at this point the star becomes a black hole.

We will be interested in analogues in these more complicated geometries of the various measures of quantum effects discussed above.  In particular, the general three-charge geometries can be arranged to have a deep $AdS_2$ throat much like the special solutions based on round supertubes discussed above, and one can again compare the various quantum scales that arise with the quantum scale associated to the extremal black hole.
Our claim is that the quantum scale of $AdS_2$ black holes and the non-abelian scale of fivebrane stars are one and the same, in the sense that they correspond to the same $AdS_2$ throat length ${L_{\rm throat}}$.

Another, different scale which has been considered is the ``stretched horizon'' scale of two-charge states, where the surface area surrounding the fivebrane blob is of order the two-charge BPS entropy~\rcite{Mathur:2005zp}.  This scale corresponds to fivebrane random walks of size
\be
\label{dx smallBH}
|\delta\sfx| \sim \mu (n_p^\perp n_5)^{1/4} \sim \mu (n_p n_5)^{1/6}~.
\ee
This result follows from the fact that the volume of the angular $\bS^3$ in these solutions scales as ${\it vol}_{\bS^3}\sim n_5^{3/2}$, while the proper size of the $y$-circle scales as ${\it vol}_{\bS^1}\sim(n_p^\perp/n_5)^{1/4}$, and the dilaton at the source saturates at $e^{2\Phi}\sim (n_5/n_p^\perp)^{1/2}V_4$.  Putting these together to calculate the area $A_b$ in Planck units of the surface surrounding the source blob, one finds
\be
\frac{A_b}{G_N} \sim \bigg[n_5^{3/2}\Big(\frac{n_p^\perp}{n_5}\Big)^{1/4} V_4\bigg] \bigg[\Big(\frac{n_p^\perp}{n_5}\Big)^{1/2}\frac1{V_4}\bigg] = \big(n_p^\perp n_5\big)^{3/4} ~.
\ee
Demanding that this area be approximately the same as the two-charge entropy $(n_pn_5)^{1/2}$ yields the second relation in~\eqref{dx smallBH}.  Thus, provided $n_p\gg n_5^{1/2}$, the typical nearest-neighbor separation between fivebrane strands is larger than the magnetic string length $\mu$, and the nonabelian fivebrane excitations are typically decoupled; the effective string picture of the bound states is self-consistent, and a finite fraction of the entropy is contained within a surface whose area is what one might call the ``geometric entropy'' \`a la Bekenstein-Hawking.  However, in order to make the fivebrane blob this small, one has to scale ${n_p^\perp}/{n_p}\sim (n_pn_5)^{-1/3}$, and so one is very far from equipartition of the momentum budget.  One expects that a small perturbation will push the system to expand back to the typical size $\mu(n_pn_5)^{1/4}$ of such configurations, which is much larger than the scale~\eqref{dx smallBH} (and of course, much larger than the quantum/non-abelian scale).

\section{BPS geometries with both NS5-P and F1-P sources}
\label{sec:STsolns}

In this section we derive a class of three-charge 1/8-BPS geometries by forming multi-center solutions for which each charge center is a two-charge 1/4-BPS solution (NS5-P or F1-P).  They are solutions to the variational equations derived by coupling supergravity to explicit string and fivebrane sources.  The supergravity, string, and brane effective actions are simultaneously varied to obtain the equations of motion coupled to string and fivebrane sources.  We follow a systematic procedure for solving these equations~\rcite{Bena:2011dd}, borrowing in the final stages from the work of~\rcite{Niehoff:2012wu}.  Readers who are less interested in the details might want to skip to the final result, which we summarize in Appendix~\ref{app:soln sum}.

We work in asymptotically flat spacetime, but suppress any $\flat$ decorations on the coordinates.  Later, we will take various decoupling limits as in Section~\ref{sec:blackness2}. 
We look for solutions of the supergravity field equations having the BPS form~\eqref{AnsatzIIB}, with classical string and fivebrane sources, generalizing~\eqref{H5soln}-\eqref{Hpsoln}.  The bosonic part of the action of Type II supergravity, together with the effective action of strings and fivebranes, is given by (in the absence of R-R fields and in the string frame)
\begin{align}
\label{actions}
\cI_{\rm bulk} =~& \frac{1}{2\kappa_6^2} \int d^6 x\, e^{-2\Phi}\sqrt{-G} \left( R+ 4 G^{\mu \nu} \partial_{\mu} \Phi \partial_{\nu} \Phi  -\frac{1}{12} H_{\alpha \beta \gamma} H^{\alpha \beta \gamma} \right)~.
\nn\\[.2cm]
\mathcal{S}_{\text{source}} = ~& -\frac{\tau_{\sst\rm F1}}{2}\int d^2 \sigma \Big[\Big(\sqrt{-\gamma} \gamma^{ab} G_{\mu \nu} \partial_a \sfX^\mu \partial_{b} \sfX^\nu+\epsilon^{ab} B_{\mu \nu} \partial_a \sfX^\mu \partial_{b} \sfX^\nu  \Big)
\\
&\hskip 2cm
+ \Big(\sqrt{-\gamma} \gamma^{ab} G_{IJ} \partial_a \sfZ^I \partial_{b} \sfZ^J+\epsilon^{ab} B_{IJ} \partial_a \sfZ^I \partial_{b} \sfZ^J  \Big) \Big]
\nn\\[.2cm]
&-\frac{ \tau_{\sst\rm NS5}}{2}\int d^2 \tilde{\sigma} \left(\sqrt{-\tilde{\gamma}} \tilde{\gamma}^{ab} e^{-2\Phi}G_{\mu \nu} \partial_a \sfF^\mu \partial_{b} \sfF^\nu+\epsilon^{ab} \widetilde{B}_{\mu \nu} \partial_a \sfF^\mu \partial_{b} \sfF^\nu  \right)~.
\nn
\end{align}
Here $\gamma$ is an intrinsic metric on the string worldsheet, while for the fivebranes reduced on $\bT^4$ one has an effective string with intrinsic metric $\tilde{\gamma}$.  The gravitational coupling $\kappa_0$ is given by~\eqref{kappa0}; the fivebrane tension in~\eqref{tauNS5}; and the string tension is
\begin{equation}
\label{tauF1}
 \tau_{\sst\rm F1} = \frac{1}{2\pi \alpha'}~.
\end{equation}

While the fivebranes wrap $\bT^4$ and are naturally reduced along it, the strings do not wrap $\bT^4$.   However, we will assume that the string sources fill the $\bT^4$ uniformly, so that only their average is felt by the geometry, such that the $\bT^4$ metric and $B$-field are constant.  For simplicity, we set $B_{IJ}=0$ and $G_{IJ} = \delta_{IJ}$, with the torus volume
\be
V_4 = \Big(\frac{L}{2\pi}\Big)^4 ~,
\ee
with the torus coordinates having periodicity $\sfZ^I\sim\sfZ^I+L$.

We focus on solutions with left-moving momentum waves along $\bS^1_y$, so that $\sfF^i = \sfF^i(v)$, $\sfX^i=\sfX^i (v)$, and $\sfZ^I=\sfZ^I (v)$.
The stress energy tensor in the six-dimensional target space is given by
\begin{align}
\label{Tmunu}
	T^{\mu \nu} = \frac{2}{\sqrt{-G}}\frac{\delta \mathcal{S}_{\text{source}}}{\delta G_{\mu \nu}}
    &=-\frac{1}{\sqrt{-G}}\tau_{\sst\rm F1} \sqrt{-\gamma}\gamma^{ab}  \partial_a \sfX^\mu \partial_b \sfX^\nu  \delta^4 (\sfx-\sfX) \nonumber\\
	& \hskip .5cm
    -\frac{1}{\sqrt{-G}}\tau_{\sst\rm NS5} e^{-2\Phi}  \sqrt{-\tilde{\gamma}}\tilde{\gamma}^{ab} \partial_a \sfF^\mu \partial_b \sfF^\nu\delta^4 (\sfx-\sfF)~.
\end{align}
The components of the stress energy tensor in the torus dimensions are
\begin{align}
\label{TIJ}
	T^{IJ} = \frac{2}{\sqrt{-G}}\frac{\delta \mathcal{S}_{\text{source}}}{\delta G_{IJ}}
    &=-\frac{1}{\sqrt{-G}}\tau_{\sst\rm F1} \sqrt{-\gamma}\gamma^{ab}  \partial_a \sfZ^I \partial_b \sfZ^J  \delta^4 (\sfx-\sfX) ~.
\end{align}
Off-diagonal components of this tensor in the $I$-$\mu$ directions are given by
\begin{align}
\label{TImu}
	T^{I\mu} 
    &=-\frac{1}{\sqrt{-G}}\tau_{\sst\rm F1} \sqrt{-\gamma}\gamma^{ab}  \partial_a \sfZ^I \partial_b \sfX^{\mu}  \delta^4 (\sfx-\sfX) ~.
\end{align}
Below, we smear the stress energy tensor about the torus, supplanting each component of $T$ by its integral along this compact manifold, normalized by the volume $L^4=(2\pi)^4 V_4$. Focusing on classical string solutions where the bosonic embeddings of the string $Z^I$ are chiral and have no winding about the torus, we conclude that the smeared components of $T$ in the torus vanish: 
\begin{equation} 
 T^{IJ} = 0 ~,~ T^{I \mu}=0~.
\end{equation}

The stress energy tensor components transverse to the torus are
\begin{align}
\label{Tmunu compts}
	T^{vv}&=0~~,~~~~
	T^{vi}=0 ~~,~~~~
    T^{ij}=0 ~~,~~~~
\nn\\
	T^{uv} &= \frac{1}{\sqrt{-G}}\tau_{\sst\rm F1}  \delta^4 (\sfx-\sfX(v)) +\frac{1}{\sqrt{-G}}\tau_{\sst\rm NS5} e^{-2\Phi}  \delta^4 (\sfx-\sfF(v))~.  
\nn\\
	T^{uj} &= \frac{1}{\sqrt{-G}}\Big[ \tau_{\sst\rm F1}  \partial_v X^j \delta^4 (\sfx-\sfX(v))+\tau_{\sst\rm NS5} e^{-2\Phi} \partial_v F^j \delta^4 (\sfx-\sfF(v)) \Big]~.
\\
	T^{uu} &=	\frac{2}{\sqrt{-G}}\left[ \tau_{\sst\rm F1} \partial_v \sfX^u \delta^4 (\sfx-\sfX(v)) + \tau_{\sst\rm NS5} e^{-2\Phi} \partial_v F^u \delta^4 (\sfx-\sfF(v))\right]~.
\nn
\end{align}
The field equations now have a layered structure much like the two-charge system sketched above, wherein at each layer one solves harmonic equations for a subset of the supergravity field components, with sources that are determined by the harmonic forms/functions found in previous layers, as well as new sources determined by the brane configurations $\sfF(v),\sfX(v)$.

\subsection{Layer 0}
\label{sec:layer0}

First, we pick the ``base'' space transverse to the fivebranes to be $\bR^4$; the one-form $\beta$ in~\eqref{AnsatzIIB} is self-dual:
\begin{equation}
d\beta = *d\beta~,
\end{equation}
where the Hodge star is taken with respect to the four-dimensional flat transverse base space. 
A trivial solution of the self-duality condition is 
\begin{equation}
\label{Beta0}
 \beta =0 ~.
\end{equation}
This choice is exemplified by the known two-charge BPS solutions, F1-P and separately NS5-P.%
\footnote{The T-dual NS5-F1 frame does in general feature a nonzero $\beta$, as a result of the Buscher rules for T-duality starting from an NS5-P solution having $\beta=0$.  In the NS5-P duality frame we choose to work in, the possibility to choose $\beta=0$ simplifies the equations of motion considerably.}

\subsection{Layer 1}
\label{sec:layer1}

The goal of the first layer is to determine the harmonic functions $Z_1,Z_2$ and a pair of two-forms $\Theta_{1},\Theta_{2}$, whose components appear in the field strengths of $B,\widetilde B$ and are defined as
\begin{subequations}
\label{Thetadefs}
\begin{align}
\label{Theta1def}
 \Theta_1 &= da_1 + \partial_v \gamma_2 
\\[.1cm]
\label{Theta2def}
 \Theta_2 &= da_2 + \partial_v \gamma_1 ~.
\end{align}
\end{subequations}

The Polyakov action describing string propagation is%
\footnote{Because our analysis is classical, We neglect the coupling of the dilaton to the worldsheet scalar curvature, which is a one-loop effect in the inverse of the string tension.}
\begin{align}
\label{SF1}
	&\mathcal{S}_{\sst\rm F1} = -\frac{\tau_{\sst\rm F1}}{2}\int d^2 \sigma \left(\sqrt{-\gamma} \gamma^{ab} \big(G_{\mu \nu} \partial_a \sfX^\mu \partial_{b} \sfX^\nu
    + \partial_a\sfZ^I\partial_b\sfZ^I\big)
    +\epsilon^{ab} B_{\mu \nu} \partial_a \sfX^\mu \partial_{b} \sfX^\nu  \right)~,
\end{align}
where $\tau_{\sst\rm F1}$ is the tension of the strings, $\sigma$ is a worldsheet coordinate, $\gamma_{ab}$ is the worldsheet instrinsic metric and $\sfX^\mu (\sigma)$ is the embeddings of the strings in target-space.  Since we are interested in classical solutions on a worldsheet cylinder, we can ignore the dilaton contribution.

The NS5-P source effective action can be written~\rcite{Martinec:2024emf}
\begin{align}
\label{SNS5}
	&\mathcal{S}_{\sst\rm NS5} =
	-\frac{\tau_{\sst\rm NS5}}{2}\int d^2 \tilde{\sigma} \left(\sqrt{-\tilde{\gamma}} \tilde{\gamma}^{ab} e^{-2\Phi}G_{\mu \nu} \partial_a \sfF^\mu \partial_{b} \sfF^\nu+\epsilon^{ab} \widetilde{B}_{\mu \nu} \partial_a \sfF^\mu \partial_{b} \sfF^\nu  \right)~,
\end{align}
where $\tau_{\sst\rm NS5}$ is the tension of the NS5-branes, $\tilde{\sigma}$ is parametrizes the effective worldsheet, $\tilde{\gamma}_{ab}$ is the worldsheet intrinsic metric, and $\sfF^\mu (\tilde{\sigma})$ specifies the embedding of the fivebranes in target-space.

The $\sfF^\mu$ equations of motion are
\begin{align}
\label{Feom}
		& \nabla^2 \sfF^\mu = -\widetilde{\Gamma}^{\mu} _{\alpha \beta} \gamma^{ab} \partial_a \sfF^{\alpha} \partial_b \sfF^{\beta}+ \frac{1}{2} (*H)^{\mu}_{~~\alpha \beta} \epsilon^{ab} \partial_a \sfF^{\alpha} \partial_b \sfF^{\beta}~,
\end{align}
where $\nabla^2$ is the worldsheet Laplacian in the metric $\tilde\gamma$, and $\widetilde{\Gamma}^{\mu} _{\alpha \beta}$ is the Christoffel symbol of the metric $e^{-2\Phi} G_{\mu \nu}$.  
The $\sfX^\mu$ equations of motion are
\begin{align}
\label{Xeom}
	& \nabla^2 \sfX^\mu = -\Gamma^{\mu} _{\alpha \beta} \gamma^{ab} \partial_a \sfX^{\alpha} \partial_b \sfX^{\beta} + \frac{1}{2} H^{\mu}_{~~\alpha \beta} \epsilon^{ab} \partial_a \sfX^{\alpha} \partial_b \sfX^{\beta}~,
\end{align}
and the torus coordinates $\sfZ^I$ are free fields.
As in \cite{Martinec:2024emf}, we choose the conformal gauge for the intrinsic metrics
\begin{equation}
\gamma_{ab} =\eta_{ab}~~,~~~~ \tilde{\gamma}_{ab} = \eta_{ab}~.
\end{equation}
Since $\partial_u$ is a Killing vector in the backgrounds in question, it follows that $\sfF^v$ and $\sfX^v$ are free fields. Reparametrization gauge redundancy can be used to set $\sfF^v = \sigma^+$ and $\sfX^v=\sigma^+$. In turn, any $\sigma^+$-dependent smooth profile functions describing the transverse locations of the fivebranes and strings 
\begin{equation}
 \sfX^i = \sfX^i (\sigma^+) 
 ~~,~~~~ 
 \sfZ^I = \sfZ^I (\sigma^+) 
 ~~,~~~~ 
 \sfF ^i = \sfF^i(\sigma^+)
\end{equation}
are solutions.
The determination of $\sfX^u,\sfF^u$ requires the equations of motion derived from varying the source actions with respect to the intrinsic metric variables. In conformal gauge, these are given by%
\footnote{We ignore the contribution to the stress tensor from the dilaton, as it is suppressed by a factor of $\alpha'$, and we are focusing on classical solutions.}
\begin{align}
\label{constraints}
\begin{split}
	G_{\mu \nu} \partial_v \sfX^\mu \partial_v \sfX^\mu + G_{IJ} \partial_v \sfZ^I \partial_v \sfZ^J &= 0
\\[.2cm]
e^{-2\Phi}G_{\mu \nu} \partial_v \sfF^\mu \partial_v \sfF^\mu &= 0~.
\end{split}
\end{align}
The first equation in (\ref{constraints}) implies that 
\begin{equation}
\label{Virasoro0}
	-\frac{2 }{Z_1} \partial_v \sfX^u \partial_v \sfX^v -2\frac{ \omega_i}{Z_1} \partial_v \sfX^i \partial_v \sfX^v  -\frac{ \mathcal{F}}{Z_1} \partial_v \sfX^v \partial_v \sfX^v + Z_2 \partial_v \sfX^i \partial_v \sfX^i + \partial_v\sfZ^I\partial_v\sfZ^I =0~.
\end{equation}
For $\partial_v \sfX^v=1$, this implies
\begin{equation}
	 \partial_v \sfX^u = \frac{1}{2}Z_1 \Big(Z_2 \partial_v \sfX^i \partial_v \sfX^i + \partial_v\sfZ^I\partial_v\sfZ^I\Big) -\frac{\mathcal{F}}{2}-\omega_i \partial_v \sfX^i ~.
\end{equation}
Similarly,
\begin{equation}
	 \partial_v \sfF^u = \frac{1}{2}Z_1 Z_2 \partial_v \sfF^i \partial_v \sfF^i -\frac{\mathcal{F}}{2}-\omega_i \partial_v \sfF^i ~.
\end{equation}
The dilaton equation of motion is
\begin{equation}
\label{Z2 eom}
	\nabla^2 Z_2 = 
    -2\kappa_6 ^2 \tau_{\sst\rm NS5} \sum_{\sfm =1} ^{\nfive}\delta^4 (\sfx-\sfF_\sfm(v))~,
\end{equation}
having the solution
\begin{equation}
\label{Z2soln}
 Z_2(\sfx,v) = 1+\frac{\kappa_6 ^2 \tau_{\sst\rm NS5}}{2\pi^2} \sum_{\sfm =1} ^{\nfive} \frac{1}{|\sfx-\sfF_\sfm(v)|^2} ~.
\end{equation}
The $B_{uv}$ equation of motion is 
\begin{equation}
\label{Z1 eom}
\nabla^2 Z_1 = 
 -2\kappa_6 ^2 \tau_{\sst \rm F1} \sum_{\sfn =1} ^{\none} \delta^4 (\sfx-\sfX_\sfn (v))~.
\end{equation}
Consequently,
\begin{equation}
\label{Z1soln}
 Z_1(\sfx,v) = 1+\frac{\kappa_6^2 \tau_{\sst\rm F1}}{2\pi^2} \sum_{\sfn =1} ^{\none} \frac{1}{|\sfx-\sfX_\sfn (v)|^2}~. 
\end{equation}
Strings are electrically charged under $B$, and magnetically charged under $\tilde B$; conversely, fivebranes are electrically charged under $\tilde B$ and magnetically charged under $B$.  These two potentials are related by Hodge duality between $B$ and $\widetilde B$, Eq.~\eqref{Hodge}.
In appendix~\ref{app:hodge}, we calculate the components of this equation and show the following.
The $i$-$j$-$k$ components of this equation imply
\begin{equation}
\label{dgamma1}
     	d\gamma_1 = *_4 dZ_1~,
\end{equation}
whereas the $u$-$v$-$i$ components give rise to
\begin{equation}
\label{dgamma2}
     	d\gamma_2 = *_4 dZ_2~,
\end{equation}
where $\gamma_2,\gamma_1$ describe the purely transverse components of $B,\tilde B$ respectively. 
The $v$-$i$-$j$ components of Eq.~(\ref{Hodge}) are
\begin{align}
	\label{Omega2b} 
	(d\omega)_{ij} +(*d\omega)_{ij} =Z_1 (* \Theta_1)_{ij} +Z_2 \Theta_{2ij} ~.   
\end{align}
Subtracting the R.H.S of the last equation with its Hodge dual gives
\begin{equation}
     \frac{1}{Z_2} (\Theta_1 - *\Theta_1 ) = \frac{1}{Z_1} (\Theta_2 - *\Theta_2)~.
\end{equation}
The L.H.S depends only on the fivebrane oscillations about their center of mass, whereas the R.H.S depends only on the string profiles. Since one can consider independent motions for both sources, both the R.H.S and L.H.S vanish. Thus:
\begin{subequations}
\begin{align}
\label{selfDuality1}
     	\Theta_1 &= *\Theta_1~,
\\
\label{selfDuality2}
     	\Theta_2 &= *\Theta_2~.
\end{align}
\end{subequations}

The $B_{vi}$ equation is
\begin{align}
\label{Bvi}
&\frac{1}{Z_2^2} (d\omega+*d\omega)_{ij} \partial_j Z_2-\frac{1}{Z_2}\left(\nabla^2 \omega_i -\partial_i (\partial_j \omega_j)\right)
\nonumber\\
&\hskip 1cm
+\partial_v \partial_i Z_1 + \frac{Z_1}{Z_2}\partial_j \Theta_{1ij} +\frac{1}{Z_2}\Theta_{1ij} \partial_j Z_1 -\frac{Z_1}{ Z_2 ^2} \Theta_{1ij} \partial_j Z_2
\\[.2cm]
&\hskip 3cm
= 2\kappa_6 ^2 \tau_{\sst\rm F1} \sum_{\sfn =1} ^{\none}\partial_v X_{in} \delta^4 \big(\sfx-\sfX_\sfn (v)\big)~;
\nn
\end{align}
a combination of the $v$-$i$ Einstein equations and Eq.~(\ref{Omega2b}) yields
\begin{align}
\label{OmegaEom3}
&\nabla^2 \omega_i -\partial_i (\partial_j \omega_j)  - Z_1 \partial_v \partial_i Z_2 - Z_2\partial_i \partial_v Z_1 - \Theta_{2ij}\partial_j Z_2-\Theta_{1ij} \partial_j Z_1 
\nonumber\\[.2cm]
&\hskip 2cm 
= -2\kappa_6^2 \Big( \tau_{\sst\rm F1} Z_2 \sum_{\sfn =1} ^{\none}\partial_v \sfX^i _{\sfn} \delta^4 (\sfx-\sfX_\sfn (v))+ \tau_{\sst\rm NS5}Z_1\sum_{\sfm =1} ^{\nfive}\partial_v \sfF^i_{\sfm} \delta^4 \big(\sfx-\sfF_\sfm(v)\big)\Big)~.
\end{align}
A simplification occurs when summing these two equations: 
\begin{align}
\label{Bvi+gvi}
	\partial_j \Theta_{1ij}-\partial_v \partial_i Z_2 =
    -2\kappa_6 ^2 \tau_{\sst\rm NS5} \sum_{\sfm =1} ^{\nfive}\partial_v \sfF_\sfm ^i \delta^4 \big(\sfx-\sfF_\sfm(v)\big)~.
\end{align}
Using the definition (\ref{Theta1def}) of $\Theta_{1}$, one has
\begin{align}
	\label{Combvi}
	\nabla^2 a_{1i} -\partial_{i} (\partial_j a_{1j})+\partial_v \partial_j \gamma_{2ij}-\partial_v \partial_i Z_2 =
    -2\kappa_6 ^2 \tau_{\sst\rm NS5} \sum_{\sfm =1} ^{\nfive} \partial_v \sfF^i _{\sfm} \delta^4 \big(\sfx-\sfF_\sfm(v)\big)~.
\end{align}
It is convenient to pick the gauges
\begin{equation}
\label{gauge1}
	\partial_j a_{1j} = -\partial_v Z_2~,
\end{equation}
and
\begin{equation}
\label{gauge2}
	\partial_j \gamma_{2ij} =0 ~.
\end{equation}
(the latter choice was made implicitly in~\rcite{Martinec:2024emf}).
The gauge choices in Eqs.~(\ref{gauge1}), (\ref{gauge2}) imply that Eq.~(\ref{Combvi}) is simplified as follows
\begin{align}
	\nabla^2 a_{1i} =
    -2\kappa_6 ^2 \tau_{\sst\rm NS5} \sum_{\sfm =1} ^{\nfive}\partial_v \sfF_\sfm^i \delta^4 \big(\sfx-\sfF_\sfm(v)\big)~.
\end{align}
Therefore,
\begin{equation}
\label{a1result}
	a_{1i} = 
    \frac{\kappa_6 ^2 \tau_{\sst\rm NS5}}{2\pi^2} \sum_{\sfm =1} ^{\nfive} \frac{\partial_v \sfF_{\sfm i}}{|\sfx-\sfF_\sfm(v)|^2} + b_{1i}~,
\end{equation}
where $b_{1}$ is a possible ``magnetic'' contribution, which we will determine below.
Similarly, for the gauge choices
\begin{equation}
\label{gauge3}
 \partial_j a_{2j} = -\partial_v Z_1
\quad,\qquad
 \partial_j \gamma_{1ij} =0~, 
\end{equation}
the linear combination of the $g_{vi}$, $\widetilde{B}_{vi}$ and the $v$-$i$-$j$ Hodge duality relation Eq.~(\ref{Omega2b}) is
\begin{align}
\label{a2Poisson}
	\nabla^2 a_{2i} = 
    -2\kappa_6 ^2 \tau_{\sst\rm F1} \sum_{\sfn =1} ^{\none} \partial_v \sfX_{\sfn i} \delta^4 \big(\sfx-\sfX_\sfn (v)\big)~.
\end{align}
This equation implies
\begin{equation}
\label{a2result}
	a_{2i} = 
    \frac{\kappa_6 ^2 \tau_{\sst\rm F1}}{2\pi^2} \sum_{\sfn =1} ^{\none} \frac{\partial_v \sfX_{\sfn i}}{|\sfx-\sfX_\sfn (v)|^2} + b_{2i}~,
\end{equation}
where again $b_2$ is a ``magnetic'' contribution.
To find an expression for $\Theta_1$, one wants to satisfy the requirement of self-duality in Eq.~(\ref{selfDuality1}) and Eq.~(\ref{Combvi}).  An ansatz based on the form of $da_1$, which is extracted from Eq.~(\ref{a1result}), is
\begin{equation}
\label{Theta1}
	\Theta_{1ij} = 
    -\frac{\kappa_6^2 \tau_{\sst\rm NS5}}{\pi^2}\sum_{\sfm =1} ^{\nfive} \frac{\partial_v \sfF_{\sfm i} (\sfx_j-\sfF_{\sfm j}) - \partial_v \sfF_{\sfm j} (\sfx_i-\sfF_{\sfm i}) + \epsilon_{ijkl} \partial_v \sfF_{\sfm k} (\sfx_l-\sfF_{\sfm l}) }{|\sfx-\sfF_\sfm|^4}~.
\end{equation}
The first two terms on the R.H.S are contributions to $(da_1)_{ij}$.  However, one has the relations
\be
\label{gamdot-da}
\partial_v \gamma_2 = *da_1
\quad,\qquad
\partial_v \gamma_1 = *da_2 ~.
\ee
The application of an exterior derivative and a Hodge star on \eg\ the first of these gives Eq.~(\ref{Combvi}) outside sources, using~\eqref{dgamma2}; integrating up away from sources, the result follows.  

From this result, one can deduce expressions for $\gamma_2$ and the ``magnetic'' piece $b_1$ of the one-form~$a_1$:
\begin{align}
\label{gam2b1}
\gamma_{2} &= \frac16  \frac{\kappa_6 ^2 \tau_{\sst\rm NS5}}{2\pi^2}  \sum_{\sfm =1}^{\nfive} \frac{\epsilon_{ijkl}(\sfx^i-\sfF_{\sfm }^i)(\sfx^j-\sfF_{\sfm }^j)\,d\sfx^k\wedge d\sfx^l}{|\sfx-\sfF_\sfm|^2\big((\sfx_j-\sfF_{\sfm j})^2+(\sfx_l-\sfF_{\sfm l})^2\big)} ~.
\nn\\[.2cm]
b_{1} &= \frac16  \frac{\kappa_6 ^2 \tau_{\sst\rm NS5}}{2\pi^2}  \sum_{\sfm =1}^{\nfive} \frac{\epsilon_{ijkl}(\sfx^i-\sfF_{\sfm }^i)(\sfx^j-\sfF_{\sfm }^j)\big(\partial_v\sfF_{\sfm }^k\,d\sfx^l-\partial_v\sfF_{\sfm }^l\,d\sfx^k\big)}{|\sfx-\sfF_\sfm|^2\big((\sfx_j-\sfF_{\sfm j})^2+(\sfx_l-\sfF_{\sfm l})^2\big)} 
\end{align}
If we set $\sfF(v)=0$, the expression for $\gamma_2$ becomes that of static fivebranes~-- the CHS solution~\rcite{Callan:1991at}.  The above expressions then simply result from the substitutions $\sfx\to \sfx-\sfF(v)$, $d\sfx\to d\sfx-\partial_v\sfF\,dv$.

A similar solution is found for $\Theta_2$:
\begin{equation}
\label{Theta2}
	\Theta_{2ij} =
     -\frac{\kappa_6^2 \tau_{\sst\rm F1}}{\pi^2}\sum_{\sfn =1} ^{\none} \frac{\partial_v \sfX_{\sfn i} (\sfx_j-\sfX_{\sfn j}) - \partial_v \sfX_{\sfn j} (\sfx_i-\sfX_{\sfn i}) + \epsilon_{ijkl} \partial_v \sfX_{\sfn k} (\sfx_l-\sfX_{\sfn l}) }{|\sfx-\sfX_\sfn |^4}~,
\end{equation}
from which one deduces
\begin{align}
\label{gam1b2}
\gamma_{1} &= \frac16 \frac{\kappa_6 ^2 \tau_{\sst\rm F1}}{2\pi^2} \sum_{\sfn =1}^{\none} \frac{\epsilon_{ijkl}(\sfx^i-\sfX_{\sfn }^i)(\sfx^j-\sfX_{\sfn }^j)\,d\sfx^k\wedge d\sfx^l}{|\sfx-\sfX_\sfn |^2\big((\sfx_j-\sfX_{\sfn j})^2+(\sfx_l-\sfX_{\sfn l})^2\big)}
\nn\\[.2cm]
b_{2} &= \frac16 \frac{\kappa_6 ^2 \tau_{\sst\rm F1}}{2\pi^2} \sum_{\sfn =1}^{\none} \frac{\epsilon_{ijkl}(\sfx^i-\sfX_{\sfn }^i)(\sfx^j-\sfX_{\sfn }^j)\big(\partial_v\sfX_{\sfn }^k\,d\sfx^l-\partial_v\sfX_{\sfn }^l\,d\sfx^k\big)}{|\sfx-\sfX_\sfn |^2\big((\sfx_j-\sfX_{\sfn j})^2+(\sfx_l-\sfX_{\sfn l})^2\big)} ~.
\end{align}

Note that thus far, neither type of source talks to the other~-- the strings and fivebranes source different harmonic forms and functions in the supergravity background.  The presence of the strings has no effect on the layer 1 metric and $B$-field coefficients seen by the fivebranes, and vice versa.  It is in the next and final layer that one finds new effects that are not present if one has only one kind of source but not the other, and the strings and fivebranes feel one another's presence.

\subsection{Layer 2}
\label{sec:layer2}

The second layer equations determine the angular momentum one-form $\omega$ and the momentum function~$\mathcal{F}$.
The first step is to solve the $\cG_{vi}$ Einstein equation for the one-form $\omega$ in the BPS metric~\eqref{AnsatzIIB}. 

It is useful to define several quantities that appear frequently:
\begin{equation}
\label{deltas}
\widetilde\sfR_\sfm \equiv \sfx-\sfF_\sfm (v) ~~,~~~~ 
\sfR_\sfn  \equiv \sfx-\sfX_\sfn (v)
~~,~~~~
\sfD_{\sfm\sfn} \equiv \sfF_\sfm - \sfX_\sfn  = \sfR_\sfn  - \widetilde\sfR_\sfm 
~.
\end{equation}
\begin{equation}
\label{CurlyA}
	\mathcal{A}_{ij} \equiv\widetilde\sfR_{\sfm i}\sfR_{\sfn j} -\sfR_{\sfn i}\widetilde\sfR_{\sfm j} -\epsilon_{ijkl}\widetilde\sfR_{\sfm k}\sfR_{\sfn l}~.
\end{equation}
For convenience, we copy again the $v$-$i$ metric equations:
\begin{align}
\label{OmegaEom111}
	&\nabla^2 \omega_i -\partial_i (\partial_j \omega_j)  - Z_1 \partial_v \partial_i Z_2 - Z_2\partial_i \partial_v Z_1 - \Theta_{2ij}\partial_j Z_2-\Theta_{1ij} \partial_j Z_1 \nonumber\\
&\hskip2cm
=-2\kappa_6 ^2\Big( \tau_{\sst\rm F1} \sum_{\sfn =1} ^{\none} Z_2 \partial_v \sfX^i _{\sfn} \delta^4 (\sfx-\sfX_{\sfn}(v))+\sum_{\sfm=1} ^{\nfive}Z_1\tau_{\sst\rm NS5} \partial_v \sfF^i _{\sfm} \delta^4 (\sfx-\sfF_{\sfm}(v)) \Big)~.
\end{align}
Picking a gauge 
\begin{equation}
\label{gauge4.5}
  \partial_j \omega_j = -\partial_v (Z_1 Z_2)~,
\end{equation} 
Eq.~(\ref{OmegaEom111}) becomes
\begin{align}
\label{OmegaEom100}
	&\nabla^2 \omega_i 
    = -\partial_i Z_1 \partial_v  Z_2 - \partial_i Z_2\partial_v Z_1 + \Theta_{2ij}\partial_j Z_2 +\Theta_{1ij} \partial_j Z_1 
\nonumber\\
&\hskip2cm 
-2\kappa_6 ^2\Big( \tau_{\sst\rm F1} \sum_{\sfn =1} ^{\none} Z_2 \partial_v \sfX^i _{\sfn} \delta^4 (\sfx-\sfX_{\sfn}(v))+\tau_{\sst\rm NS5}\sum_{\sfm=1} ^{\nfive}Z_1 \partial_v \sfF^i _{\sfm} \delta^4 (\sfx-\sfF_{\sfm}(v)) \Big)~.
\end{align}
The angular momentum vector $\omega$ is thus sourced directly by contributions from the brane sources, given by the second line of the above equation, as well as by bulk terms quadratic in harmonic forms/functions $Z_i,\Theta_i$ determined by the layer 1 equations.

This feature is similar to what one encounters in four-dimensional gauge theory in the presence of both electrically and magnetically charged point sources.  The electric charges directly (and linearly) source the gauge potential $A$, while the magnetic charges directly (and linearly) source the dual potential $\widetilde A$.  But the momentum (and angular momentum) carried by the configuration receives delta function source contributions from both the particles themselves, as well as from the Poynting flux ${\bf E}\times{\bf B}$ of the bulk gauge field, which is present only when both kinds of charge are present.  We see exactly the same phenomenon for electrically and magnetically charged strings coupling to the two-form gauge potentials $B,\widetilde B$ in six dimensions.

Following reference~\cite{Niehoff:2012wu}, one can find an explicit solution.
From the previous solutions appearing in the first layer, one can compute
\begin{align}
\label{Theta1dZ1-dvZ2dZ1}
	&\Theta_{1ij} \partial_j Z_1-	\partial_v Z_2 \partial_i Z_1 = 
    \frac{\kappa_6^4 \tau_{\sst\rm NS5} \tau_{\sst\rm F1}}{\pi^4} \sum_{\sfm =1} ^{\nfive} \sum_{\sfn =1} ^{\none} \frac{\partial_v \sfF_{\sfm i} \widetilde\sfR_\sfm  \cdot \sfR_\sfn -\mathcal{A}_{ij} \partial_v \sfF_{\sfm j}}{\sfR_\sfn ^4 \widetilde\sfR_\sfm ^4}~.
\end{align}
\begin{align}
\label{Theta2dZ2-dvZ1dZ2}
	&\Theta_{2ij} \partial_j Z_2-	\partial_v Z_1 \partial_i Z_2 =\frac{\kappa_6^4 \tau_{\sst\rm NS5} \tau_{\sst\rm F1}}{\pi^4} \sum_{\sfm =1} ^{\nfive} \sum_{\sfn =1} ^{\none} \frac{\partial_v \sfX_{\sfn i} \widetilde\sfR_\sfm  \cdot \sfR_\sfn +\mathcal{A}_{ij} \partial_v \sfX_{\sfn j}}{\sfR_\sfn ^4 \widetilde\sfR_\sfm ^4}~.
\end{align}
As in \cite{Niehoff:2012wu}, we make the ansatz
\begin{align}
\label{omegaAnsatz}
	\omega_i &= \omega_{0i}+\omega_{1i}+\omega_{2l,i}+\omega_{2r,i}~,
\nn\\[.2cm]
	\omega_{0i} &\equiv  \frac{\kappa_6 ^2 \tau_{\sst\rm NS5}}{2\pi^2} \sum_{\sfm =1} ^{\nfive} \frac{\partial_v \sfF_{\sfm i}}{\widetilde\sfR_\sfm  ^2} + \frac{\kappa_6^2 \tau_{\sst\rm F1}}{2\pi ^2} \sum_{\sfn =1} ^{\none} \frac{\partial_v \sfX_{\sfn i}}{\sfR_\sfn ^2}~, 
\nn\\
	\omega_{1i} &\equiv \frac{\kappa_6 ^4 \tau_{\sst\rm NS5} \tau_{\sst\rm F1}}{8\pi^4 } \sum_{\sfm =1} ^{\nfive} \sum_{\sfn =1} ^{\none} \frac{\partial_v \sfF_{\sfm i} + \partial_v \sfX_{\sfn i}}{\sfR_\sfn  ^2 \widetilde\sfR_\sfm ^2}~,
\\
	\omega_{2l,i} &\equiv \frac{\kappa_6 ^4 \tau_{\sst\rm NS5}\tau_{\sst\rm F1}}{8\pi^4} \sum_{\sfm =1} ^{\nfive} \sum_{\sfn =1} ^{\none} \frac{\partial_v \sfD_{\sfm\sfn, i}}{|\sfR_\sfn -\widetilde\sfR_\sfm |^2} \bigg(\frac{1}{\widetilde\sfR_\sfm ^2}-\frac{1}{\sfR_\sfn  ^2}\bigg)~,
\nn\\
	\omega_{2r,i} &\equiv  \frac{\kappa_6 ^4 \tau_{\sst\rm NS5} \tau_{\sst\rm F1}}{4\pi^4} \sum_{\sfm =1} ^{\nfive} \sum_{\sfn =1} ^{\none} \frac{\partial_v \sfD_{\sfm\sfn, j} }{|\sfR_\sfn  - \widetilde\sfR_\sfm |^2}\frac{\mathcal{A}_{ij}}{\sfR_\sfn  ^2 \widetilde\sfR_\sfm ^2}~.
\nn
\end{align}
One can then verify that
\begin{align}
\label{LaplacianOmega}
	 \nabla^2 \omega_i = &-2\kappa_6 ^2 \tau_{\sst\rm NS5} \sum_{\sfm =1} ^{\nfive} Z_1 (\sfx) \partial_v \sfF_{\sfm i} \delta^4 (\sfx-\sfF_\sfm)-2\kappa_6 ^2 \tau_{\sst\rm F1} \sum_{\sfn =1} ^{\none} Z_2(\sfx) \partial_v \sfX_{\sfn i} \delta^4 (\sfx-\sfX_\sfn )\nonumber\\
	 &+\frac{\kappa_6^4 \tau_{\sst\rm NS5}\tau_{\sst\rm F1}}{\pi^4}\sum_{\sfm =1} ^{\nfive} \sum_{\sfn =1} ^{\none} \frac{\partial_v \sfF_{\sfm j} \left(\delta_{ij}\sfR_\sfn  \!\cdot \widetilde\sfR_\sfm  - \mathcal{A}_{ij}\right)}{\sfR_\sfn  ^4 \widetilde\sfR_\sfm ^4} \nonumber\\
	  &+ \frac{\kappa_6^4 \tau_{\sst\rm NS5}\tau_{\sst\rm F1}}{\pi^4}\sum_{\sfm =1} ^{\nfive} \sum_{\sfn =1} ^{\none} \frac{\partial_v \sfX_{jm} \left(\delta_{ij}\sfR_\sfn  \!\cdot \widetilde\sfR_\sfm  + \mathcal{A}_{ij}\right)}{\sfR_\sfn  ^4 \widetilde\sfR_\sfm ^4}~.
\end{align}
Calculating the contributions to $\partial_i \omega_i$,
\begin{align}
	\partial_i \omega_{0i} &= -\frac{\kappa_6^2 \tau_{\sst\rm NS5}}{\pi^2}\sum_{\sfm =1} ^{\nfive} \frac{\partial_v \sfF_{\sfm }\!\cdot \widetilde\sfR_{\sfm }}{\widetilde\sfR_\sfm ^4}-\frac{\kappa_6^2 \tau_{\sst\rm F1}}{\pi^2}\sum_{\sfn =1} ^{\none} \frac{\partial_v \sfX_{\sfn }\!\cdot\sfR_{\sfn }}{\sfR_\sfn ^4}~.
\nn\\[.2cm]
	\partial_i \omega_{1i} + \partial_i \omega_{2i} &= -\frac{\kappa_6^4 \tau_{\sst\rm NS5} \tau_{\sst\rm F1}}{2\pi^4}\sum_{\sfm =1} ^{\nfive} \sum_{\sfn =1} ^{\none}  \frac{(\partial_{v} \sfF_{\sfm }\!\cdot\widetilde\sfR_{\sfm }) \sfR_\sfn  ^2 + (\partial_v \sfX_{\sfn }\!\cdot\sfR_{\sfn }) \widetilde\sfR_\sfm ^2}{ \widetilde\sfR_\sfm ^4 \sfR_\sfn ^4}~.
\end{align}
It follows that the solution ansatz has selected the gauge choice
\begin{equation}
\label{gauge5}
	\partial_i \omega_i = -\partial_v (Z_1 Z_2)~.
\end{equation}
As a result, Eq.~(\ref{OmegaEom111}) can be rewritten as
\begin{align}
\label{OmegaEom9}
	&\nabla^2 \omega_i + \partial_i Z_1 \partial_v  Z_2 +\partial_i Z_2 \partial_v Z_1 - \Theta_{2ij}\partial_j Z_2-\Theta_{1ij} \partial_j Z_1 
\nonumber\\[.1cm]
	&\hskip 2cm
     =-2\kappa_6 ^2\left( \tau_{\sst\rm F1} \sum_{\sfn=1} ^{\none} Z_2 \partial_v \sfX^i _{\sfn} \delta^4 (\sfx-\sfX_{\sfn}(v))+ \tau_{\sst\rm NS5}\sum_{\sfm=1} ^{\nfive} Z_1 \partial_v \sfF^i _{\sfm} \delta^4 (\sfx-\sfF_{\sfm}(v)) \right)~.
\end{align}
Eqs.~(\ref{Theta1dZ1-dvZ2dZ1}), (\ref{Theta2dZ2-dvZ1dZ2}), and (\ref{LaplacianOmega}) imply that the $v$-$i$ Einstein equations (\ref{OmegaEom9}) are satisfied. 

We now turn to the $v$-$v$ Einstein equation, which determines $\cF$ in the metric~\eqref{AnsatzIIB}:
\begin{align}
\label{vv}
	\nabla^2 \mathcal{F} -2\partial_v (\partial_j \omega_j) &=
	2Z_1 \partial_v ^2 Z_2 + 2Z_2 \partial_v ^2 Z_1+2\partial_v Z_1 \partial_v Z_2-\frac{1}{2}\Theta_{1ij}\Theta_{2ij}
\nn\\
&\hskip .5cm
+2\kappa_6^2\bigg[ \tau_{\sst\rm F1} \sum_{\sfn=1} ^{\none} \Big(Z_2  \big|\partial_v \sfX_{\sfn}\big|^2 + \big|\partial_v\sfZ _{\sfn}\big|^2  \Big)\delta^4 (\sfx-\sfX_{\sfn}(v)) 
\\
&\hskip 2cm
+\tau_{\sst\rm NS5} \sum_{\sfm=1} ^{\nfive} Z_1 \big|\partial_v \sfF_{\sfm}\big|^2  \delta^4 (\sfx-\sfF_{\sfm}(v)) \bigg]~.
\nn
\end{align}
Much like $\omega$, the momentum harmonic function $\cF$ is sourced explicitly by the stress tensors of the brane sources,%
\footnote{On a winding effective string in the sort of static gauge solution $\partial_v \sfX^v\tight=\partial_u \sfX^u\tight=1$, $\partial_v \sfF^v\tight=\partial_u \sfF^u\tight=1$ we are working with here, the $v$-$v$ component of the spacetime stress tensor $T_{vv}$ is identical to the stress tensor $\sfT_{++}$ on the worldsheet.  The latter can be simply read off of the effective actions~\eqref{SF1}, \eqref{SNS5}.  }
given by the second line, as well as by bulk terms quadratic in forms/functions determined by the layer 1 equations.  Again using~\eqref{gauge5}, one has
\begin{align}
\label{Einsteinvv}
\nabla^2 \mathcal{F} &=
	-2\partial_v Z_1 \partial_v Z_2-\frac{1}{2}\Theta_{1ij}\Theta_{2ij}
\nn\\
&\hskip .5cm
+2\kappa_6^2\Big[ \tau_{\sst\rm F1} \sum_{\sfn=1} ^{\none} \Big(Z_2  \big|\partial_v \sfX_{\sfn}\big|^2 + \big|\partial_v\sfZ _{\sfn}\big|^2  \Big)\delta^4 (\sfx-\sfX_{\sfn}(v)) 
+\tau_{\sst\rm NS5} \sum_{\sfm=1} ^{\nfive} Z_1 \big|\partial_v \sfF_{\sfm}\big|^2  \delta^4 (\sfx-\sfF_{\sfm}(v)) \Big]~.
\end{align}
After some simplifications, the distributed sources for $\cF$ are given by
\begin{align}
\label{Intermediatevv}
&- 2 \partial_v Z_1 \partial_v Z_2-\frac{1}{2} \Theta_{1ij}\Theta_{2ij} 
= \frac{2\kappa_6^4 \tau_{\sst\rm NS5}\tau_{\sst\rm F1}}{\pi^4} \sum_{\sfm =1} ^{\nfive} \sum_{\sfn =1} ^{\none}\frac{-(\partial_v \sfF_{\sfm }\!\cdot \partial_v \sfX_\sfn ) (\widetilde\sfR_{\sfm }\cdot \sfR_\sfn ) +\mathcal{A}_{ij}\partial_v \sfF_{\sfm j} \partial_v \sfX_{\sfn i}}{\sfR_\sfn  ^4 \widetilde\sfR_\sfm ^4}~.
\end{align}
Then the following ansatz%
\footnote{Note that this ansatz differs slightly from that of~\rcite{Niehoff:2012wu}, which solved the bulk field equations without regard to matching onto sources, and so simply included a term $Q_p/r^2$ in $\cF$ to carry an independently specified momentum charge.  Because we have taken care to incorporate the brane effective actions into the variational principle, we are guaranteed to match the bulk solution onto the correct source terms arising from the branes, which (1) are not constant, but rather depend explicitly on $v$; and (2) determine the total momentum $Q_p$ in terms of the contributions of the individual brane stress tensors $\sfT_{++}$ (see Eq.~\eqref{CFT stress} below).}
\begin{align}
\label{CurlyF}
	\mathcal{F} &=  \cF_0 + \cF_1 + \cF_{2l} + \cF_{2r} 
\nn\\[.2cm]
\cF_0 &=
    -\frac{\kappa_6^2 \tau_{\sst\rm NS5}}{2\pi^2}\sum_{\sfm =1} ^{\nfive} \frac{|\partial_v \sfF_{\sfm } |^2}{|\sfx-\sfF_\sfm|^2} -\frac{\kappa_6^2 \tau_{\sst\rm F1}}{2\pi^2}\sum_{\sfn =1} ^{\none} \frac{|\partial_v \sfX_{\sfn } |^2+|\partial_v\sfZ_{\sfn}|^2}{|\sfx-\sfX_\sfn|^2}
\nn\\[.1cm]
\cF_1 &=
    -\frac{\kappa_6 ^4 \tau_{\sst \rm NS5}\tau_{\sst \rm F1}}{4\pi^4}\sum_{\sfm=1} ^{\nfive} \sum_{\sfn=1} ^{\none} \frac{\partial_v\sfF_{\sfm}\cdot \partial_v \sfX_{\sfn }}{|\sfx-\sfF_\sfm|^2|\sfx-\sfX_\sfn|^2}
\nn\\[.1cm]
\cF_{2l} &= 
    -\frac{\kappa_6^4 \tau_{\sst\rm NS5} \tau_{\sst\rm F1}}{4\pi^4}\sum_{\sfm =1} ^{\nfive} \sum_{\sfn =1} ^{\none}  \frac{\partial_v \sfD_{\sfm\sfn}^i}{|\sfR_\sfn  - \widetilde\sfR_\sfm |^2}\Big(\frac{\partial_v\sfF_\sfm^i}{\widetilde\sfR_\sfm ^2} - \frac{\partial_v\sfX_\sfn^i}{\sfR_\sfn ^2} \Big) 
\\[.1cm]	
    \cF_{2r} &=- \frac{\kappa_6^4 \tau_{\sst\rm NS5} \tau_{\sst\rm F1}}{2\pi^4}\sum_{\sfm =1} ^{\nfive} \sum_{\sfn =1} ^{\none}\frac{\mathcal{A}_{ij} \partial_v \sfF_{\sfm i} \partial_v \sfX_{\sfn j}}{\widetilde\sfR_\sfm  ^2 \sfR_\sfn ^2|\sfR_\sfn - \widetilde\sfR_\sfm |^2} 
\nn
\end{align}
solves~\eqref{Einsteinvv}, since applying the Laplacian yields
\begin{align}
\label{delsqF}
\nabla^2 \mathcal{F} & = 2\kappa_6^2 \tau_{\sst\rm NS5} \sum_{\sfm =1} ^{\nfive} |\partial_v \sfF_\sfm|^2 Z_1 \delta^4 (\sfx-\sfF_\sfm) + 2\kappa_6^2 \tau_{\sst\rm F1} \sum_{\sfn =1} ^{\none} \Big( Z_2|\partial_v \sfX_\sfn |^2 + |\partial_v\sfZ_{\sfn}|^2\Big)  \delta^4 (\sfx-\sfX_\sfn )
\\
&-\frac{2\kappa_6^4 \tau_{\sst\rm NS5} \tau_{\sst\rm F1}}{\pi^4} \sum_{\sfm =1} ^{\nfive} \sum_{\sfn =1} ^{\none} \frac{(\partial_v \sfF_\sfm \!\cdot \partial_v \sfX_\sfn ) (\sfR_\sfn  \!\cdot \widetilde\sfR_\sfm )}{\widetilde\sfR_\sfm ^4 \sfR_\sfn ^4}+\frac{2\kappa_6 ^4 \tau_{\sst\rm NS5} \tau_{\sst\rm F1}}{\pi^4}\sum_{\sfm =1} ^{\nfive} \sum_{\sfn =1} ^{\none} \frac{\mathcal{A}_{ij} \partial_v \sfF_{\sfm i} \partial_v \sfX_{\sfn j}}{\widetilde\sfR_\sfm ^4  \sfR_\sfn ^4}~.\nn
\end{align}
Therefore, the $v$-$v$ Einstein equation (\ref{Einsteinvv}) is satisfied, and the conclusion is that supergravity plus the worldvolume effective field theory admits such classical solutions. 

For the convenisnce of the reader and future reference, we collect the expressions for the harmonic forms and functions $Z_2$~\eqref{Z2soln}, $Z_1$~\eqref{Z1soln}, $\Theta_1$~\eqref{Theta1}, $\Theta_2$~\eqref{Theta2}, $\omega$~\eqref{omegaAnsatz}, and $\cF$~\eqref{CurlyF} in Appendix~\ref{app:soln sum}.

\subsection{(Non)singularities of the solution}
\label{sec:singularity}

The geometry derived above appears to have double pole singularities in the metric coefficients and other supergravity fields at the source locations $\sfx=\sfF_\sfm$, $\sfx=\sfX_\sfn$, as well as when the string and fivebrane sources collide $\sfX_\sfn=\sfF_\sfm$.  However, none of these is an actual singularity in string theory~-- they are all artifacts of the supergravity approximation.  

As discussed in the introduction, the singularities at $\sfx=\sfF_\sfm$ are resolved by $\alpha'$ effects~-- string wavefunctions don't fit down isolated fivebrane throats.  This phenomenon was originally observed in a particular solvable example of separated NS5-branes (in the absence of string or momentum charge), where the fivebranes are placed in a $\bZ_\nfive$-symmetric array on a circle of radius $\mfa$ on their Coulomb branch.  In this special source configuration, worldsheet string theory is exactly solvable in the decoupling limit~\eqref{decoupling}~\rcite{Giveon:1999px,Giveon:1999tq}, and the loop expansion is well-behaved provided the radius $\mfa$ is sufficiently large (as discussed in the introduction and in section~\ref{sec:ST scales}, sufficiently large that D-branes stretched between the fivebranes are sufficiently heavy and thus suppressed).  In the construction of~\rcite{Giveon:1999px,Giveon:1999tq}, the worldsheet conformal field theory is a gauged Wess-Zumino-Witten (WZW) model for the coset
\be
\frac\cG\cH = \bigg(\frac{\sltwo_\nfive}{\uone}\times\frac{\sutwo_\nfive}{\uone}\bigg)\Big/\bZ_\nfive ~,
\ee
which is completely well-behaved.  In particular, the first factor $\frac\sltwo\uone$ has the cigar-like geometry
\begin{align}
ds^2 &= \nfive\alpha'\Big(\frac{dr^2+r^2d\phi^2}{\mfa^2+r^2}\Big)
~~,~~~~
e^{2\Phi} = \frac{\nfive\alpha'}{\mfa^2+r^2} ~.
\end{align}
Strings see a geometry which caps off at $r=0$, where the string coupling saturates at a value set by the fivebrane separation $\mfa$.

Strings don't see the double pole singularities in the fivebrane harmonic function $Z_2$~\eqref{Z2soln}, because they can't get close enough to the fivebranes.  At the scale $\mfa$, the naive geometry of the supergravity approximation transitions from a linear dilaton throat of radius $\sqrt{\nfive\alpha'}$ into $\nfive$ separate ``little throats'', each of size $\sqrt{\alpha'}$.  But to the extent that one can believe such a stringy geometry, strings don't fit down such a little throat, and so they never see a string coupling stronger than the value of the dilaton at the top of these little throats; instead, the effective geometry smoothly caps off at this scale.

A more nuanced perspective is provided by the recent construction~\cite{Dei:2025ilx} of worldsheet string theory in the decoupled throat of a single fivebrane.  The WZW model at $\nfive=1$ has only a single affine $\sltwo$ representation in its physical spectrum, whose wavefunction is essentially constant (it lies at the threshold $j=1/2$ of the continuous series representations).  With such a limited representation content, one cannot build local wavepackets traveling up or down such a throat.  Since strings represent the small fluctuations of the background, we see that the geometry of such a throat is entirely rigid.  

Strings traveling down the thoat of $\nfive$ separated fivebranes reach the cap, and reflect back with unit probability~\cite{Giveon:1999tq}.  There is no transmission amplitude into the little throats.  There is however a reflection phase shift, which codes the depth to which a string penetrates before it reflects, which grows with the radial momentum~\cite{Giveon:2015cma,Giveon:2016dxe}.  Thus, while strings don't fit into the little throats, one can try to shove them down these throats, but they can only get so far before they turn around and head back out. 

When we compactify and add momentum to the fivebrane to turn it into an NS5-P supertube, a very similar result holds~\cite{Martinec:2017ztd,Martinec:2018nco}.  The circular supertube that arises when only a single Fourier mode of the fivebrane is excited, once again corresponds to an exactly solvable worldsheet theory (see Section~\ref{sec:superstrata} and Appendix~\ref{app:GWZW}).  Now the scale $\mfa$ is dynamically set by the momentum charge $n_p$ carried by the fivebranes; the fivebranes are well-separated at large $n_p$, and string perturbation theory is valid.  There are now more singularities in the supergravity approximation~-- not only the transverse metric coefficient $Z_2$~\eqref{Z2soln}, but also the momentum harmonic function $\cF$~\eqref{CurlyF} and the angular momentum harmonic function $\omega$~\eqref{omegaAnsatz} (for two-charge backgrounds, only the contributions $\cF_0$ and $\omega_0$ are non-zero).  All of these apparent singularities are artifacts of classical supergravity, and are totally benign in perturbative string theory, provided $n_p/n_5$ is large.  The worldsheet theory is again a gauged WZW model~\cite{Martinec:2017ztd} with a slightly different embedding of the gauge group, and so is just as sensible as the worldsheet theory for Coulomb branch fivebranes.

These features of the exactly solvable examples continue to hold for a large class of deformations along the Coulomb branch moduli space away from this point, and deformations in the NS5-P supertube configuration space~\cite{Martinec:2020gkv,Martinec:2024emf}, for which a worldsheet dual Landau-Ginsburg CFT describes the near-source region.  One sees quite clearly in this dual description why perturbative string dynamics remains sensible, and how it fails due to strong coupling effects whenever the fivebranes come close to intersecting.

When we quantize strings in the background of the fivebranes, they live in some wavefunction which is spread out over the string scale $\alpha'$ or more, and the singularities at $\sfx=\sfX_\sfn$ in the harmonic forms/functions are smeared over and resolved (otherwise string perturbation theory would be pathological; string theory became a theory of quantum gravity precisely because it resolves the short-distance singularities of particles and strings in general relativity).  In the gauged WZW model for the NS5-P supertube, physical string wavefunctions are a non-singular gauge projection of those on a group manifold, and are thus manifestly non-singular; perturbative string amplitudes are built out of correlations of string dynamics on a group manifold, and so are again manifestly non-singular.

Finally, there are the apparent singularities in the momentum and angular momentum harmonic fuctions $\cF,\omega$ when a string approaches a fivebrane, $\sfX_\sfn\to \sfF_\sfm$, via the contributions $\cF_{2l},\cF_{2r},\omega_{2l},\omega_{2r}$ in~\eqref{omegaAnsatz}, \eqref{CurlyF}.  But as we have already established, strings don't see this geometry because (a) the fivebrane background is smeared over the scale of its self-separation; and (b) perturbative strings can't get close enough to the fivebranes to see any such pathology.

Thus, all of the apparent singularities in the geometries we have constructed are artifacts of the classical supergravity approximation; perturbative string theory will be entirely well-behaved as long as the fivebranes are sufficiently self-separated that there are no regions in the background where strong-coupling effects begin to take hold, that perturbative strings can visit.  The geometry seen by strings effectively caps off at the scale set by the fivebrane separation, which for generic fivebrane source profiles is the scale~\eqref{walksize}.

\subsection{Conserved charges}
\label{sec:charges}

The expectation value of the $\bS^1_y$ energy-momentum tensor can be read off the asymptotic form of $g_{vv}$
\be
\label{Tdef}
g_{vv} = -\frac{\cF}{Z_1} ~\sim~ \frac{\kappa_6^2 \cT(v)}{2\pi^3 R_y^2 r^2} ~.
\ee
We have
\be
\label{CFT stress}
\frac{1}{\pi R_y^2}\cT = \tau_{\sst\rm F1}\sum_{\sfn =1} ^{\none} \Big(|\partial_v \sfX_\sfn |^2 + |\partial_v\sfZ_{\sfn}|^2\Big) +  \tau_{\sst\rm NS5} \sum_{\sfm =1}^\nfive |\partial_v\sfF_\sfm|^2
+\frac{\kappa_6 ^2 \tau_{\sst\rm F1} \tau_{\sst\rm NS5}}{4\pi^4} \sum_{\sfm =1} ^{\nfive} \sum_{\sfn =1} ^{\none} \frac{|\partial_v\sfX_\sfn -\partial_v\sfF_\sfm|^2}{|\sfX_\sfn -\sfF_\sfm|^2} ~,
\ee
which agrees with the source stress tensor derived from the joint effective action~\eqref{JointAction} of the brane sources, to be discussed in the next subsection.
We also have the total momentum charge
\be
\label{Pcharge}
Q_p =  \oint\!dv\, \cT(v) ~.
\ee

Similarly, the asymptotics of $g_{vi}$ can be put in the form
\be
\label{omasymp}
g_{vi} = -\frac{\omega_i}{Z_1}
\sim -\frac{2\kappa_D ^2}{\Omega_{D-2}} \bigg(\frac{\cP^i (v)}{r^{D-3}}
+ \frac{\cJ^{ij} \sfx^j}{r^{D-1}} \bigg)
~,
\ee
(with $D=5$), from which one reads off the momentum and angular momentum currents $\cP^i(v)$ and $\cJ^{ij}(v)$.

The solution~(\ref{omegaAnsatz}) for $\omega$ above does not quite have the requisite form:
\begin{align}
\label{omegaMult}
 \omega_i \sim~& \frac{1}{r^2} \Big( \frac{\kappa_6 ^2 \tau_{\sst\rm  NS5}}{ 2\pi^2} \sum_{\sfm =1} ^{\nfive} \partial_v \sfF_{\sfm i} +\frac{\kappa_6 ^2 \tau_{\sst\rm  F1}}{ 2\pi^2} \sum_{\sfn =1} ^{\none} \partial_v \sfX_{\sfn i}   \Big)
 \nonumber\\
 &+\frac{\sfx^j}{r^4} \bigg( \frac{\kappa_6 ^2 \tau_{\sst\rm  NS5}}{ \pi^2} \sum_{\sfm =1} ^{\nfive} \partial_v \sfF_{\sfm i}~  \sfF_{\sfm j}  +\frac{\kappa_6 ^2 \tau_{\sst\rm  F1}}{ \pi^2} \sum_{\sfn =1} ^{\none} \partial_v \sfX_{\sfn i} \sfX_{\sfn j} +\frac{\kappa_6 ^2}{4\pi^3} \sum_{\sfm =1} ^{\nfive} \sum_{\sfn =1} ^{\none} \frac{ \sfD_{\sfm\sfn} \cdot \partial_v \sfD_{\sfm\sfn}}{|\sfD_{\sfm\sfn}|^2} \delta_{ij}   
 \nonumber\\
 &\qquad\quad +\frac{\kappa_6 ^2}{4\pi^3} \sum_{\sfm =1} ^{\nfive} \sum_{\sfn =1} ^{\none} \frac{   \sfD_{\sfm\sfn,j}\partial_v \sfD_{\sfm\sfn,i}-\sfD_{\sfm\sfn,i} \partial_v \sfD_{\sfm\sfn,j} -\epsilon_{ijkl} \sfD_{\sfm\sfn,k}\partial_v \sfD_{\sfm\sfn,l}}{| \sfD_{\sfm\sfn} |^2} \bigg)~,
\end{align}
where $\sfD_{\sfm\sfn}$ is defined in Eq.~\eqref{deltas}.
The terms symmetric in $i,j$ on the second line come at the same order as $\cJ^{ij}$, but can be removed by a coordinate transformation
\begin{equation}
 \omega_{i} \to \omega_{i} ' = \omega_{i} - \partial_v \xi_i -\partial_i \xi_v ~,
\end{equation}
where we choose
\begin{align}
\label{reparam}
  \xi_i  &= 
  \frac{\sfx_j}{2r^4} \bigg( \frac{\kappa_6 ^2 \tau_{\sst\rm  NS5}}{ \pi^2} \sum_{\sfm =1} ^{\nfive} \sfF_{\sfm i}~  \sfF_{\sfm j}+\frac{\kappa_6 ^2 \tau_{\sst\rm  F1}}{ \pi^2} \sum_{\sfn =1} ^{\none} \sfX_{\sfn i}~  \sfX_{\sfn j} \bigg) ~,
\nn\\[.2cm]
  \xi_v &= \frac{\kappa_6 ^2}{8\pi^3 r^2} \sum_{\sfm =1} ^{\nfive} \sum_{\sfn =1} ^{\none} \frac{ \sfD_{\sfm\sfn}\cdot \partial_v \sfD_{\sfm\sfn}}{ |\sfD_{\sfm\sfn}|^2}~.
\end{align}
The transverse momentum current is thus
\begin{align}
\label{Pi}
  \frac{1}{2\pi R_y}\cP^i (v) = \frac{1}{2} \Big( \tau_{\sst\rm NS5} \sum_{\sfm=1} ^{\nfive} \partial_v \sfF_{\sfm} +\tau_{\sst\rm F1} \sum_{\sfn=1} ^{\none} \partial_v \sfX_{\sfn} \Big) ~,
\end{align}
and the angular momentum current is
\begin{align}
\label{Jij}
   \frac{1}{2\pi R_y} \cJ^{ij} &=  \frac{1}{2} \bigg(\tau_{\sst\rm NS5} \sum_{\sfm =1} ^{\nfive} \big( \sfF_{\sfm i} \partial_v \sfF_{\sfm j} - \sfF_{\sfm j} \partial_v \sfF_{\sfm i} \big) + \tau_{\sst\rm F1} \sum_{\sfn =1} ^{\none} \big( \sfX_{\sfn i} \partial_v \sfX_{\sfn j}-\sfX_{\sfn j} \partial_v \sfX_{\sfn i}  \big) \bigg)\nonumber\\[.1cm]
    &\hskip .5cm + \frac{1}{4 \pi} \sum_{\sfm =1} ^{\nfive} \sum_{\sfn =1} ^{\none} \frac{  \sfD_{\sfm\sfn,i}\partial_v \sfD_{\sfm\sfn,j}-\sfD_{\sfm\sfn,j}\partial_v \sfD_{\sfm\sfn,i}  +\epsilon_{ijkl} \sfD_{\sfm\sfn}\partial_v \sfD_{\sfm\sfn,k}}{|\sfD_{\sfm\sfn,l}|^2}~.
\end{align}

For completeness, we can also write currents whose integrals provide the quantized number of string winding and fivebrane wrapping around the $y$-circle:
\begin{align}
    \cJ_{\rm\sst F1} =  \frac{1}{2} \tau_{\sst \rm NS5}\sum_{\sfn=1} ^{\none}\partial_v \sfX_{\sfn} ^y 
    ~~,~~~~
    \cJ_{\rm\sst NS5} = \frac{1}{2} \tau_{\sst \rm F1} \sum_{\sfm=1} ^{\nfive}\partial_v \sfF_{\sfn} ^y~.
\end{align}
We have used residual gauge symmetries to choose a static gauge, in which these currents are constant, and the charges are trivially $\none$ and $\nfive$ respectively.

Similar to supersymmetric unsmeared two-charge NS5-P solutions, and F1-P solutions smeared on $\bT^4$, the above three-charge solutions admit harmonic functions with quadratic singularities at the source positions. As mentioned in the introduction, stringy effects are expected to resolve these singularities, a feature based on known exact worldsheet constructions~\cite{Giveon:1999px,Martinec:2017ztd}. For more discussion, see \rcite{Martinec:2024emf}.  For fundamental string sources, quantum fluctuations distribute the source over the scale set by the string tension such that the singularity is resolved on that scale.

In addition, the functions $\omega_i,\mathcal{F}$ possess single pole singularities as sources are approached.  These milder singularities are expected to be resolved by $\alpha'$ corrections as well.

We checked that curvature invariants such as the Ricci scalar, the Ricci tensor squared, and the Kretschmann scalar (and as a byproduct, the Weyl tensor squared as well) all depend solely on $Z_1,Z_2$ and their derivatives. Hence, the behavior of $\mathcal{F}$ and $\omega$ is completely irrelevant to the regularity of lack thereof of the geometry in question. This is true both in the string frame and in the Einstein frame. 
Physically, one is simply moving around a string or fivebrane source, so the singularity is locally that of the string or fivebrane.
Another aspect of the benign nature of the aforementioned pole singularities, is that the metric function $g_{vv}=-\frac{\mathcal{F}}{Z_1}$, and $g_{vi}$ and also $B_{vi}$ depend on the ratio $\frac{\omega_i}{Z_1}$. In the vicinity of fundamental strings, these ratios lack a singularity that one might think arises from the pole singularities of $\omega,\mathcal{F}$. Ultimately, we expect that an $\alpha'$ exact description of the three-charge system would bring about regular $\omega$ and $\mathcal{F}$, as is seen in the particular examples for which an exact string worldsheet theory is available~\rcite{Martinec:2017ztd}.

\section{The joint effective action}
\label{sec:joint}

Having in hand the supergravity solution, one can plug it into the effective action for the strings and fivebranes.  The fundamental strings of course see the background sourced by the fivebranes through the first layer equations, and the fivebranes see the geometry sourced by the strings through their first layer equations; and both see the parts of the geometry resulting from the second layer equations.  Overall, one finds a joint effective action, which we expect to be the leading contribution to the effective dynamics of the brane sources.

We begin with the string action, specialized to the background~\eqref{AnsatzIIB} with $\beta=0$
\begin{align}
\cS_{\sst\rm F1} &=\frac{1}{2\pi \alpha'}\int d^2 \sigma \left(G_{\mu \nu} \partial_{+} \sfX^\mu \partial_- \sfX^\nu+ G_{IJ} \partial_+\sfZ^ I \partial_- \sfZ^J  + B_{\mu \nu}  \partial_+ \sfX^\mu \partial_- \sfX^\nu \right)
\nonumber\\
\label{Polyakov3}
&=\tau_{\sst\rm F1}\int d^2 \sigma \left( -\frac{2}{Z_1} \partial_+ \sfX^u \partial_- \sfX^v -\frac{2\omega_i}{Z_1}\partial_+ \sfX^i \partial_- \sfX^v  -\frac{\mathcal{F}}{Z_1} \partial_+ \sfX^v \partial_- \sfX^v + Z_2\, \partial_+ \sfX^i \partial_- \sfX^i\right.\nonumber\\
	&\hskip 3cm \left. +G_{IJ} \partial_+ \sfZ^ I \partial_- \sfZ^J    +a_{1i} \partial_+ \sfX^i \partial_- \sfX^v - a_{1i} \partial_- \sfX^i \partial_+ \sfX^v +\gamma_{2kl} \partial_+ \sfX^k \partial_- \sfX^l \right)~.
\end{align}
The Virasoro constraint $\sfT_{++}=0$ is 
\begin{equation}
	\label{Vir}
	 -\frac{2}{Z_1}\partial_+ \sfX^u  =  \frac{2\omega_i}{Z_1} \partial_+ \sfX^i  +\frac{\mathcal{F}}{Z_1} \partial_+ \sfX^v -\frac{Z_2}{\partial_+ \sfX^v} \partial_+ \sfX^i\partial_+ \sfX^i-\frac{1}{\partial_+ \sfX^v}  G_{IJ} \partial_+ \sfZ^I \partial_+ \sfZ^J ~.
\end{equation}
Plugging this relation into Eq.~(\ref{Polyakov3}) yields
\begin{align}
\label{Polyakov5}
\cS_{\sst\rm F1}=\tau_{\sst\rm F1}\int d^2 \sigma &\bigg(  -\frac{Z_2}{\partial_+ \sfX^v} \partial_+ \sfX^i\partial_+ \sfX^i \partial_- \sfX^v-\frac{1}{\partial_+ \sfX^v}  G_{IJ} \partial_+ \sfX^I \partial_+ \sfX^J\partial_- \sfX^v +Z_2 \partial_+ \sfX^i \partial_- \sfX^i  
\nn\\
&\hskip 1cm
+G_{IJ} \partial_+ \sfZ^I \partial_- \sfZ^J+a_{1i} \partial_+ \sfX^i \partial_- \sfX^v - a_{1i} \partial_- \sfX^i \partial_+ \sfX^v +\gamma_{2kl} \partial_+ \sfX^k \partial_- \sfX^l \bigg)~.
\end{align}
The fivebrane effective action similarly reads
\begin{align}
\label{Polyakov6}
	\cS_{\sst\rm NS5}&=\tau_{\sst\rm NS5}\int d^2 \sigma \bigg( -\frac{2  }{Z_2} \partial_+ \sfF^u \partial_- \sfF^v -\frac{2 \omega_i}{Z_2}\partial_+ \sfF^i \partial_- \sfF^v  -\frac{\mathcal{F}}{Z_2} \partial_+ \sfF^v \partial_- \sfF^v 
\\
	&\hskip 3cm   +Z_1 \partial_+ \sfF^i \partial_- \sfF^i   +a_{2i} \partial_+ \sfF^i \partial_- \sfF^v - a_{2i} \partial_- \sfF^i \partial_+ \sfF^v +\gamma_{1kl} \partial_+ \sfF^k \partial_- \sfF^l \bigg)~.
\nn
\end{align}
The Virasoro constraint on the fivebrane effective worldsheet is 
\begin{equation}
\label{Vir2}
    -\frac{2 }{Z_2} \partial_+ \sfF^u=\frac{2 \omega_i}{Z_2} \partial_+ \sfF^i + \frac{ \cF}{Z_2} \partial_+ \sfF^v - \frac{Z_1}{\partial_+ \sfF^v} \partial_+ \sfF^i \partial_+ \sfF^i ~.
\end{equation}
Substituting Eq.~(\ref{Vir2}) into Eq.~(\ref{Polyakov6}) gives
\begin{align}
\label{Polyakov7}
\cS_{\sst\rm NS5}=\tau_{\sst\rm NS5}\int d^2 \sigma &\bigg( -\frac{Z_1}{\partial_+ \sfF^v} \partial_+ \sfF^i \partial_+ \sfF^i \partial_- \sfF^v   +Z_1 \partial_+ \sfF^i \partial_- \sfF^i   
\nn\\
&\hskip1cm
+a_{2i} \partial_+ \sfF^i \partial_- \sfF^v - a_{2i} \partial_- \sfF^i \partial_+ \sfF^v +\gamma_{1kl} \partial_+ \sfF^k \partial_- \sfF^l \bigg)~.
\end{align}
Next, we use the gauge conditions $\partial_+ \sfX^v = \partial_+ \sfF^v=1$ and $\partial_- \sfX^v = \partial_- \sfF^v =0$ and plug them in the joint effective action, which is that sum of Eqs.~(\ref{Polyakov5}) and (\ref{Polyakov7}):
\begin{align}
\label{Joint}
  \cS_{\text{joint}} &= \tau_{\sst\rm F1}\sum_{\sfn =1} ^{\none}\int d^2 \sigma \Big( Z_2 \partial_+ \sfX^i _{\sfn} \partial_- \sfX^i_{\sfn} +G_{IJ} \partial_+ \sfZ^I \partial_- \sfZ^J-a_{1i} \partial_- \sfX^i_{\sfn} +\gamma_{2kl} \partial_+ \sfX^k_{\sfn} \partial_- \sfX^l_{\sfn}\Big) \nonumber\\ 
  &\hskip .5cm
  +\tau_{\sst\rm NS5}\sum_{\sfm =1} ^{\nfive} \int d^2 \sigma \Big( Z_1 \partial_+ \sfF^i _{\sfm} \partial_- \sfF^i _{\sfm} -a_{2i} \partial_- \sfF^i _{\sfm} +\gamma_{1kl} \partial_+ \sfF^k _{\sfm} \partial_- \sfF^l _{\sfm}\Big)~. 
\end{align}
In this action, $Z_2,a_1$ and $\gamma_2$ are evaluated at the embedding of the string and $Z_1,a_2,\gamma_1$ are evaluated at the embedding of the fivebrane.  One finds after some algebra
\begin{align}
\label{JointAction}
  \cS_{\text{joint}} = \int d^2 \sigma &\bigg[   \tau_{\sst\rm NS5}\sum_{\sfm =1}^{\nfive}\partial_+ \sfF^i _{\sfm} \partial_- \sfF^i _{\sfm}+ \tau_{\sst\rm F1}\sum_{\sfn =1}^{\none}\partial_+ \sfX^i _{\sfn} \partial_- \sfX^i _{\sfn}+\tau_{\sst \rm F1}G_{IJ} \partial_+ \sfZ^I \partial_- \sfZ^J   
\nn\\[.1cm]
    &\hskip.5cm
    +\frac{\kappa_6 ^2 \tau_{\sst\rm NS5} \tau_{\sst\rm F1}}{2\pi^2} \sum_{\sfn =1} ^{\none} \sum_{\sfm =1} ^{\nfive} \frac{\partial_+ (\sfX_\sfn - \sfF_\sfm )\cdot \partial_- (\sfX_\sfn  - \sfF_\sfm )}{|\sfX_\sfn  - \sfF_\sfm|^2}
    \\[.1cm]
    &\hskip1cm 
     + \tau_{\sst\rm F1}\sum_{\sfn =1} ^{\none}\Big( b_{1i} (\sfX_\sfn) \partial_- \sfX^i_\sfn  + \gamma_{2kl}(\sfX_\sfn)\, \partial_+ \sfX^k_{\sfn} \partial_- \sfX^l_{\sfn} \Big)
\nn\\[.1cm]
    &\hskip1.5cm 
    + \tau_{\sst\rm NS5}\sum_{\sfm =1} ^{\nfive}\Big( b_{2i} (\sfF_\sfm) \partial_- \sfF^i_\sfm +\gamma_{1kl}(\sfF_\sfm)\, \partial_+ \sfF^k _{\sfm} \partial_- \sfF^l_\sfm\Big)\bigg]~.
\nn
\end{align}
A similar effective action arose in an analysis of the S-dual D1-D5 system in~\rcite{Papadopoulos:2000hb}, where due to the amount of supersymmetry, agreement was found between (1) an evaluation of the Coulomb branch moduli space metric of the branes, and (2) the gauge theory effective action obtained by integrating out 1-5 strings at one loop.  Here we find a similar result in a somewhat different context through an analysis of 1/8-BPS supergravity solutions. Our result also includes explicit torsion terms expressed in the last two lines of Eq.~(\ref{JointAction}), and generalizes it to branes carrying arbitrary left-moving wave profiles.

The joint effective action governs slightly non-BPS excitations.  Indeed, it was arrived at in~\rcite{Papadopoulos:2000hb} as the moduli space metric for Coulomb branch excitations of the onebrane-fivebrane system, which by construction governs the leading non-BPS excitations.  The full nearly-BPS effective action must include the contributions of the bulk supergravity fields, but this contribution reduces to a boundary term~\rcite{Tseytlin:1988tv,Chen:2021dsw}, which vanishes for asymptotically flat horizonless solutions like the ones we have derived in this section.

Extending the analogy between our setup and that of electric/magnetic charges in 4d gauge theory, the near-BPS dynamics of monopoles and dyons is governed by a 0+1d sigma model on the BPS moduli space~\rcite{Manton:1981mp}.  Similarly here, the near-BPS dynamics is a 1+1d sigma model for the string sources, that is determined by the BPS solutions and captures the leading interactions between the sources when they have a small relative (right-moving) velocity in addition to their collection of left-moving mutually BPS excitations.

This effective dynamics has some similarity to the treatment of near-BPS excitations of three-charge black holes in~\rcite{Callan:1996dv,Dhar:1996vu,Das:1996wn}, where the string absorbing/emitting energy is an effective approximation to the little string sitting inside the fivebranes (oriented {\it along}~$\bT^4$); here one is instead absorbing energy onto the effective strings describing motion of the sources in their {\it transverse} space.

\section{The decoupling limit}
\label{sec:decoupling}

As in section~\ref{sec:blackness2}, we will be interested in taking the fivebrane decoupling limit by
making a similar rescaling of the transverse space coordinates (along the lines of~\eqref{decoupling})%
\footnote{We do not include factors of $n_1n_5$ or $R_y$ as in~\eqref{rhat defn}, which are specific to the $AdS_3$ decoupling limit considered there.}
\be
\label{rhat defn4}
\runsc = |\sfx_\flat| = \mu\,\rhat
~~,~~~~
\mu^2 =\frac{g_s^2(\alpha')^3}{V_4} ~.
\ee
With this rescaling, radial scales are referred to the scale set by the fivebrane tension~\eqref{tauNS5}.
The $\bT^4$ coordinates are also made dimensionless by referring them to the string length scale:
\begin{equation}
   \sfZ =  \sqrt{\alpha'} \,\hat \sfZ~. 
\end{equation}
We make this substitution, and subsequently drop the hat decoration to avoid notational clutter.

This rescaling has the effect of eliminating all prefactors involving the string tension $\tau_{\rm\sst F1}$, fivebrane tension $\tau_{\rm\sst NS5}$, and gravitational coupling $\kappa_6^2$ in the various harmonic functions, making the interpretation of the physics particularly transparent.  One exception is that the non-abelian scale of the fivebrane dynamics in rescaled variables is 
\be
\label{Lnonab}
\hat\ell_{na}^2 = \bigg[\frac{V_4}{(\alpha')^2}\bigg]^{\frac12} ~.
\ee
This scale can of course also be set to unity by choosing the torus size to be the string scale. 

The harmonic functions and forms are rescaled according to
\begin{align}
\label{newbless}
  	  	Z^\flat_1 &=  Z_1
        \quad~~,\qquad
  	  	Z^\flat_2 = \frac{\alpha'}{\mfa^2} Z_2
        \quad,\quad~ \mathcal{F}^\flat = \alpha' \mathcal{F}
        ~~~,~~~~~
  	  \omega^\flat_i = \frac{ \alpha'}{ \mfa} \omega_i
      ~~,~~~~
      \beta^\flat_i = \frac{ \alpha'}{\mfa} \beta_i~~.
\nn\\[.2cm]
  	  a^\flat_{1i} &= \frac{\alpha'}{\mfa} a_{1i}
      ~~,~~~~
  	  \gamma^\flat_{2ij}= \frac{\alpha'}{\mfa^2} \gamma_{2ij}
    ~~,~~~~
      a^\flat_{2i} = \mfa ~a_{2i} 
    ~~,~~~~    
  	  \gamma^\flat_{1ij} =  \gamma_{1ij}  ~~,
\end{align}
where again the $\flat$ appellation denotes asymptotically flat spacetime quantities (the variables we have been using until now in this section); after making the transformation, we reuse the same variables to write the rescaled solution, dropping the hat and $\flat$ decorations.  We hope this does not cause confusion.

The background functions and forms are now given by
\begin{subequations}
\label{Z2solne}
\begin{align}
 Z_2(\sfx,v) &=  \sum_{\sfm =1} ^{\nfive} \frac{1}{|\sfx-\sfF_\sfm(v)|^2} 
\\[.3cm]
\label{Z1solne}
 Z_1(\sfx,v) &=  1+ \sum_{\sfn =1} ^{\none} \frac{1}{|\sfx-\sfX_\sfn (v)|^2}  
\\[.3cm]
\label{Theta1e}
	\Theta_{1ij} &=
     -2\sum_{\sfm =1} ^{\nfive} \frac{\partial_v \sfF_{\sfm i} (\sfx_j-\sfF_{\sfm j}) - \partial_v \sfF_{\sfm j} (\sfx_i-\sfF_{\sfm i}) + \epsilon_{ijkl} \partial_v \sfF_{\sfm k} (\sfx_l-\sfF_{\sfm l}) }{|\sfx-\sfF_\sfm|^4}
\\[.3cm]
\label{Theta2e}
	\Theta_{2ij} &=
     -2\sum_{\sfn =1} ^{\none} \frac{\partial_v \sfX_{\sfn i} (\sfx_j-\sfX_{\sfn j}) - \partial_v \sfX_{\sfn j} (\sfx_i-\sfX_{\sfn i}) + \epsilon_{ijkl} \partial_v \sfX_{\sfn k} (\sfx_l-\sfX_{\sfn l}) }{|\sfx-\sfX_\sfn |^4}
\end{align}
\end{subequations}
for the first layer; and for the second layer,
\begin{subequations}
\label{Z2solnf}
\begin{align}
\label{omegaAnsatze}
	\omega_i &=  ~\sum_{\sfm =1} ^{\nfive} \frac{\partial_v \sfF_{\sfm i}}{\widetilde\sfR_\sfm  ^2} + \frac{1}{2} \sum_{\sfm =1} ^{\nfive} \sum_{\sfn =1} ^{\none} \frac{\partial_v \sfF_{\sfm i} + \partial_v \sfX_{\sfn i}}{\sfR_\sfn  ^2 \widetilde\sfR_\sfm ^2}
\\
    &\qquad-\frac{1}{2} \sum_{\sfm =1} ^{\nfive} \sum_{\sfn =1} ^{\none} \frac{\partial_v \sfF_{\sfm i}- \partial_v \sfX_{\sfn i}}{|\sfR_\sfn -\widetilde\sfR_\sfm |^2} \left(\frac{1}{\sfR_\sfn  ^2}-\frac{1}{\widetilde\sfR_\sfm ^2}\right)
+\sum_{\sfm =1} ^{\nfive} \sum_{\sfn =1} ^{\none} \frac{\partial_v \sfF_{\sfm j} - \partial_v \sfX_{\sfn j}}{|\sfR_\sfn  - \widetilde\sfR_\sfm |^2}\frac{\mathcal{A}_{ij}}{\sfR_\sfn  ^2 \widetilde\sfR_\sfm ^2}
\nn
\\[.3cm]
\label{CurlyFe}
	\mathcal{F} &=  
    -\sum_{\sfm =1} ^{\nfive} \frac{Z_1 (\sfF_\sfm)|\partial_v \sfF_{\sfm } |^2}{|\sfx-\sfF_\sfm|^2}- \sum_{\sfn =1} ^{\none} \frac{Z_2(\sfX_\sfn )|\partial_v \sfX_\sfn |^2+|\partial_v \sfZ_\sfn|^2}{|\sfx-\sfX_\sfn |^2} 
\nonumber\\
    &\qquad-\sum_{\sfm =1} ^{\nfive} \sum_{\sfn =1} ^{\none} \partial_v\sfF_{{m}}\cdot \partial_v \sfX_{\sfn }\bigg[ \frac{1}{\widetilde\sfR_\sfm ^2 R_{\sfn} ^2}  - \frac{1}{|\sfR_\sfn  - \widetilde\sfR_\sfm |^2}\Big(\frac{1}{\widetilde\sfR_\sfm ^2} + \frac{1}{\sfR_\sfn ^2} \Big) \bigg]
\\	
    &\qquad- 2\sum_{\sfm =1} ^{\nfive} \sum_{\sfn =1} ^{\none}\frac{\mathcal{A}_{ij} \partial_v \sfF_{\sfm i} \partial_v \sfX_{\sfn j}}{\widetilde\sfR_\sfm  ^2 \sfR_\sfn ^2|\sfR_\sfn - \widetilde\sfR_\sfm |^2}~. 
\nn
\end{align}
\end{subequations}

In the decoupling limit, the joint effective action becomes
\begin{align}
\label{JointAction2}
  \cS_{\text{joint}} = \frac{1}{2\pi}\int d^2 \sigma &\bigg(    \sum_{\sfm =1}^{\nfive}\partial_+ \sfF^i _{\sfm} \partial_- \sfF^i _{\sfm}+G_{IJ} \partial_+ \sfZ^I \partial_- \sfZ^J   
    + \sum_{\sfn =1} ^{\none} \sum_{\sfm =1} ^{\nfive} \frac{\partial_+ (\sfX_\sfn - \sfF_\sfm )\cdot \partial_- (\sfX_\sfn  - \sfF_\sfm )}{|\sfX_\sfn  - \sfF_\sfm|^2}
\nonumber\\
  &\hskip 0.5cm+ \sum_{\sfn =1} ^{\none} \big(b_{1i} (\sfX_\sfn) \partial_- \sfX^i_\sfn+\gamma_{2kl}(\sfX_{\sfn})\, \partial_+ \sfX^k_{\sfn} \partial_- \sfX^l_{\sfn} \big) 
\\
  &\hskip 1cm 
  + \sum_{\sfm =1} ^{\nfive} \big( b_{2i} (\sfF_\sfm) \partial_- \sfF^i_\sfm+\gamma_{1kl}(\sfF_{\sfm})\, \partial_+ \sfF^k _{\sfm} \partial_- \sfF^l _{\sfm} \big) \bigg)~.
\nn
\end{align}
In other words, the terms involving the quadratic kinetic term of the transverse string coordinates $\sfX$ scale away in the fivebrane decoupling limit $\sfX\sim O(g_s)$, $g_s\to 0$ because the normalization of the kinetic term only involves $\alpha'$, whereas the corresponding scaling $\sfF\sim O(g_s)$ is compensated by a factor of $1/g_s^2$ in the fivebrane tension; and the torus coordinates $\sfZ$ of the strings are not rescaled when taking the limit.

The stress energy current is given by
\be
\label{CFT stress2}
\frac{1}{\pi R_y^2}\cT = 
\frac{1}{2\pi} \sum_{\sfn =1} ^{\none}  |\partial_v\sfZ_{\sfn}|^2 + 
\frac{1}{2\pi} \sum_{\sfm =1}^\nfive |\partial_v\sfF_\sfm|^2
+\frac{1}{4\pi^3} \sum_{\sfm =1} ^{\nfive} \sum_{\sfn =1} ^{\none} \frac{|\partial_v\sfX_\sfn -\partial_v\sfF_\sfm|^2}{|\sfX_\sfn -\sfF_\sfm|^2} ~,
\ee
The momentum current is
\begin{align}
\label{Pi2}
  \frac{1}{2\pi R_y}\cP^i (v) = \frac{1}{4\pi}   \sum_{\sfm=1} ^{\nfive} \partial_v \sfF_{\sfm}^i ~,
\end{align}
The angular momentum current is
\begin{align}
\label{Jij2}
   \frac{1}{2\pi R_y} \cJ^{ij} =  \frac{1}{4\pi} \Bigg( &\sum_{\sfm =1} ^{\nfive} \big( \sfF_{\sfm }^i \partial_v \sfF_{\sfm }^j - \sfF_{\sfm }^j \partial_v \sfF_{\sfm }^i \big) 
\nn\\
&\hskip .5cm
   +  \sum_{\sfm =1} ^{\nfive} \sum_{\sfn =1} ^{\none} \frac{  \sfD_{\sfm\sfn}^i\partial_v \sfD_{\sfm\sfn}^j-\sfD_{\sfm\sfn}^j\partial_v \sfD_{\sfm\sfn}^i  +\epsilon_{ijkl} \sfD_{\sfm\sfn}^k\partial_v \sfD_{\sfm\sfn}^l}{|\sfD_{\sfm\sfn}|^2} \Bigg)~.
\end{align}
In terms of these, one has the angular momentum of left- and right-handed rotations of the angular three-sphere
\be
\label{JLR defs}
J_L = \cJ_{12} + \cJ_{34}
~~,~~~~
J_R = \cJ_{12} - \cJ_{34} ~.
\ee

\section{Circular superstrata and their stringy siblings}
\label{sec:superstrata}

A special case of the above construction is a semi-classical, T-dual version of superstratum geometries.  Superstrata are three-charge BPS supergravity solutions that are obtained as non-linear deformations of two-charge onebrane-fivebrane backgrounds, upon which a fully back-reacted supergravity wave is added to incorporate the third (momentum) charge.  For a review, see for instance~\rcite{Shigemori:2020yuo}.  In the semi-classical (WKB) limit, this supergravity wave localizes along a null geodesic and is well-approximated as a point particle source~\rcite{Bena:2025uyg}.  Starting with a momentum wave added to an NS5-F1 supertube background, T-duality yields a winding string condensate in an NS5-P supertube background, which is a special case of our construction above where we impose that the winding strings carry no momentum excitations themselves.  In this way, we develop a picture of generic superstrata.

When the charge quanta have the hierarchy $n_p < \none$ in the original duality frame of P excitations on top of an NS5-F1 supertube, this T-dual frame (where the sources are NS5-P and F1) is in fact the appropriate one for describing the near-source structure of superstrata in this regime of charge quanta.  It was pointed out in~\rcite{Martinec:2023xvf} that the NS5-P frame describes the near-source structure of generic two-charge BPS geometries.  For small enough values of the third (F1) charge in this duality frame, the geometry is not sufficiently deformed by the backreaction of the string winding to change this conclusion, and so continues to be the appropriate frame to describe the three-charge near-source structure.  The geometry will be that of Eq.~\eqref{AdS2two} until one reaches the outer envelope of the fivebrane star and beyond.

Regardless of the connection to the existing superstratum literature, it is of interest to consider the simplest cases of the general solution~\eqref{soln sum}.  We thus specialize the solution to round source profiles, having only a single momentum mode excited on each string or fivebrane.  We work in the fivebrane decoupling limit, to avoid the clutter of brane tensions and gravitational couplings in the expressions.

\subsection{Round source profiles}
\label{sec:helices}

\begin{figure}[ht]
\centering
\includegraphics[scale=0.4]{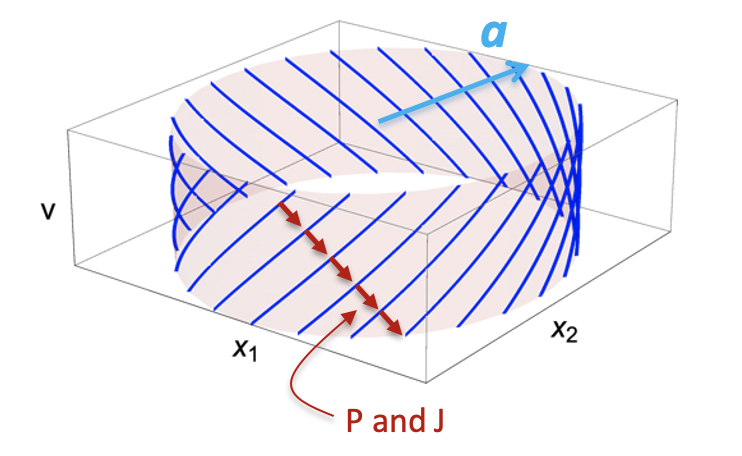}
\caption{\it A fivebrane source coherently spinning in a transverse plane. The BPS condition is satisfied if the net momentum is locally perpendicular to the brane; the brane thus carries both momentum along $y$ and angular momentum in the $\sfx^1$-\,$\sfx^2$ plane, which dynamically determines the radius $\mfa$. }
\label{fig:helix}
\end{figure}

Explicitly constructed superstrata are built on the helical fivebrane supertube profile depicted in figure~\ref{fig:helix}
\be
\label{roundF}
\sfF^1+i\sfF^2 = \mfa\, \exp\Big[\frac{i\sfp(t+y)}{n_5 R_y}\Big]
~~,~~~~
\sfF^3+i\sfF^4 =0 ~.
\ee
%
%
Similarly, we will consider helical F1 string profiles
\be
\label{roundX}
\sfX^1+i\sfX^2 = \mfb\, \exp\Big[ \frac{ik(t+y)}{w_y R_y} \Big]
~~,~~~~
\sfX^3+i\sfX^4 = 0 
\ee
with winding number $w_y$ and mode number $k$.  We take the number of such strings to be $n_1/w_y\in\bZ$, so that the total string winding is $n_1$.  We will work in the fivebrane decoupling limit of the previous section.

Evaluating the conserved charges $n_p, J_L, J_R$ associated to the currents~\eqref{CFT stress2}, \eqref{Jij2} and \eqref{JLR defs} of this background, one finds
\begin{align}
\label{round chgs}
n_p &= \frac{\sfp^2\mfa^2}{n_5\mu^2} + 
\frac{\none}{\nfive w_y^2|\mfb^2-\mfa^2|}\Big(\sfp^2 \mfa^2 w_y^2 + \nfive^2 k^2 \mfb^2 -{2\,}\nfive k\,\sfp w_y\times\text{min}(\mfa^2,\mfb^2)\Big)~~,
\nn\\[.2cm]
J_R &= \frac{\sfp \mfa^2}{2\mu^2}~~,
\\[.2cm]
J_L &= \frac{\sfp \mfa^2}{2\mu^2} + 
\frac{\none}{ w_y|\mfb^2-\mfa^2|}\Big(\sfp w_y \mfa^2  + \nfive k \mfb^2 -(\nfive\, k + \sfp w_y)\times\text{min}(\mfa^2,\mfb^2)\Big)
~.\nn
\end{align}
The seeming non-analyticity of these expressions as a function of parameters is an illusion; when evaluated in terms of the quantum numbers of the string, the expression is in fact analytic, as we show in Appendix~\ref{app:matching}.

We can compare the back-reacted geometry of the strings and fivebranes with exact worldsheet results for quantized BPS fundamental strings in the round NS5-P supertube background associated to the source~\eqref{roundF}, which we analyze in Appendix~\ref{app:GWZW}.  
For the round supertube, the windings of the fivebrane (and similarly for the string source) are very finely spaced on the string scale, and are not resolved in the supergravity approximation (strings resolve them, but only at the non-perturbative level in $\alpha'$~\rcite{Sfetsos:1998xd,Martinec:2017ztd}).  One can thus average the source over the circle of radius $\mfa$, or equivalently average over $v$.  In the following, we will carry out such a smearing of our solutions for round, single-mode sources.

The smeared sources form a ring in the transverse space.  It proves useful to describe them using Gibbons-Hawking (GH) coordinates on $\bR^4$
\begin{equation}
  \begin{aligned}
 \sfx^1 + i \sfx^2 ~&=~   2 \,\sqrt{\hat r} \,  \sin \bigg(\frac{\hat \theta}{2}\bigg)  \  e^{i (\hat\psi - \hat\phi)/ 2}\,, \\
 \sfx^3 + i \sfx^4 ~&=~   2 \,\sqrt{\hat r}\,  \cos \bigg(\frac{\hat \theta}{2}\bigg) \  e^{i (\hat\psi + \hat\phi)/ 2}\,, 
\end{aligned}
\label{GH coords}
\end{equation}
where $\bR^4$ is described as a circle fibration over $\bR^3$, with the angular $\bS^3$ of $\bR^4$ written as the Hopf fibration of $\bS^1_{\hat\psi}$ over the angular $\bS^2$ of $\bR^3$. 
The source rings lie at
\be
\label{abrels}
\mfa^2 = 4\hat a
~~,~~~~
\mfb^2 = 4\hat b ~.
\ee
What were double pole source singularities in the harmonic forms/functions appearing in the geometry before smearing, become after smearing single-pole singularities in the Gibbons-Hawking $\bR^3$ base.

The smeared $Z_2$ is given by
\begin{align}
\label{Z2 ave}
     		\bar{Z}_2 &=	\frac{n_5}{2\pi} \int_{0} ^{2\pi} d\alpha \frac{1}{|\sfx |^2 + \mfa^2 -2\mfa \sqrt{\sfx_1 ^2 + \sfx_2 ^2} \cos(\alpha-\phi )}
\nn\\[.2cm]
&= \frac{\nfive}{\sqrt{(|\sfx |^2 + \mfa^2)^2 - 4\mfa^2 (\sfx_1 ^2 + \sfx_2 ^2)}}=\frac{\hat Q_5}{\hat{r}_S}~,
\end{align} 
where the factor of four in~\eqref{abrels} makes it convenient to define rescaled charges as follows
\begin{equation}
\label{rescaledCharges}
   \hat{Q}_5 = \frac{1}{4} \nfive ~~,~~~~ \hat{Q}_1 = \frac{1}{4} \none~.
\end{equation}
Similarly, the smeared one-brane harmonic function is
\begin{align}
\label{Z1 ave}
     		\bar{Z}_1 &= 1+\frac{n_1}{2\pi } \int_0 ^{2\pi } d\alpha~\frac{1}{|\sfx |^2 + \mfb^2 -2\mfb \sqrt{\sfx_1 ^2 + \sfx_2 ^2} \cos(\alpha-\phi )}
\nn\\[.2cm]
&= 1+\frac{\none}{\sqrt{(|\sfx |^2 + \mfb^2)^2-4\mfb^2 (\sfx_1 ^2 + \sfx_2 ^2)}}=1+\frac{\hat Q_1}{\hat{r}_1}~.
\end{align}
The one-form $\beta$ vanishes, by construction:
\begin{equation}
  \beta=0~.
\end{equation}
The one-form $a_1$ contains the one-form $A_1$ defined as
\begin{equation}
     		(A_1)_i \equiv \sum_{\sfm=1} ^{n_5}\frac{\partial_v \sfF_{\sfm,i}}{|\sfx-\sfF_{\sfm}(v)|^2}~. 
\end{equation}
The nonzero component of the smeared $A_{1}$ is
\begin{align}
\label{Aphi1b}
     		\bar{A}_{1,\phi}&= \frac{1}{\sqrt{2}\pi R_y} \int_0 ^{\sqrt{2}\pi R_y} dv A_{1,\phi}   
\nn\\
     		& =\frac{\sfp}{\sqrt{2}\pi R_y} \int_0 ^{2\pi} d\tilde{v}  \frac{\mfa\sqrt{x_1 ^2+ x_2 ^2}\cos(\sfp\tilde{v} -\phi)}{\mfa^2 + |\sfx|^2-2\mfa \sqrt{\sfx_1 ^2 +\sfx_2^2} \cos(\sfp\tilde{v}-\phi)}
\nn\\[.2cm]            
            &=-\frac{\sfp}{\sqrt{2}R_y} \Big(1- \frac{\mfa^2+|\sfx|^2}{\sqrt{(\mfa^2+|\sfx|^2)^2-4\mfa^2 (\sfx_1 ^ 2+ \sfx_2 ^2)}}\Big)
\\[.2cm]
            & = -\frac{2\sqrt{2}\sfp}{R_y\hat{r}_S} \Big( \hat{r}+\hat{a}-\hat{r}_S\Big)~.
\nn
\end{align}
The ``magnetic piece'' of $a_1$, which we denoted by $b_1$, is defined in Eq.~(\ref{gam2b1}). 
The only nonzero component of the smeared $b_{1}$ is
\begin{align}
     	\bar{b}_{1\psi} &=  \frac{ \sfp}{\sqrt{2}\pi R_y} \int_0 ^{2\pi}d\tilde{v}\frac{\mfa\sqrt{\sfx_1 ^2+ \sfx_2^2}\cos(\sfp \tilde{v}-\phi)-\mfa^2}{|\sfx|^2+ \mfa^2 -2\mfa\sqrt{\sfx_1 ^2 +\sfx_2^2} \cos(\sfp \tilde{v}-\phi)} 
    \nonumber\\[.2cm]
        &=\frac{ \sfp}{\sqrt{2}R_y\sqrt{(|\sfx|^2 + \mfa^2)^2-4\mfa^2 (\sfx_1 ^2 + \sfx_2 ^2)}} \Big(|\sfx|^2-\mfa^2-\sqrt{(|\sfx|^2 + \mfa^2)^2-4\mfa^2 (\sfx_1 ^2 + \sfx_2 ^2)}\Big)
    \nonumber\\[.2cm]
        & =\frac{2\sqrt{2}\sfp}{ R_y \hat{r}_S} \Big(\hat{r} -  \hat{a} - \hat{r}_S \Big)~.
\end{align}
We thus have
\begin{align}
  a_{1\hat{\phi}} = -\frac{2\sqrt{2} \sfp \hat a }{ R_y \hat{r}_S}
  ~~,~~~~
  a_{1\hat{\psi}} = -\frac{2\sqrt{2}\sfp}{R_y \hat{r}_S} (\hat{r}-\hat{r}_S)~.
\end{align}
Similarly, 
\begin{align}
  a_{2\hat{\phi}} = -\frac{2\sqrt{2} k \none\hat{a} }{ R_y \hat{r}_1}
~~,~~~~
  a_{2\hat{\psi}} = -\frac{2\sqrt{2}k \none}{ R_y \hat{r}_1} (\hat{r}-\hat{r}_1)~.
\end{align}
One can evaluate the smeared, nonzero components of the two-forms $\gamma_1,\gamma_2$, with the results
\begin{align}
\label{gam ave}
\begin{split}
     		\bar{\gamma}_{2\hat{\phi} \hat{\psi}} &= \frac{\nfive}{\hat{r}_S} \bigg[ \hat{r}_S -\big(\hat{r}\cos(\hat{\theta})+\hat a \big)\,\bigg]~,
\\[.2cm]
     		 \bar{\gamma}_{1\hat{\phi} \hat{\psi}} &= \frac{\none}{\hat{r}_1} \bigg[ \hat{r}_1-\big(\hat{r} \cos(\hat{\theta})+\hat b \big)\,\bigg]~.
\end{split}
\end{align}
Next, we write contributions to the smeared one-form $\omega$. To simplify the calculation, we smear over both the azimuthal angle $\phi$ and the parameter $v$ in the solution. The first contribution is
\begin{align}
\label{om0 ave}
     		\bar{\omega}_{0,\phi} &= -\frac{\sfp}{\sqrt{2} R_y} \Big(1- \frac{\mfa^2+|\sfx|^2}{\sqrt{(\mfa^2+|\sfx|^2)^2-4\mfa^2 (\sfx_1 ^ 2+ \sfx_2 ^2)}}\Big)
    \nonumber\\[.2cm]
            & = \frac{2\sqrt{2}\sfp}{ R_y \hat{r}_S} \Big(\hat{r}_S- \hat a -\hat{r}\Big)~.
\end{align}
The second contribution to the smeared $\omega$ is given by
\begin{align}
\label{om1 ave}
     	\bar{\bar{\omega}}_{1\phi} &= \frac{k\none \nfive (\mfb^2 + |\sfx|^2)+\sfp\none  (\mfa^2 + |\sfx|^2)}{2\sqrt{2}  R_y \sqrt{(\mfa^2+ |\sfx|^2)^2-4\mfa^2 (\sfx_1 ^2+ \sfx_2 ^2)}\sqrt{(\mfb^2+ |\sfx|^2)^2-4\mfb^2 (\sfx_1 ^2+ \sfx_2 ^2)}}
\nonumber\\
     	&\quad-\frac{k \none \nfive}{2\sqrt{2} R_y w_y \sqrt{(|\sfx|^2 + \mfa^2)^2 - 4 \mfa^2 (\sfx_1 ^2 + \sfx_2 ^2)}}-\frac{ \sfp \none }{2\sqrt{2}R_y \sqrt{(|\sfx|^2 + \mfb^2)^2 - 4 \mfb^2 (\sfx_1 ^2 + \sfx_2 ^2)}}
\nonumber\\[.2cm]
        & =\frac{k\hat{Q}_1 \hat{Q}_5 (\hat b +\hat{r})+\sfp \hat{Q}_1 w_y(\hat a +\hat{r}) }{2\sqrt{2}  R_y\, \hat{r}_S \, \hat{r}_1}-\frac{k \hat{Q}_1 \hat{Q}_5}{2\sqrt{2} R_y w_y \hat{r}_S}-\frac{\sfp \hat{Q}_1}{2\sqrt{2} R_y \hat{r}_1}~.
\end{align}
The third contribution reads
\begin{align}
\label{om2l ave}
\bar{\bar{\omega}}_{2l,\phi} &=\frac{\none}{\sqrt{2} R_y w_y}\frac{  \nfive k \mfb-w_y \sfp \mfa}{|\mfa^2 - \mfb^2|}\bigg[\bigg( \frac{|\sfx|^2+\mfa^2}{\mfa\sqrt{(|\sfx|^2+\mfa^2)^2-4\mfa^2 (\sfx_1 ^2 + \sfx_2 ^2)}}-\frac{1}{\mfa}\bigg)
\nonumber\\[.1cm]
&\hskip 5cm
+\bigg(\frac{|\sfx|^2+\mfb^2}{\mfb\sqrt{(|\sfx|^2+\mfb^2)^2-4\mfb^2 (\sfx_1 ^2 + \sfx_2 ^2)}}-\frac{1}{\mfb}\bigg)\bigg]
\nonumber\\[.2cm]
& = \frac{\none}{ \sqrt{2}R_y w_y}\frac{  \nfive k \sqrt{\hat b}-w_y \sfp \sqrt{\hat a}}{|\hat{a}-\hat{b}|}\bigg[\bigg( \frac{\hat{r}+\hat{a}}{\sqrt{\hat a} \hat{r}_S}-\frac{1}{\sqrt{\hat{a}}}\bigg)
+\bigg(\frac{\hat{r}+\hat{b}}{\sqrt{\hat{b}}\hat{r}_1}-\frac{1}{\sqrt{\hat{b}}}\bigg)\bigg]
\end{align}
The results for the fourth contribution, $\omega_{2r,\phi}$, is quite complicated, and relegated to Appendix~\ref{app:om-F aves}.

We next write contributions to the scalar momentum harmonic function, the first of which is given by
\begin{align}
\label{F0 ave}
     		\bar{\mathcal{F}}_0 = -\frac{2\sfp^2 \mfa^2}{\nfive R_y ^2} \frac{1}{\sqrt{(|\sfx|^2 + \mfa^2)^2 -4\mfa^2 (\sfx_1 ^2 + \sfx_2 ^2)}}=-\frac{\sfp^2 \hat{a}}{2\hat{Q}_5 R_y ^2}\frac{1}{\hat{r}_S}~.
\end{align}
The second contribution to the smeared $\mathcal{F}$ is
\begin{align}
\label{F1 ave}
     		\bar{\bar{\mathcal{F}}}_1 =
     		& \frac{ \sfp k \none}{2w_y R_y^2 (\mfa^2\!+\mfb^2)(\sfx_1^2+\sfx_2^2) } \bigg[1-\frac{|\sfx|^2+\mfa^2}{\sqrt{(|\sfx|^2\tight+\mfa^2)^2-4\mfa^2 (\sfx_1^2\tight+\sfx_2^2)}}\bigg]\bigg[1-\frac{|\sfx|^2+\mfb^2}{\sqrt{(|\sfx|^2\tight+\mfb^2)^2-4\mfb^2(\sfx_1^2\tight+\sfx_2^2)}}\bigg]\nonumber\\
            & = \frac{ \sfp k \hat{Q}_1}{8w_y R_y^2 (\hat{a}+\hat{b})\hat{r} \sin^2 \big(\frac{\hat{\theta}}{2}\big) } \Big(1-\frac{\hat{r}+\hat{a}}{\hat{r}_S}\Big)\Big(1-\frac{\hat{r}+\hat{b}}{\hat{r}_1}\Big)~.
\end{align}
The third contribution to $\mathcal{F}$ is
\begin{align}
\label{F2 ave}
\bar{\bar{\mathcal{F}}}_{2l} &= \frac{\none}{w_y}\frac{2}{\nfive w_y R_y^2 |\mfa^2-\mfb^2|} \Big(\frac{ \sfp^2 \mfa^2 w_y^2}{\sqrt{(|\sfx|^2+\mfa^2)^2-4\mfa^2 (\sfx_1^2+\sfx_2^2)}}+\frac{k^2 \mfb^2 \nfive^2}{\sqrt{(|\sfx|^2+\mfb^2)^2-4\mfb^2 (\sfx_1^2+\sfx_2^2)}}\Big)
\nonumber\\
&\quad +\frac{2k \sfp \none \text{min}(\mfa^2,\mfb^2)}{w_yR_y^2 |\mfa^2-\mfb^2|}\Big(\frac{1}{\sqrt{(|\sfx|^2+\mfa^2)^2-4\mfa^2 (\sfx_1^2+\sfx_2^2) }}+\frac{1}{\sqrt{(|\sfx|^2+\mfb^2)^2-4\mfb^2 (\sfx_1^2+\sfx_2^2) }}\Big)
\nonumber\\[.2cm]
&=\frac{2\hat{Q}_1}{\hat{Q}_5 w_y^2 R_y^2 |\hat{a}-\hat{b}|}\Big(\frac{\sfp^2  w_y^2 \hat{a}}{4\hat{r}_S}+\frac{4k^2 \hat{Q}_5^2 \hat{b}}{\hat{r}_1} \Big)+\frac{2k \sfp \hat{Q}_1 \text{min}(\hat{a},\hat{b})}{w_yR_y^2 |\hat{a}-\hat{b}|}\Big( \frac{1}{\hat{r}_S}+\frac{1}{\hat{r}_1}\Big)~.
\end{align}
Finally, the fourth contribution $\bar{\bar{\mathcal{F}}}_{2r}$ is written in Appendix~\ref{app:om-F aves}. In the special case $k=0$, $\mathcal{F}_{2r}=0$.

\subsection{Comparing to single-mode effective superstrata}
\label{sec:singlemode}

We can compare the preceding solution to the effective description of round superstrata derived in~\rcite{Bena:2025uyg}.
These solutions have, in addition to the NS-NS fields we have restricted our consideration to, a profile $\mfa_0(v)$ of the R-R scalar component of the tensor multiplet on the fivebrane, which sources R-R fields in the bulk (see~\rcite{Martinec:2024emf} for a discussion in the present context).  However, in the analysis~\rcite{Bena:2025uyg}, the polarization state of the momentum excitation was largely irrelevant~-- only the overall momentum flux mattered when the solutions were coarse-grained by averaging over their oscillations along the $y$-circle.  We will thus proceed, under the assumption that the similarities with our solutions largely outweigh the differences.  As mentioned above, this seems a reasonable assumption, since in the WKB limit we employ, the polarization state of the F1-P sources in our solution is a minor effect compared to the overall momentum/winding carried by the strings.

Because we work with an NS5-P supertube, whereas the usual superstratum construction starts with an NS5-F1 supertube, we need to process the solutions through a T-duality along $\bS^1_y$ in order to compare them.
For these ``T-dual'' superstrata, we are looking for the third BPS charge to consist of strings winding the T-dual $\ytil$ circle that carry no momentum $p_\ytil$, and as a consequence of the Virasoro constraints, no oscillator excitations.  The advantage of this T-dual frame is that we can be quite explicit, because we have in hand the exact classical solution.  
This solution is a special case of our construction above, in which
\begin{align}
\label{straight}
\sfX^i_\perp(v)={\it const.}
~~,~~~~
|\sfX_\perp|^2 = \mfb^2
~~,~~~~
\sfX^v = v
~~,~~~~
\sfX^u = u ~. 
\end{align}
We consider a gas of such strings.  Under T-duality, they map to massless particles traveling null geodesics at the same location in the transverse space, located at a constant position along $v$ and moving only along the null Killing coordinate $u$.

In the background~\eqref{AnsatzIIB}, the T-duality transformation takes one to a supergravity solution of the same functional form, with a coefficient map
\begin{align}
\label{Tduality map}
\hskip .5cm 
Z_1' &= 1-\hf \cF
&Z_2' &= Z_2
&\cF' &= 2(1-Z_1)
&\nn\\[.2cm]
\beta' &=\frac{a_1}{2} 
&\omega' &= \omega -\frac{a_1}{2} 
&a_1' &= 2\beta
\qquad\qquad
\gamma_2' = \gamma_2+\frac{a_1\wedge \beta}{Z_1}  ~,
\end{align}
where the primed quantities on the LHS denote the NS5-F1 frame for the fivebrane source and the unprimed quantities on the R.H.S. refer to the NS5-P frame.
In particular, $\beta'$ in the NS5-F1 frame is given by the one-form $a_1$ in the NS5-P frame (whose explicit form is Eq.~\eqref{a1result}), smeared over $v$.  The $\beta$ in the NS5-P frame has been chosen to vanish, therefore $a_1'=0$ and $\gamma_2'=\gamma_2$.
We denote the radius and string coupling on the T-dual $y$-circle by
\be
R_\ytil = \frac{\alpha'}{R_y}
~~,~~~~
\tilde g_s^2 = \frac{g_s^2\alpha'}{R_y^2}
~.
\ee


The averaged solutions of the preceding subsection, being independent of $y$, can be dimensionally reduced to five dimensions.  BPS 5d supergravity solutions are again characterized by a set of harmonic forms and functions, which now have single poles at sources in their three transverse spatial dimensions.  The 5d metric can be written
\be
\label{5dgeom}
ds_5^2 = -Z^{-2}\big(dt+{\bf k}\big)^2+Z\,ds_4^2(\cB) ~,
\ee
where $ds_4^2(\cB)$ is a multi-center Gibbons-Hawking (GH) metric
\be
ds_4^2(\cB) = V^{-1}\big(d\hat\psi+A\big)^2+V\,d\hat\sfx \cdot d\hat\sfx  ~;
\ee
here $\hat\sfx \in \bR^3$ are Cartesian coordinates on the transverse space and $\hat\psi\simeq\hat\psi+4\pi$ parametrizes a fibered circle.

The various harmonic forms/functions appearing in the 5d solution are written in terms of a basic set
\begin{align}
V = \sum_i\frac{\hat q^{(i)}}{|\hat\sfx -\hat\sfx ^{(i)}|}
~~&,~~~~ 
K^I = \sum_i\frac{\hat \kappa^I_{(i)}}{|\hat\sfx -\hat\sfx ^{(i)}|}
\nn\\
L_I = l_I^{(0)} + \sum_i\frac{\hat Q_I^{(i)}}{|\hat\sfx -\hat\sfx ^{(i)}|}
~~&,~~~~ 
M = \hat m^{(0)} + \sum_i\frac{\hat m^{(i)}}{|\hat\sfx -\hat\sfx ^{(i)}|}  ~,
\end{align}
in terms of which one has for instance the warp factor  
\be
Z = \Big(\coeff16C^{IJK}Z_IZ_JZ_K\Big)^{\frac13}
~~,~~~~
Z_I = L_I + \frac{C_{IJK}K^JK^K}{2V} 
\ee
(for more details, see for instance~\rcite{Bena:2025uyg}).
The angular momentum one-form $\bf k$ is expressed as
\be
{\bf k} = \mu\big(d\hat\psi+A\big)+\varpi
~~,~~~~
\mu = \frac16 C_{IJK} \frac{K^IK^JK^K}{V^2} + \frac{K^IL_I}{2V} + M 
\ee
with $\varpi$ also expressed in terms of the basic set; $\bf k$ is related to the angular momentum one-forms $\beta,\omega$ of the 6d solution via 
\be
{\bf k} = \frac{\beta+\omega}{\sqrt 2} ~.
\ee

The underlying round NS5-F1 supertube in this duality frame still corresponds to the profile function
\be
\label{round NS5-F1}
\sfF^1+i\sfF^2 = a\, \exp\bigg[\frac{ i\kappa (t+y)}{\nfive R_y}\bigg]
~~,~~~~
\sfF^3+i\sfF^4 = 0 ~,
\ee
processed through smearing and the T-duality transformation~\eqref{Tduality map}.  We will identify the parameter $a$ of superstrata with the parameter $[x\frac{\alpha'}{R_{\tilde{y}}^2}]^{\half}\mfa$ of our circular source solutions in Eq.~(\ref{a rel}) below; and of course the pitches match, 
\be
\label{p eq kappa}
\sfp=\kappa ~.
\ee

It is convenient to work with either (or both) of two polar coordinate systems~-- spheroidal coordinates $(r, \theta, \phi, \psi)$ appropriate to the 6d geometry of the round supertube; and Gibbons-Hawking coordinates $\hat r,\hat\theta,\hat\phi,\hat\psi$ appropriate to the 5d solution~\eqref{5dgeom} obtained after dimensional reduction along $y$, in which $\bR^4$ is written as a circle fibration over $\bR^3$.  The two are related by
\begin{equation}
  \begin{aligned}
 \sfx^1 + i \sfx^2 ~&=~  \sqrt{r^2+a^2}\,\sin\theta\,e^{i \phi}~=~ 2 \,\sqrt{\hat r} \,  \sin \bigg(\frac{\hat \theta}{2}\bigg)  \  e^{i (\hat\psi - \hat\phi)/ 2}\,, 
\\
 \sfx^3 + i \sfx^4 ~&=~  r \,\cos\theta\,e^{i \psi}~=~ 2 \,\sqrt{\hat r}\,  \cos \bigg(\frac{\hat \theta}{2}\bigg) \  e^{i (\hat\psi + \hat\phi)/ 2}\,, 
\end{aligned}
\label{GH-bipolar}
\end{equation}
In spheroidal coordinates, the round supertube sits at $r=0$, $\theta=\pi/2$, and runs along $\phi$ ($\psi$ degenerates at this locus). 
In GH coordinates, the fivebrane supertube locus is $\hat r = \coeff14 a^2= \hat \sfa$, $\hat\theta=\pi$, and can be taken to run along the fiber coordinate $\hat\psi$ ($\hat\psi+\hat\phi$ is degenerate at this locus).  The momentum wave in superstrata runs along $y$ and $\hat\psi$, and in the WKB limit localizes along the GH $\bR^3$ base to a point along the latitudinal circle on the two-sphere of radius $\hat r_P$ at the polar angle $\hat\theta_P$.  One can then dimensionally reduce along both the $y$-circle and the $\bS^1$ GH fiber (the $\hat\psi$-circle).  

The single-mode superstratum solutions are derived in the spheroidal coordinates adapted to the underlying fivebrane supertube.
The 4d harmonic functions smeared along $y$ become harmonic functions in the $\bR^3$ GH base, as discussed in~\rcite{Bena:2025uyg}.  
The harmonic functions determining the supergravity solution in the NS5-F1 supertube frame were written in section 3.2 of~\rcite{Bena:2025uyg}:
\begin{equation}
\begin{aligned}
V&=\frac1{\hat r}
~~,~~~~
K^1=K^2=0
~~,~~~~
K^3 = \hat \kappa_3\Big(\frac1{\hat r_{\sst S}}-\frac1{\hat r}\Big)
\\[.2cm] 
Z_1 = L_1 &= 1 +  \frac{\hat Q_1}{\hat r_{\sst S}}  ~~,~~~~   
Z_2 = L_2 =~ 1 +  \frac{\hat Q_5}{\hat r_{\sst S}}  ~~,~~~~ 
Z_3 =L_3 = 1 +  \frac{\hat Q_{\sst P}}{\hat r_{\sst P}}  ~,
\\[.2cm]   
M &= \frac{ {\hat m}_{\sst S}}{\hat r_{\sst S}}  +  \frac{{\hat m}_{\sst P}}{\hat r_{\sst P}}
~~,~~~~
\mu = \frac{\hat \kappa_3}{2}\, \bigg(\frac{ \hat r}{\hat r_{\sst S}} -1\bigg)\bigg( 1 +  \frac{\hat Q_{\sst P}}{\hat r_{\sst P}}\bigg)  +  M ~, 
\label{Zmufns} ~.   
\end{aligned}
\end{equation}
Here $\hat r=|\hat\sfx|$ is the radius of a given point $\hat\sfx$ in the GH base; $\hat r_S$ is the Cartesian distance $|
\hat\sfx-\hat\sfx_S|$ between the given point $\hat\sfx$ and the supertube center at $\hat\sfx_S$; and $\hat r_P=|\hat\sfx-\hat\sfx_P|$ is the distance to the momentum center at $\hat\sfx_P$.  The 5d charges $\hat Q_I$ are related to the 6d charges~\eqref{charges} by
\be
\widetilde Q_I=4\hat Q_I ~,
\ee
where we use a tilde to denote T-dual charges
\be
\label{dual chgs}
\widetilde Q_5 \tight= \nfive \alpha' \tight= Q_5
~~,~~~~
\widetilde Q_1 \tight= \frac{\tilde g_s^2 \tilde n_1 (\alpha')^3}{V_4} = \frac{g_s^2 n_p (\alpha')^4}{V_4 R_y^2} \tight= Q_p
~~,~~~~
\widetilde Q_p \tight= \frac{\tilde g_s^2 \tilde n_p (\alpha')^4}{V_4 R_\ytil^2} = \frac{g_s^2 n_1 (\alpha')^3}{V_4 } \tight= Q_1
~.
\ee

Extracting the contribution of $\beta=\frac{1}{2}\hat\kappa(\hat r/\hat r_S-1)$ from $\mu$,
one finds
\begin{align}
\beta &= \frac{R_\ytil \kappa a^2}{\sqrt{2} \Sigma} \big(\sin^2\!\theta\, d\phi - \cos^2\!\theta\, d\psi \big)   = \frac{\hat{\kappa}}{\sqrt{2}} \Big(  \frac{\hat{r}-\hat{r}_S}{\hat{r}_S} d\hat{\psi} -\frac{\hat{\sfa}}{\hat{r}_S}  d\hat{\phi}\Big)~. 
\nn\\[.2cm]
\label{ompsihat}
\omega_{\hat\psi} &=  \frac{\hat\kappa}{\sqrt{2}}\frac{(\hat r-\hat r_S)\hat Q_P}{\hat r_S\hat r_P} + \frac{\hat m_S \sqrt{2}}{\hat r_S} + \frac{\hat m_P \sqrt{2}}{\hat r_P}  ~,
\end{align}
where $\hat\kappa = \frac{R_\ytil}{2}\kappa$.
After some processing, one finds that the angular momentum harmonic function $M$ of Eqs.~\eqref{Zmufns} can be written as 
\be
\label{M1mode}
16 M =16\Big(\frac{\hat m_S}{\hat r_S} + \frac{\hat m_P}{\hat r_P}\Big) 
= 
\frac{\kappa R_\ytil a^2}{\hat r_S} 
+  \frac{\kappa R_\ytil\, (m/\ell)b^2}{\hat r_P} 
+ \frac{\kappa R_\ytil \,a^2 \hat Q_P}{|\hat\sfx_P-\hat\sfx_S|}\Big(\frac{1}{\hat r_S} - \frac{1}{\hat r_P}\Big)   ~,
\ee
where the parameter $m$ in the middle term in~\eqref{M1mode} specifies the left angular momentum carried by the momentum wave, and the parameter $b^2$ is determined by the overall string winding budget (equivalently, momentum budget in the T-dual NS5-P frame)
\be
\label{strandbudget}
a^2+\frac{b^2}2 = \frac{\widetilde Q_1\widetilde Q_5}{\kappa^2 R_\ytil^2} ~,
\ee

The angular momentum charges $\cJ_{L,R}$ are given by
\be
\label{ST J}
\cJ_R = \frac{\mu^2\alpha'}{\kappa R_\ytil  }J_R = \frac{\kappa R_\ytil }2 a^2
~~,~~~~
\cJ_L = \frac{\mu^2\alpha'}{\kappa R_\ytil  }J_L = \frac{\kappa R_\ytil}2 \Big(a^2 + \frac{m}{\ell} \frac{b^2}2\Big) ~.
\ee
and the momentum charge is
\be
\label{Pstratum}
\widetilde Q_P = \frac{\mu^2\alpha'}{\kappa^2 R_\ytil^2} \tilde n_p = \frac{m\tight+n}{\ell} \,\frac{b^2}{2} ~.
\ee
Thus $m+n$ parametrizes momentum along the $y$-circle carried by a wave quantum, while $m$ parametrizes its left angular momentum.
The comparison of the right angular momentum with~\eqref{round chgs} implies
\be
\label{a rel}
\mfa^2 = \frac{R_\ytil^2}{\alpha'}\, a^2 
~~,~~~~
\sfp = \kappa
~.
\ee

Note that there is an unfortunate clash of notation: Unlike the relation $\hat \sfa=|\hat\sfx_S|=\frac14 a^2$, the parameter $b^2$ is {\it not} related to the radial distance $\hat \sfb=|\hat\sfx_P|$ to the momentum center defined in~\rcite{Bena:2025uyg}.  Rather, $b^2$ codes the {\it number} of momentum quanta, not their radial location.
Therefore, the parameter $\mfb$ characterizing the radius of the F1-P supertube~\eqref{roundX} is also unrelated to the parameter~$b$.  We do, however, have the relations~\eqref{abrels} between spheroidal and Gibbons-Hawking parametrizations of the ring radii.
In the translation between Gibbons-Hawking and spheroidal coordinates, and between the parametrizations~\eqref{roundF}, \eqref{round NS5-F1} of the sources, one has the following relations 
\begin{align}
\label{locations}
|\hat \sfx_S| = \hat \sfa = \frac{\alpha'\mfa^2}{4R_\ytil^2}  = \frac{a^2}{4}
~~,~~~~
|\hat \sfx_P| = \hat \sfb = \frac{\alpha'\mfb^2}{4R_\ytil^2} = \frac{\ell\tight+n\tight-m}{4\ell} a^2
~~,~~~~
|\hat\sfx_P-\hat\sfx_S| = \frac{\alpha'|\sfX-\sfF|^2}{4R_\ytil^2} 
= \frac{m+n}{4\ell} a^2 ~,
\end{align}
where we have dropped the subscript labels $\sfm,\sfn$ on the source locations $\sfF,\sfX$ since there is only one kind of source of each type (also, note that the momentum and angular momeentum quantum numbers $m,n$ are distinct from these omitted labels $\sfm,\sfn$).  These relations are consequences of the ``bubble equations'' of 5d supergravity~\cite{Bena:2025uyg}; they can also be independently derived via an analysis of worldsheet string theory in circular supertube background in the NS5-F1 duality frame~\rcite{Martinec:2025xoy,Bena:2025uyg}; we repeat this analysis in the T-dual NS5-P frame in Appendix~\ref{app:GWZW}.
Here $\ell,m,n$ are the center-of-mass quantum numbers of each of the momentum quanta in the coherent condensate at radius $|\sfx_P|=\hat\sfb$
(which under T-duality become wound strings located at $|\sfX|=\mfb$).

We now compare these results to the smeared circular source solution of the previous subsection.  Superstratum excitations carry no string winding, so we set $k=0$ in the string source~\eqref{roundF}.
Under T-duality, the roles of momentum P and F1 string winding are interchanged as in~\eqref{dual chgs}, and the harmonic forms/functions are related as in~\eqref{Tduality map}. 
The momentum harmonic function $Z_1$ of~\eqref{Z1 ave} is thus related to the effective superstratum harmonic function $\cF=2(1-Z_3)$ in~\eqref{Zmufns}.  Of course, the fivebrane harmonic functions $Z_2$ of~\eqref{Z2 ave} and~\eqref{Zmufns} match.

Consider the momentum charge $Q_P$; in the NS5-F1 superstratum, it is related to the positions of the charge centers by equation 3.24 of~\rcite{Bena:2025uyg}, which we can rearrange to
\begin{equation}
\widetilde Q_1 \widetilde Q_5 = \bigg( 1+\frac{\widetilde Q_P}{\big|\sfX-\sfF\big|^2}\bigg) \kappa^2R_\ytil^2 a^2  ~.
\label{bubble eqn}
\end{equation}
In addition to being arrived at by taking the WKB limit of superstratum solutions, this relation was also derived in~\rcite{Bena:2025uyg} in terms of 5d multicenter harmonic functions, and generalized to F1-P sources in the perturbative limit, where the triangle of center locations continues to satisfy~\eqref{locations}.  Using~\eqref{Pstratum}, \eqref{locations}, we see that this is simply a rewriting of the strand budget relation~\eqref{strandbudget}.

Under T-duality to the NS5-P supertube frame, one has the charge map~\eqref{dual chgs}, and this expression reads
\be
\label{dualbubble}
Q_{p} = \bigg( 1+\frac{Q_1}{\big|\sfX-\sfF\big|^2}\bigg) \frac{\sfp^2 \mfa^2}{\nfive}  ~,
\ee
which is essentially~\eqref{CFT stress}, \eqref{Pcharge} in the special case where the strings carry no oscillator excitation.
Correspondingly, in the expression for the smeared momentum harmonic function $\cF$ in the NS5-P frame, one has $\cF_{1}=\cF_{2r}=0$ (see~\eqref{F1 ave} and~\eqref{F2r smeared}); using~\eqref{F0 ave}, \eqref{F2 ave}
\be
\bar{\bar\cF} = \bar{\bar\cF}_0+\bar{\bar\cF}_{2l} = -\frac{\sfp^2 \mfa^2}{2n_5 R_y^2\,\hat r_S}\Big(1+\frac{n_1}{4|\hat \sfx_1-\hat \sfx_S|}\Big) = -\frac{\hat Q_{P}}{\hat r_S} ~. 
\ee
We thus find complete agreement between the smeared momentum harmonic function (when the strings themselves carry no momentum), and the string winding harmonic function of the effective superstratum.

The superstratum momentum harmonic function $Z_3$~\eqref{Zmufns} maps to the smeared string harmonic function~\eqref{Z1 ave}; we simply need to relate the charges.
The momentum (winding) charge of the system in the NS5-F1 (NS5-P) frame is~\eqref{Pstratum}
\be
\label{Qp 1mode}
\widetilde Q^{~}_P = \frac{(m+n) b^2}{2\ell }  = Q_1
~~,~~~~
\tilde n_p = {(m+n)N_0} = n_1
\ee
$N_0\in\bZ$ is the number of momentum carrying R-R excitations of the underlying fivebrane supertube in the superstratum construction.

As discussed in section~\ref{sec:strata scales}, the geometry has a long $AdS_2$ throat when $b\gg a$.  The location $r^2\approx \widetilde Q_P$ of the top of the $AdS_2$ throat is determined by the momentum charge, while all of the sources are located at scales~\eqref{locations} set by $a^2\ll b^2$; the ratio of these scales determines the length of the $AdS_2$ throat, as we saw in Eq.~\eqref{AdS2 len}.

Turning to the angular momentum harmonic function $\omega$, we can identify the first term on the R.H.S. of~\eqref{M1mode} as the contribution $\omega_0$ to the angular momentum harmonic function coming from the underlying fivebrane supertube~\eqref{omegaAnsatz}; the second and third terms in~\eqref{M1mode} have the form of $\omega_{2l}$.  To see this, rewrite the terms in parenthesis in~\eqref{om2l ave} as 
\be
\bigg( \frac{\hat{r}+\hat{a}}{ \hat{r}_S}-1\bigg)
= \frac{-2\hat a\sin^2(\hat\theta/2)}{\hat r_S}
\ee
Next, we rewrite the pole structure in~\eqref{om2l ave} as 
\be
\label{pole diff}
\omega_{2l,\phi} = \frac{n_1}{\sqrt2 R_y} \frac{2\sfp \, \sin^2(\hat\theta/2)}{|\hat b-\hat a|}\bigg(\frac{\hat a}{\hat r_S} + \frac{\sqrt{\!\hat a\hat b}}{\hat r_1}\bigg) 
~.
\ee
In order to match the pole structure, one needs $\omega_{2r}$ to have a similar form
\be
\omega_{2r} \supset - \frac{n_1}{\sqrt2 R_y} \frac{2\sfp \, \sin^2(\hat\theta/2)}{|\hat b-\hat a|}\bigg(\frac{\hat b}{\hat r_1} + \frac{\sqrt{\!\hat a\hat b}}{\hat r_S}\bigg)
\ee
in order to reproduce the structure of the angular momentum charge $J_L$ of~\eqref{round chgs}.  

However, as one sees from the results in Appendix~\ref{app:om-F aves}, there are many subleading terms in $\omega_{2r}$ that do not contribute to the charge $J_L$.
This contribution to $\omega$ in the smeared NS5-P frame solution is not an exact match between the two T-dual frames.  One can trace the discrepancy to the bulk source terms on the R.H.S. of the $\omega$ equations of motion~\eqref{OmegaEom9}.  For circular superstrata, there is only a contribution to this bulk source from the R-R fields $Z_4,\Theta_4$ and the momentum harmonic function $\cF$, both associated to the momentum source; the fivebrane harmonic function $Z_2$ has already been ``pre-smeared'' by the T-duality to the NS5-F1 frame of the fivebrane supertube, and $\Theta_1$ vanishes; as a result, the further averaging involved in the effective solution only gets a contribution from the momentum wave harmonics $Z_4,\Theta_4$.  In contrast, there are non-zero source terms coming from $Z_1,\Theta_2$ and $Z_2,\Theta_1$ having a more complicated spatial structure that depends on both sources, which does not vanish upon smearing, leading to the rather complicated form of $\omega_{2r}$ presented in Appendix~\ref{app:om-F aves}.

At this point, in order complete the comparison, we need to further relate the parameters of the smeared brane source solution of the previous subsection $\mfa^2,\mfb^2,\sfp,k,w_y,$~\etc, to the parameters of the effective superstratum $\hat a,\hat b, \kappa,m,n,\ell,$ \etc.  One way to bridge the two is to study BPS F1-P solutions in the NS5-P supertube using the exact worldsheet methods of~\rcite{Martinec:2017ztd,Martinec:2018nco,Martinec:2020gkv,Martinec:2022okx,Martinec:2025xoy}, as was done for strings in the NS5-F1 supertube in~\rcite{Bena:2025uyg}.  We carry out this exercise in Appendix~\ref{app:GWZW}, and use the results to complete the match between the two solutions in Appendix~\ref{app:matching}.

\section{Discussion}
\label{sec:discussion}

The 5+1d BPS supergravity field equations, sourced by electric and magnetic strings carrying arbitrary chiral wave profiles, yield a large class of solutions which can be explicitly computed in closed form.  The field equations boil down to a hierarchy of harmonic equations with delta-function sources, whose solutions have pole singularities at the charged sources whose locations and residues are determined in terms of the chiral wave profiles. 

These solutions arise from type II string theory compactified on $\bT^4$, with the electric sources being fundamental strings and the magnetic sources being NS5-branes wrapped on $\bT^4$.  Provided the magnetic sources are sufficiently well-separated, strong-coupling effects are suppressed and string perturbation theory is valid.  

While classical supergravity sees a geometry that appears to be singular due to the presence of explicit string and 
fivebrane sources, string theory sees a geometry that is effectively smoothed and capped at the scale of the brane separation, because strings cannot penetrate the throats of isolated fivebranes (see the discussions in~\rcite{Martinec:2020gkv,Martinec:2024emf} for an explanation).  This capping off of the geometry also caps the running of the dilaton toward strong coupling.

The strong coupling effects that arise when two fivebrane strands approach one another are associated to the liberation of D-branes stretching between the NS5 strands. The results~\eqref{MD1}, \eqref{tauD2} for the mass/tension of such branes continue to hold for three charge geometries~\eqref{AnsatzIIB} with the coefficients summarized in~\eqref{soln sum}.  There is a redshift factor $Z_1^{-1}$ in the metric, and a factor of $Z_1$ in the dilaton; as well as a warp factor $Z_2$ in the transverse geometry, and a factor $Z_2^{-1}$ in the dilaton.  These effects precisely cancel one another, and thus the mass/tension of stretched D-branes in the decoupled fivebrane throat is {\it always} governed by the Cartesian separation of the fivebranes in the transverse $\bR^4$ in units of $\alpha'$.

Without an understanding of the underlying brane dynamics at work, it can seem odd that there should be dramatic quantum fluctuations of the three-charge geometry.  In duality frames where the fivebrane star is described as a bubbling geometry, in the effective supergravity nothing peculiar seems to be happening as the throat lengthens~\rcite{Bena:2007qc}.%
\footnote{The usual D5-D1-P duality frame considered in the literature is related to ours by a T-duality along $\bS^1_y$, so that the explicit fivebrane sources turn into non-singular KK-monopole structures; followed by S-duality.}
The cap in the geometry is large and smooth, with curvature of order the $AdS$ scale.  Truncating the dynamics to just the pole locations of the spherical bubbles, angular momentum quantization leads to the head-scratching result that the depth of the throat in the classical geometry changes from finite to infinite when the last unit of angular momentum is removed~\rcite{deBoer:2008zn,deBoer:2009un}.  We now understand that such large fluctuations in the depth of the throat are ubiquitous near extremality, and provide an effective IR cutoff on the dynamics.  The non-abelian brane dynamics associated to this cutoff is hidden in the effects of branes wrapping the bubbles~\rcite{Martinec:2014gka,Martinec:2015pfa}.  The introduction of additional BPS charges, and the map to another duality frame, has simply obscured this dynamics; but it is still there.

As far as non-abelian NS5-brane dynamics is concerned, the addition of fundamental string and momentum charges is just window dressing, and has no effect on W-brane energetics~-- it is as if one were simply working with fivebranes on the Coulomb branch.  As a result, in the fivebrane decoupling limit~\eqref{decoupling} the non-abelian scale is reached when the scaled fivebrane separation is of order $\hat\ell_{na}$, Eq.~\eqref{Lnonab} (\ie~order one if the $\bT^4$ is string scale in size), so that the tension of little strings~\eqref{tauD2} is order one.

The introduction of F1 charge (and additional P charge) to NS5-P backgrounds modifies the geometry of the throat above the scale set by the fivebrane separation.  Focusing on the region well away from the sources, strings attract the geometry to $AdS_3$, and stabilize the dilaton at a value governed by the ratio $n_5/n_1$.  Non-abelian excitations which were light because the string coupling was growing large with smaller fivebrane separation, are now light because of the increasing redshift as the NS5-F1 throat deepens.

The effect of momentum charge is to stabilize the circle $\bS^1_y$ wrapped by both strings and fivebranes, and thus generate an $AdS_2$ throat.  As discussed in section~\ref{sec:blackness2}, the stabilized value of the radius~\eqref{Ryhor} depends on the ratio $n_p/n_1$, and so whether the fivebrane near-source region is described using NS5-F1 or NS5-P language depends on whether this ratio is larger or smaller than one. 

Initial analyses of two-charge NS5-brane backgrounds focused on the NS5-F1 duality frame, as the smooth, capped nature of such backgrounds is apparent in supergravity in this frame.  T-duality from NS5-P turns fivebranes in an NS5-P bound state into an NS5-F1 bubbling geometry where the brane charges are carried by fluxes through the bubbles~\rcite{Lunin:2001fv,Lunin:2002iz} rather than explicit (and naively singular) brane sources.  This observation spawned a detailed investigation of such bubbling geometries (see~\rcite{Bena:2007kg,Shigemori:2019orj,Warner:2019jll,Bena:2025pcy} for reviews).

But again, string theory doesn't care.  It knows about T-duality, which is simply a field redefinition on the string worldsheet, and therefore knows that there are still branes localized on the T-dual circle~\rcite{Gregory:1997te,Martinec:2019wzw,Martinec:2022okx}; this structure is hidden from supergravity in the NS5-F1 frame in T-dual variables that supergravity doesn't have access to.  It doesn't really matter if we consider NS5-P sources or NS5-F1 sources, as they are simply two different descriptions of the same object in string theory.  In particular, as discussed in~\rcite{Martinec:2019wzw,Martinec:2022okx}, the D-branes which stretch between NS5-branes in the NS5-P description T-dualize to D-branes wrapping the bubbling geometry of the NS5-F1 description; their energetic cost is the same, and so the criterion for the onset of non-abelian fivebrane dynamics is the same in both frames.  We have chosen to work with the NS5-P description, as it allows for a completely explicit solution of the supergravity field equations.

The structure and properties of the geometry are particularly transparent in the frame we have been working with.  The NS5-P source configuration sets the self-separation of the fivebrane.  As far as non-abelian dynamics of the fivebranes is concerned, the addition of fundamental strings is something of a sideshow, that generates an $AdS_2$ throat outside the source but doesn't particularly affect the degree to which non-abelian fivebrane dynamics is decoupled.  Remarkably, the BPS supergravity field equations in the presence of both F1-P and NS5-P sources remain exactly solvable, leading to the hope that we might be able to say a lot about the near-BPS dynamics.

\subsection{The quantum scale and the non-abelian scale}
\label{sec:quscale}

The near-horizon region of a nearly-BPS black hole is subject to large quantum fluctuations.  A universal sector of the excitation spectrum of such black holes is the set of Schwarzian modes of the nearly-$AdS_2$ throat (see for instance~\cite{Ghosh:2019rcj,Heydeman:2020hhw,Iliesiu:2020qvm,Lin:2022rzw,Kolanowski:2024zrq,Castro:2025itb,Acito:2025hka}).  As a BTZ black hole approaches extremality, these modes undergo large quantum fluctuations.  The Schwarzian excitations are concentrated very near the horizon (see \eg\ the explicit solution in~\rcite{Kolanowski:2024zrq}), and are thus parametrically lighter than other modes in the throat; they largely decouple at low energies from other modes, and can be regarded as fluctuations in the length of the throat. 

The large quantum fluctuations in the depth of the throat at extremality lead to a finite effective depth of the throat, and an effective ``stretched horizon'' at a radius we have called the quantum scale $r_\qu$.
Quantum fluctuations of fivebrane stars become large when fivebrane strands approach within the non-abelian scale $\hat\ell_{na}^2 = \big(\frac{V_4}{{\alpha'}^2}\big)^{\!\half}$.  Since the onset of black hole formation is associated to a deconfinement transition in the underlying brane dynamics, it is natural to identify these two scales, and indeed we have seen in examples that the $AdS_2$ throat lengths associated to these scales match.

There are many potential excitation modes of the fivebrane star.  One may ask whether there is a {\it universal} sector that plays the same role as the Schwarzian, and if there is, whether it decouples from other excitations for the longest throats.  There is indeed an obvious candidate collective mode which controls fluctuations in the depth of the $AdS_2$ throat of fivebrane stars.   

In fivebrane stars, the depth and redshift of the cap in the geometry scales with the typical separation of its string and fivebrane sources.  A rescaling of the source profiles by a factor $e^{-\lambda}$ rescales by a factor $e^{2\lambda}$ the harmonic functions $Z_1$ and $Z_2$ that determine the redshift of, and proper distance to, the cap in the geometry.  Such a rescaling moves the cap in the geometry up or down the fivebrane throat depending on the sign of $\lambda$.   

Thus the analogue of the Schwarzian should be a set of ``breathing modes'' of the sources.  
Approaching the extremal black hole along the configuration space of BPS fivebrane stars, fluctuations in the length of the throat are fluctuations in the harmonic function $Z_2$, which as we have seen are fluctuations in the size of the fivebrane random walk, Eq.~\eqref{walksize}.  These fluctuations become highly quantum at the scale of the inverse fivebrane tension~\rcite{Martinec:2023gte}, which is the magnetic string length scale $\mu$, Eq.~\eqref{tauNS5}, of the effective 6d magnetic string that it becomes upon dimensional reduction on~$\bT^4$.  When the fivebrane separation becomes of order this scale, the relative quantum fluctuations in the fivebrane positions become large, the nonabelian fivebrane excitations become unsuppressed, and one starts to enter the black hole phase.%
\footnote{As discussed in section~\ref{sec:ST scales}, the non-abelian scale in fact differs from the scale $\mu$ by a factor of $(\frac{V_4}{{\alpha'}^2})^{1/2}$; non-abelian effects therefore kick in at a somewhat larger bound state size, unless the the $\bT^4$ size is of order the string scale.}
Similarly, the redshift to the cap is governed by the harmonic function $Z_1$ of the strings.  Fluctuations in the size of the string configurations are fluctuations in the peak of $Z_1$ and thus generate fluctuations in the redshift to the cap.

Consider therefore a common rescaling of all the sources
\be
\sfF_\sfm = e^{-\lambda} \bF_\sfm 
~~,~~~~
\sfX_\sfn = e^{-\lambda} \bX_\sfn ~.
\ee
The effect of this rescaling on the joint effective action is a Liouville-like kinetic term
\begin{align}
\delta\cS_{\rm joint} &\sim \frac{n_1n_5}{2\pi}\int\! d^2\sigma \Big( \partial_+\lambda\,\partial_-\lambda +\lambda\big(\partial_+\cD_- + \partial_-\cD_+\big)\Big)
\nn\\[.2cm]
\cD_\pm &= \frac{1}{n_1n_5}\sum_{\sfm=1}^{n_5}\sum_{\sfn=1}^{n_1}\frac{(\bX_n-\bF_m)\cdot\partial_\pm(\bX_n-\bF_m)}{|\bX_\sfn-\bF_\sfm|^2}
\end{align}
where the divergence of the dilatation current $\cD$ of the sources plays the role of a background curvature.

In~\rcite{Lin:2022rzw}, the Schwarzian dynamics of the throat length was reduced to Liouville quantum mechanics; it would be interesting to see if one can make a further connection between these two collective fluctuations of the $AdS_2$ throat depth.  The Schwarzian approaches the extremal black hole from the direction of non-extremality, while the brane star breathing mode approaches the extremal black hole through a family of ultracompact BPS brane star geometries.  In both cases, a Liouville-like effective dynamics governs the fluctuations in the depth of the $AdS_2$ throat near extremality.

\subsection{Ensemble averages}
\label{sec:ensembles}

The fact that the string and fivebrane sources don't affect one another in the solution until the final layer in the hierarchy of BPS equations suggests that one can construct an ensemble of typical solutions by considering separately a fivebrane supertube deep in the core of the geometry, together with a surrounding cloud of strings.  

In the joint effective action, the kinetic term of the fivebranes does interact with the surrounding strings through the harmonic function $Z_1$ in its kinetic term, and the the magnetic interactions $a_2$ and $\gamma_1$ (which package into the two-form $\Theta_2$ of Eq.~\eqref{Theta2def}).  If the extent of the string cloud is much larger, then the fivebrane sees the approximately constant local value of these quantities at the origin $\sfx=0$, which appear as undetermined parameters.  One can then apply the results of our analysis of the ensemble of NS5-P solutions~\rcite{Martinec:2024emf} to determine the average harmonic function $Z_2$ and two-form $\Theta_1$ of the first layer sourced by the fivebranes.  

The surrounding strings see the geometry sourced by a highly compact fivebrane supertube at smaller radius, which sufficiently outside the source is well-approximated by the vacuum fivebrane geometry of~\rcite{Callan:1991at}.  One can then carry out in this background an ensemble average of the harmonic function $Z_1$ and two-form $\Theta_2$ sourced by the strings, and thus self-consistently determine the parameters used in the fivebrane ensemble.  
The second layer of BPS equations can then be analyzed in this joint ensemble.  In this way one gets a picture of the typical geometry in this class of solutions.  

When the two source clouds are not separated, but rather highly intermingled, one can resort to a mean field theory approximation to disentangle the sources, and find a self-consistent solution.  

We defer a detailed analysis of ensemble averages to an upcoming companion work.

\subsection{Near-BPS chaos}
\label{sec:chaos}

Even when we arrange the fivebrane sources to be well-separated, the geometry surrounding the generic source can have a large region that is exponentially close to the extremal rotating BTZ black hole.  This is perhaps not a surprise, in that one would expect a generic localized source to look like the generic vacuum solution well away from the source, and to start differing at the outer envelope of the source.  If the scale of the source distribution is well outside the horizon radius of the black hole solution with the given charges, it is perhaps more appropriate to think of it as a star rather than a black hole, and this is the case for the ensemble of horizonless BPS supergravity solutions that have been constructed to date.  Because they live in a long $AdS_2$ throat, the area of the angular sphere at the surface of these stars has essentially the same area as the corresponding black hole having the same charges, and may mimic features of the black hole for some observables.  But because they occupy a distinct region of configuration space where non-abelian fivebrane excitations are suppressed, they will not behave like black holes in all respects.

Our construction leads to a picture of the near-BPS dynamics of such a fivebrane star.  For radially infalling probes dropped from the top of the $AdS$ throat, one expects the probe to be significantly tidally disrupted wherever it encounters a geometry substantially different from the vacuum BTZ solution, due to the amplification of metric deviations by the blueshift of the probe.  In previously studied examples~\rcite{Martinec:2020cml,Ceplak:2021kgl}, the onset of this tidal disruption was located well away from the source due to the long power law tails of the metric deviation from the BTZ geometry in the particular microstate considered.%
\footnote{Roughly, tidal disruption occured at the geometric mean between the size of the source distribution and the top of the $AdS_2$ throat, due to the amplification by the probe blueshift of metric deviations from the vacuum black hole geometry.}
Here, however, we expect that, as in the two-charge ensemble considered in~\rcite{Martinec:2023xvf}, with generic source configurations the geometry shouldn't substantially differ from the vacuum BTZ solution until one gets quite close to the source.  A probe dropped from high up the $AdS$ throat will be highly blueshifted, and will go ``splat'' on the surface of the fivebrane star and thermalize into highly excited (non-BPS) strings. 

Near-BPS excitations encounter a rather different environment, and are expected to experience a very different fate.  Consider for instance a near-BPS massless particle probe.  The action for massless particles is
\begin{align}
\label{S 3chg}
\cS_{\rm ptcl} &= \int\!d\xi\, \sqrt{\gamma}\gamma^{-1}\Big( -\frac{1}{Z_1}\big(\dot v+\beta_i\dot \sfx^i\big)\big(\dot u+\omega_j\dot \sfx^j+\hf\cF(\dot v+\beta_j\dot \sfx^j)\big)+ \hf Z_2\,\dot\sfx \cdot \dot\sfx \Big) ~.
\end{align}
The worldline metric $\gamma$ enforces reparametrization invariance and enforces vanishing of the worldline Hamiltonian.  The BPS condition sets to zero the Killing momentum $p_u=\frac{\delta\cS}{\delta\dot u}$.  BPS null geodesics sit at fixed position in $v,\sfx$ and travel along $u$.  Near-BPS trajectories have a small velocity in $v,\sfx$ as well.
Sufficiently outside the source, the background is well-approximated by the extremal BTZ geometry~\eqref{BTZ metric}.  Within the source region, the probe encounters the randomly fluctuating background of the randomly walking string and fivebrane sources.  The trajectories of neighboring and initially parallel probes will deviate rapidly in this region; the probe dynamics will be highly (but perhaps not maximally) chaotic.%
\footnote{Similar behavior was found for null geodesics in 1/2-BPS $AdS_5$ backgrounds~\rcite{Berenstein:2025ese}.}

A similar result should hold for winding fundamental string probes, governed by the joint effective action~\eqref{JointAction2}.  The randomly walking fivebrane source generates a randomly fluctuating metric and NS $B$-field on the worldvolume of a probe fundamental string, and the fundamental string gas generates a randomly fluctuating environment for the magnetic string.  Again, dynamics in such a random environment will be chaotic, in this case for both the center-of-mass motion of the string, and also for its oscillator fluctuations.  If we think not of a probe string but of the collection of electric and magnetic strings sourcing the full geometry, one may attempt to model the flow and thermalization of the source distribution as some sort of ergodic dynamics on the near-BPS three-charge configuration space, governed by the sigma model~\eqref{JointAction2}.

The idea here is similar in spirit to the treatment of the dynamics of BPS solitons in other contexts, such as monopoles in 3+1d supersymmetric Yang-Mills-Higgs~\rcite{Manton:1981mp}.  There, one has a moduli space of BPS configurations, which has a metric \etc.  The near-BPS dynamics is governed by a 0+1d sigma model on the BPS moduli space.  Here, we have electrically and magnetically charged strings rather than particles, and so the near-BPS dynamics is governed by the 1+1d sigma model~\eqref{JointAction}.
Because of the random nature of the source string configurations, the terms in this effective action have a random character which one fully expects to lead to chaotic dynamics. 

It would be interesting to compare this chaotic dynamics with the strong chaos associated to a BPS black hole horizon~\rcite{Lin:2022rzw,Lin:2022zxd,Chen:2024oqv}.
The strength of chaos is characterized by the Thouless time (the time scale at which the system accesses all the available states).  In the analysis of~\rcite{Chen:2024oqv}, the strong chaos of black holes was argued to be associated to a Thouless time of order $N^0$, whereas the weak chaos of a system of gravitons was associated to a much longer Thouless time, scaling as a positive power of $N$ (where in the present context, the central charge of the spacetime CFT is of order $N$).  

The near-BPS dynamics of the two-charge case was also analyzed in~\rcite{Chen:2024oqv}. and found to be rather weakly chaotic.  From the present perspective, the difference with the three-charge case arises because the two-charge source does not experience a random environment, as we see from the joint effective action~\eqref{JointAction2}.  If we drop the string sources, this effective action is simply that of the type II string, which is free field theory and so manifestly not chaotic.  On the other hand, in the three-charge case, the F1 electric string sources generate random potentials for the NS5 magnetic strings and vice versa, leading to chaotic dynamics.

The much larger entropy of three-charge fivebrane stars relative to their two-charge counterparts, and a much more chaotic near-BPS dynamics, should result in a much shorter Thouless time as compared to the two-charge case.  The transition to maximal chaos should be associated to horizon formation, which happens when the system begins to access its fully nonabelian excitations~-- in the present context, the little strings.  The rapid reconnection of the little strings results in fast scrambling, which again is the hallmark of the system being able to rapidly access distant regions of the available configuration space~\rcite{Sekino:2008he}.  The weaker chaos of the fivebrane star dynamics is associated to the longer time a chaotically diffusing string or graviton takes to explore its environment, interact with its neighbors and make an appreciable change in the state of the system (for instance, to alter the fivebrane supertube configuration).  It would be interesting to explore this chaotic dynamics and determine its characteristics

\subsection{Absorption/emission}
\label{sec:scatt}

Immediately after the Strominger-Vafa result, absorption and emission of radiation from the little string gas describing the extremal 3-charge black hole was calculated~\rcite{Callan:1996dv,Dhar:1996vu,Das:1996wn} under the assumption that its excitations could be modeled using free field theory.  Remarkably, the leading order low-energy absorption and emission probabilities were found to agree with a bulk calculation of Hawking radiation, including greybody factors.

It would be interesting to compare and contrast this result with a corresponding calculation of emission and absorption of radiation from the fivebrane star, for instance modeling the brane dynamics using the joint effective action~\eqref{JointAction2}.  The chaotic near-BPS dynamics and large density of states of the fivebrane star should make it highly absorptive.  For instance, it would be of interest to compare the quasi-normal mode spectrum of fivebrane stars and near-BPS black holes.

Of course, the fundamental string constituents of the star lack the additional fractionation factor of $n_5$ that little strings have, and correspondingly they have a lower density of states.  Nevertheless, as discussed above, many features of the near-BPS dynamics are beginning to look like those of the corresponding black hole, and so it would be interesting to understand the degree to which the absorption, thermalization, and emission of low-energy radiation approximate the corresponding ``perfect black body'' characteristics of the black hole.

\subsection{Near-BPS gravitational collapse}
\label{sec:collapse}

In~\rcite{Martinec:2024emf}, we speculated about collapse mechanisms for the fivebrane star as it is perturbed away from the BPS bound.  In the context of two-charge solutions, the star is puffed up by transverse scalar excitations, but not by internal gauge excitations; a transfer of the excitation spectrum into internal modes would cause the star to shrink, and correspondingly the throat to deepen and eventually reach some sort of stretched horizon scale.  However, such a transfer runs counter to equipartition and is thus statistically disfavored.  In the present three-charge context, however, momentum excitations can be carried by both strings and fivebranes.  There can be strings whose winding number is parametrically larger than the fivebrane winding $\nfive$.  In such a situation, the finer momentum fractionation on the more highly wound F1 string as compared to the fivebrane can lead to a much larger density of states on the string, and now it is entropically {\it favored} to transfer momentum off of the fivebrane and onto the string, leading to an adiabatic collapse of the fivebrane star toward the black hole regime.

As the fivebrane loses transverse momentum excitations, its transverse extent decreases.  The fivebrane harmonic function $Z_2$ then becomes more highly peaked, so the depth of the fivebrane throat and the string coupling $e^{2\Phi} =\frac{V_4}{{\alpha'}^2} \frac{Z_2}{Z_1}$ at the cap grow in proportion.  Eventually the throat becomes deep enough, and the string coupling strong enough, that perturbative string theory breaks down.  If the string sources descend the throat along with the fivebrane, the peak of the harmonic function $Z_1$, grows as well.  Instead of the coupling growing strong in the cap, the redshift at the cap becomes large.  As discussed above, the effect is the same~-- the free energy cost of non-abelian fivebrane excitations becomes less and less.   

As discussed in~\rcite{Martinec:2019wzw,Martinec:2024emf}, at some point these new, stringy degrees of freedom become light enough that they can no longer be ignored.  These are the ``little strings'' of non-abelian fivebrane dynamics, which have been suppressed by self-consistently holding the fivebranes apart via their transverse excitation.  Turning off the momentum excitations that keep the fivebranes apart, or transferring them to other parts of the system, allows the fivebranes to come together and liberate the Hagedorn phase of little strings that accounts for black fivebrane entropy.  In this way, we see in the {\it bulk} description the relation between horizon formation (the Hawking-Page transition) and the deconfinement transition of the non-abelian brane dynamics~\rcite{Witten:1998zw}.

Thus, while generic superstrata, and the closely related configurations we have studied here, are not themselves black holes, they serve as a useful portal to the stringy dynamics that lies beyond supergravity, that plays a role in the resolution of singularities in other contexts~\rcite{Seiberg:1996vs}, that accounts for black hole entropy~\rcite{Strominger:1996sh,Maldacena:1996ya,Dijkgraaf:1998iz}, and whose appearance at the horizon scale during black hole formation could well resolve the puzzles associated to black hole radiance~\rcite{Hawking:1976ra}.

In the effective action approach we have been developing, the next step would seem to be to incorporate these new light degrees of freedom and re-solve the dynamics.  Low-energy supergravity suggests that at some point a trapped surface will form, and all this added structure of string theory will collapse down to a singularity and become invisible.  However, what we have seen so far suggests instead that the new light degrees of freedom dominate the dynamics rather than being subservient to the strictures of supergravity.  The hope is that we can find the dynamical mechanism that supersedes the singularity theorems of general relativity, and justifies the fuzzball picture of black hole microstructure.  As pointed out in the Introduction, the fact that when liberated, the little string is always at its correspondence point, suggests that supergravity no longer accurately characterizes the properties of the black hole interior, much as the black hole interior is superseded at the correspondence point of fundamental strings by a gravitating string gas~\rcite{Horowitz:1997jc,Chen:2021dsw}.

\bigskip\bigskip
\section*{Acknowledgements} 

We thank 
Iosif Bena,
Pierre Heidmann,
Samir Mathur,
David Turton,
and 
Nicholas Warner
for discussions.
We also thank the Simons Center for Geometry and Physics at Stony Brook University, for hospitality and support during the program and workshop ``{\it 50 years of the black hole information paradox}'' as this work was nearing completion.
The research of YZ was supported by the Blavatnik fellowship at the University of Cambridge and is supported by the Israel Science Foundation, grant number 1099/24.
The work of EJM is supported in part by DOE grant DE-SC0009924.


\vskip 2cm
\appendix

\section{Duality relations}
\label{app:details}
\subsection{S-duality}
This section exhibits the ansatz of the Type IIB Supergravity background preserving four supercharges, initially in the D5-D1-P duality frame, and then we perform S-duality to reach the NS5-F1-P frame.  
In the D5-D1-P Type IIB duality frame, the ansatz for $\frac{1}{4}$-BPS superstrata reads
\begin{align}
ds_{10} ^2 &= -2\sqrt{\frac{Z_1 Z_2}{\mathcal{P}^2}}(dv+\beta)\Big[du+\omega + \frac{\mathcal{F}}{2}(dv+\beta) \Big] + \sqrt{Z_1 Z_2} ds^2 (\mathcal{B})+ \sqrt{\frac{Z_1}{Z_2}}ds^2(\mathcal{M})~,
\nn\\
	e^{2\Phi} &= \frac{Z_1 ^2}{\mathcal{P}}~,
\nn\\
	B^{(2)} &= -\frac{Z_4}{\mathcal{P}}(du+\omega) \wedge (dv+\beta) +a_4 \wedge (dv+\beta)+\delta_2 ~, 
\nn\\
	C^{(0)} &= \frac{Z_4}{Z_1}~,
\\
	C^{(2)} &= -\frac{Z_2}{\mathcal{P}}(du+\omega) \wedge (dv+\beta) +a_1 \wedge (dv+\beta)+\gamma_2 ~,
\nn\\
	C^{(4)} &= \frac{Z_4}{Z_2} \text{vol}(\mathcal{M}) -\frac{Z_4}{\mathcal{P}}\gamma_2 \wedge (du+\omega) \wedge (dv+\beta) +\sfx_3 \wedge (dv+\beta)~,
\nn\\
	C^{(6)} &= \text{vol}(\mathcal{M}) \wedge \left[-\frac{Z_1}{\mathcal{P}}(du+\omega) \wedge (dv+\beta) +a_2 \wedge (dv+\beta)+\gamma_1\right]~,
\nn\\
	\mathcal{P} &= Z_1 Z_2 - Z_4 ^2 ~,
\nn
\end{align}
where
\begin{align}
\label{uv2}
	u = \frac{1}{\sqrt{2}} (t-y) ~,~ v= \frac{1}{\sqrt{2}} (t+y)~.
\end{align}
We also use a conventional notation below:
\begin{equation}
	\tau \equiv   C^{(0)} + i e^{-\Phi}.
\end{equation}
S-duality, whose transformation rules are written below, leads to a solution in the NS5-F1-P IIB duality frame, 
\begin{align}
	\widehat{\tau} = -\frac{1}{\tau}	~,~~~
    \widehat{C} ^{(2)} = -B^{(2)} ~,~~~
    \widehat{B} ^{(2)} = C^{(2)} ~,~~~
	\widehat{C} ^{(4)} = C^{(4)} ~,~~~
    \widehat{B}^{(6)}  = C^{(6)}~,~~~
    \widehat{G}_{\mu \nu} = |\tau|G_{\mu \nu}~.
\end{align}
The resulting S-dual line element is
\begin{align}
\label{hatds2}
	(\widehat{d{s}}_{10}) ^{2} = -\frac{2Z_2}{\mathcal{P}}(dv+\beta)\Big[du + \omega + \frac{\mathcal{F}}{2}(dv+\beta)\Big]+Z_2 ds^2 (\mathcal{B})+ds^2 (\mathcal{M})~,
\end{align}
the $B$-field is
\begin{equation}
\label{B2}
	\widehat{B}^{(2)} = -\frac{Z_2}{\mathcal{P}}(du+\omega) \wedge (dv+\beta) +a_1 \wedge (dv+\beta)+\gamma_2 ~.
\end{equation}
The dilaton is given by
\begin{equation}
	e^{2\widehat{\Phi}} = \frac{Z_2 ^2}{\mathcal{P}} ~,
\end{equation}
while the lower-rank R-R potentials are
\begin{align}
	\widehat{C}^{(0)} &= -\frac{Z_4}{Z_2}~,
\nn\\[.2cm]
	\widehat{C}^{(2)} &= \frac{Z_4}{\mathcal{P}}(du+\omega) \wedge (dv+\beta) -a_4 \wedge (dv+\beta)-\delta_2 ~, 
\\[.2cm]
	\widehat{C}^{(4)} &= \frac{Z_4}{Z_2} \text{vol}(\mathcal{M}) -\frac{Z_4}{\mathcal{P}}\gamma_2 \wedge (du+\omega) \wedge (dv+\beta) +\sfx_3 \wedge (dv+\beta)~,
\end{align}
The six-form gauge field to which NS5-branes couple is
\begin{align}
\label{B6}
	\widehat{B}^{(6)} = \text{vol}(\mathcal{M}) \wedge \Big[-\frac{Z_1}{\mathcal{P}}(du+\omega) \wedge (dv+\beta) +a_2 \wedge (dv+\beta)+\gamma_1\Big]~.
\end{align}

\subsection{Hodge Duality Relations}
\label{app:hodge}
In this subsection we write several relations between harmonic functions and form in the NS5-F1-P Type IIB duality frame, which follow from the equation
\begin{equation}
\label{BBtil duality}
d\widetilde{B}^{(6)} = e^{-2\Phi} *dB^{(2)}~.
\end{equation}
Eq.~(\ref{B6}) of the $\widetilde{B}^{(6)}$ six-form field implies the following components of the field strength $d\widetilde{B}^{(6)}$. In this appendix, we dimensionally reduce $\widetilde{B}^{(6)}$ on the four-torus down to a two-form $\widetilde{B}^{(2)}$. 
\begin{equation}
		(d\widetilde {B})_{ijk} = \partial_k \gamma_{1ij} + \partial_j \gamma_{1ki} + \partial_i \gamma_{1jk} ~.
\end{equation}
\begin{equation}
  (d\widetilde{B})_{uvi} = -\partial_i \frac{1}{Z_2}~.
\end{equation}
\begin{equation}
 (d\widetilde{B})_{vij} = \partial_j \Big( \frac{\omega_i}{Z_2}-a_{2i}\Big)-\partial_i \Big( \frac{\omega_j}{Z_2}-a_{2j}\Big)+\partial_v \gamma_{1ij} ~.
\end{equation}
The NS-NS three-form flux which follows from Eq.~(\ref{B2}) is
\begin{equation}
	 	H_{ijk} = \partial_k \gamma_{2ij} + \partial_j \gamma_{2ki} + \partial_i \gamma_{2jk}~.
\end{equation}
\begin{equation}
	 	H_{uvi} = -\partial_i \frac{1}{Z_1}~.
\end{equation}
\begin{equation}
	 	H_{vij} = \partial_j \Big( \frac{\omega_i}{Z_1}-a_{1i}\Big)-\partial_i \Big(\frac{\omega_j}{Z_1} - a_{1j}\Big)+\partial_v \gamma_{2ij}~.
\end{equation}
We will use the following equations:
\begin{equation}
	 	G^{uv} = -Z_1 ~,~ \sqrt{-G} = \frac{Z_2 ^2}{Z_1} ~,~ G^{ui} = -\frac{\omega_i}{Z_2}~,
\end{equation} 
\begin{equation}
	 	\epsilon_{vu1234}=+\sqrt{-G}~.
\end{equation} 
Next, the components of $*H$ are calculated:
\begin{equation}
	 	(*H)_{\mu_1...\mu_7} =\frac{1}{3!} \epsilon_{\mu_1...\mu_7} ^{~~~~~~~\alpha \beta \gamma} H_{\alpha \beta \gamma}~. 
\end{equation}
A dimensional reduction on the torus directions is done below, so that $*H$ is a three-form.
\begin{align}
	 	(*H)_{ijk}&=\epsilon_{ijk} ^{~~~~~~luv} H_{luv} =G^{vu} G^{uv} G^{ll} \epsilon_{ijklvu } H_{luv}\nonumber\\
	 	& =Z_1 ^2 \frac{1}{Z_2} \frac{Z_2 ^2}{Z_1}  \epsilon_{ijkl} \Big(-\partial_l \frac{1}{Z_1}\Big)=\frac{Z_2}{Z_1 }\epsilon_{ijkl} \partial_l Z_1~.
\end{align}
\begin{align}
	 	(*H)_{uvi} &=\frac{1}{3!} \epsilon_{uvi } ^{~~~~~~~jkl} H_{jkl} =\frac{1}{3!}G^{jj} G^{kk} G^{ll} \epsilon_{uvijkl } H_{jkl}= -\frac{1}{3!}\frac{1}{Z_1Z_2}\epsilon_{ijkl} (d\gamma_2) _{jkl}  ~. 
\end{align}
\begin{align}
     	(*H) _{ vij} =& \frac{1}{2}\epsilon_{vij }^{~~~~~~vkl} H_{vkl}+\frac{1}{3!}\epsilon_{vij }^{~~~~~~klm} H_{klm} +
     	\epsilon_{vij } ^{~~~~~~uvk} H_{uvk}\nonumber\\
     	&=\frac{1}{2}G^{vu} G^{kk} G^{ll} \epsilon_{vijukl} H_{vkl} +\frac{1}{3!}G^{kk} G^{ll} G^{mu}\epsilon_{ vij klu} H_{klm}+\frac{1}{3!}G^{kk} G^{lu} G^{mm}\epsilon_{ vij kum} H_{klm}\nonumber\\
     	&+\frac{1}{3!}G^{ku} G^{ll} G^{mm}\epsilon_{vij ulm} H_{klm} + G^{uv}  G^{un} G^{kk}\epsilon_{ vij nuk} H_{uvk}~.
\end{align}
Now, the $u$ index in the Levi Civita tensor is moved one place to the right of the index $v$:
\begin{align}
     (*H)_{vij}&=\frac{1}{2}G^{vu} G^{kk} G^{ll} \epsilon_{vuijkl} H_{vkl} +\frac{1}{3!}G^{kk} G^{ll} G^{mu}\epsilon_{ vuij kl} H_{klm}\nonumber\\
     	&-\frac{1}{3!}G^{kk} G^{lu} G^{mm}\epsilon_{vuij km} H_{klm}+\frac{1}{3!}G^{ku} G^{ll} G^{mm}\epsilon_{vuij lm} H_{klm} - G^{uv}  G^{un} G^{kk}\epsilon_{vuijnk} H_{uvk}~\nonumber\\
     	&=-\frac{Z_1}{2Z_2 ^2} \frac{Z_2 ^2}{Z_1}\epsilon_{ijkl}  \Big[\partial_l \Big( \frac{\omega_k}{Z_1}-a_{1k}\Big)-\partial_k \Big(\frac{\omega_l}{Z_1} - a_{1l}\Big)+\partial_v \gamma_{2kl}\Big] +\nonumber\\
     	&-  \frac{\omega_m}{3!Z_2 ^3}\frac{Z_2^2}{Z_1} \epsilon_{ij kl} (d\gamma_2)_{klm}+
     	\frac{\omega_l}{3!Z_2 ^3}\frac{Z_2^2}{Z_1} \epsilon_{ij km} (d\gamma_2)_{klm}-	\frac{\omega_k}{3!Z_2 ^3}\frac{Z_2^2}{Z_1} \epsilon_{ij lm} (d\gamma_2)_{klm}+\frac{Z_1 \omega_n}{Z_2^2}\frac{Z_2 ^2}{Z_1} \epsilon_{ijnk} \partial_k \frac{1}{Z_1}\nonumber\\
     	&= -\frac{1}{2}\epsilon_{ijkl}  \Big[\partial_l \Big( \frac{\omega_k}{Z_1}-a_{1k}\Big)-\partial_k \Big(\frac{\omega_l}{Z_1} - a_{1l}\Big)+\partial_v (\gamma_{2kl})\Big] +\nonumber\\
     	&-  \frac{\omega_m}{3!Z_1 Z_2 } \epsilon_{ij kl} (d\gamma_2)_{klm}+
     	\frac{\omega_m}{3!Z_1 Z_2 } \epsilon_{ij kl} (d\gamma_2)_{kml}-	\frac{\omega_m}{3!Z_2 Z_1} \epsilon_{ij kl} (d\gamma_2)_{mkl}- \epsilon_{ijkl} \partial_k \Big(\frac{1}{Z_1}\Big)\omega_l~.
\end{align}
The last line can be simplified by noting that three of the terms that involve $\omega \wedge d\gamma_2$ are equal. Also, the last term in the last line cancels against two of the terms in the line before the last. Then the result is
\begin{align}
     	&(*H)_{vij}= -\frac{1}{2}\epsilon_{ijkl}  \Big[\frac{1}{Z_1}\Big(\partial_l \omega_k-\partial_k \omega_l \Big) -\partial_l a_{1k}+\partial_k a_{1l}+\partial_v \gamma_{2kl}\Big] -  \frac{\omega_m}{2Z_1 Z_2 } \epsilon_{ij kl} (d\gamma_2)_{klm}~.
\end{align}
Therefore,
\begin{equation}
     	e^{-2\Phi} (*H)_{ijk} = \epsilon_{ijkl} \partial_l Z_1= (*_4dZ_1)_{ijk}~.
\end{equation}
This implies
\begin{equation}
     	d\gamma_1 = *_4 dZ_1~.
\end{equation}
For $u$-$v$-$i$,
\begin{equation}
     	e^{-2\Phi} (*H)_{uvi} = -\frac{1}{3!Z_2 ^2} \epsilon_{ijkl} (d\gamma_2) _{jkl} \Rightarrow  dZ_2 = *_4 d\gamma_2~.
\end{equation}
Alternatively,
\begin{equation}
     	d\gamma_2 = *_4 dZ_2~.
\end{equation}
For $v$-$i$-$j$,
\begin{align}
     	-\frac{Z_1}{2Z_2}\epsilon_{ijkl}  \Big[&\frac{1}{Z_1}\Big(\partial_l \omega_k-\partial_k \omega_l \Big) -\partial_l a_{1k}+\partial_k a_{1l}+\partial_v \gamma_{2kl}\Big] -  \frac{\omega_m}{ 2Z_2^2 } \epsilon_{ij kl} (d\gamma_2)_{klm}
\nonumber\\
     	& =\partial_j \Big( \frac{\omega_i}{Z_2}-a_{2i}\Big)-\partial_i \Big( \frac{\omega_j}{Z_2}-a_{2j}\Big)+\partial_v \gamma_{1ij}
\end{align}
Multiplying the equation by $Z_2$ and shifting terms from the left to the right sides, the last equation is equivalent to
\begin{align}
     	\partial_i \omega_j - \partial_j \omega_i &-\frac{1}{2}\epsilon_{ijkl}  \Big[\Big(\partial_l \omega_k-\partial_k \omega_l \Big) -Z_1\partial_l a_{1k}+Z_1\partial_k a_{1l}+Z_1\partial_v \gamma_{2kl}\Big] -  \frac{\omega_m}{ 2Z_2 } \epsilon_{ij kl} d(\gamma_2)_{klm}
\nonumber\\
     	& =\omega_i Z_2\partial_j \Big( \frac{1}{Z_2}\Big)-\omega_j Z_2\partial_i \Big( \frac{1}{Z_2}\Big)+Z_2\partial_v \gamma_{1ij}+Z_2 (\partial_i a_{2j}-\partial_j a_{2i})~.
\end{align}
Recall the definitions
\begin{equation}
     	\Theta_1 \equiv da_1  + \partial_v \gamma_2~,
\end{equation}
\begin{equation}
     	\Theta_2 \equiv da_2 + \partial_v \gamma_1 ~.
\end{equation}
Thus,
\begin{align}
     	\label{Omega} 
     	(d\omega)_{ij} +(*d\omega)_{ij} -Z_1 (* \Theta_1)_{ij} -Z_2 \Theta_{2ij} = \frac{\omega_m}{2Z_2} \epsilon_{ijkl} (d\gamma_2)_{klm} - \frac{\omega_i}{Z_2} \partial_j Z_2 + \frac{\omega_j}{Z_2}\partial_i Z_2~.   
\end{align}
Setting $i=1,j=2$ and evaluating 
\begin{align}
     	&\frac{\omega_m}{2Z_2} \epsilon_{12kl} (d\gamma_2)_{klm} = \frac{\omega_1}{Z_2} \epsilon_{1234} (d\gamma_2)_{341}+\frac{\omega_2}{Z_2} \epsilon_{1234} (d\gamma_2)_{342}=\frac{\omega_1 \partial_2 Z_2 - \omega_2 \partial_1 Z_2 }{Z_2}~.
\end{align}
It follows that
\begin{align}
         	\label{338a} 
     	d\omega +*d\omega =Z_1  *\Theta_1+Z_2 \Theta_{2} ~.   
\end{align}
This equation is consistent with Eq.~(3.45) of~\cite{Giusto:2013rxa} and Eq.~(3.38a) of \cite{Shigemori:2020yuo} under the assumptions $\beta=0,Z_4=0$.

\section{Solution summary}
\label{app:soln sum}

We gather here all the metric coefficients and other supergravity fields appearing in the three-charge BPS geometry~\eqref{AnsatzIIB}.  
\begin{subequations}
\label{soln sum}
\begin{align}
\label{Z2 recap}
 Z_2(\sfx,v) &= 1+ \frac{\kappa_6^2 \tau_{\sst\rm NS5}}{2\pi^2}\sum_{\sfm =1} ^{\nfive} \frac{1}{|\sfx-\sfF_\sfm(v)|^2} ~.
\\
\label{Z1 recap}
 Z_1(\sfx,v) &= 1+ \frac{\kappa_6^2 \tau_{\sst\rm F1}}{2\pi^2}\sum_{\sfn =1} ^{\none} \frac{1}{|\sfx-\sfX_\sfn (v)|^2}~. 
\\[.2cm]
\label{Theta1 recap}
	\Theta_{1ij} &= -\frac{\kappa_6^2 \tau_{\sst\rm NS5}}{\pi^2}\sum_{\sfm =1} ^{\nfive} \frac{\partial_v \sfF_{\sfm i} (\sfx_j-\sfF_{\sfm j}) - \partial_v \sfF_{\sfm j} (\sfx_i-\sfF_{\sfm i}) + \epsilon_{ijkl} \partial_v \sfF_{km} (\sfx_l-\sfF_{\sfm l}) }{|\sfx-\sfF_\sfm|^4}~.
\\[.2cm]
\label{Theta2 reap}
	\Theta_{2ij} &= -\frac{\kappa_6^2 \tau_{\sst\rm F1}}{\pi^2}\sum_{\sfn =1} ^{\none} \frac{\partial_v \sfX_{\sfn i} (\sfx_j-\sfX_{\sfn j}) - \partial_v \sfX_{\sfn j} (\sfx_i-\sfX_{\sfn i}) + \epsilon_{ijkl} \partial_v \sfX_{\sfn k} (\sfx_l-\sfX_{\sfn l}) }{|\sfx-\sfX_\sfn |^4}~.
\\[.2cm]
\label{omega recap}
\omega_i &= \frac{\kappa_6 ^2 \tau_{\sst\rm NS5}}{2\pi^2} \sum_{\sfm =1} ^{\nfive} \frac{\partial_v \sfF_{\sfm i}}{\widetilde\sfR_\sfm  ^2} + \frac{\kappa_6^2 \tau_{\sst\rm F1}}{2\pi ^2} \sum_{\sfn =1} ^{\none} \frac{\partial_v \sfX_{\sfn i}}{\sfR_\sfn ^2}+\frac{\kappa_6 ^4 \tau_{\sst\rm NS5} \tau_{\sst\rm F1}}{8\pi^4 } \sum_{\sfm =1} ^{\nfive} \sum_{\sfn =1} ^{\none} \frac{\partial_v \sfF_{\sfm i} + \partial_v \sfX_{\sfn i}}{\sfR_\sfn  ^2 \widetilde\sfR_\sfm ^2}
\nn\\
&\hskip .5cm
-\frac{\kappa_6 ^4 \tau_{\sst\rm NS5}\tau_{\sst\rm F1}}{8\pi^4} \sum_{\sfm =1} ^{\nfive} \sum_{\sfn =1} ^{\none} \frac{\partial_v \sfF_{\sfm i}- \partial_v \sfX_{\sfn i}}{|\sfR_\sfn -\widetilde\sfR_\sfm |^2} \left(\frac{1}{\sfR_\sfn  ^2}-\frac{1}{\widetilde\sfR_\sfm ^2}\right)
\\
&\hskip 1cm
+\frac{\kappa_6 ^4 \tau_{\sst\rm NS5} \tau_{\sst\rm F1}}{4\pi^4} \sum_{\sfm =1} ^{\nfive} \sum_{\sfn =1} ^{\none} \frac{\partial_v \sfF_{\sfm j} - \partial_v \sfX_{\sfn j}}{|\sfR_\sfn  - \widetilde\sfR_\sfm |^2}\frac{\mathcal{A}_{ij}}{\sfR_\sfn  ^2 \widetilde\sfR_\sfm ^2}~.
\nn\\[.2cm]
\label{F recap}
	\mathcal{F} =&  -\frac{\kappa_6^2 \tau_{\sst\rm NS5}}{2\pi^2}\sum_{\sfm =1} ^{\nfive} \frac{Z_1 (\sfF_\sfm)|\partial_v \sfF_{\sfm } |^2}{|\sfx-\sfF_\sfm|^2}-\frac{\kappa_6^2 \tau_{\sst\rm F1}}{2\pi^2}\sum_{\sfn =1} ^{\none} \frac{Z_2(\sfX_\sfn )|\partial_v \sfX_\sfn |^2}{|\sfx-\sfX_\sfn |^2} \nonumber\\
&\hskip .5cm
-\frac{\kappa_6^4 \tau_{\sst\rm NS5} \tau_{\sst\rm F1}}{4\pi^4}\sum_{\sfm =1} ^{\nfive} \sum_{\sfn =1} ^{\none} \partial_v\sfF_{{m}}\cdot \partial_v \sfX_{\sfn }\left[ \frac{1}{\widetilde\sfR_\sfm ^2 \sfR_\sfn  ^2}  - \frac{1}{|\sfR_\sfn  - \widetilde\sfR_\sfm |^2}\Big(\frac{1}{\widetilde\sfR_\sfm ^2} + \frac{1}{\sfR_\sfn ^2} \Big) \right]
\\	
&\hskip 1cm
- \frac{\kappa_6^4 \tau_{\sst\rm NS5} \tau_{\sst\rm F1}}{2\pi^4}\sum_{\sfm =1} ^{\nfive} \sum_{\sfn =1} ^{\none}\frac{\mathcal{A}_{ij} \partial_v \sfF_{\sfm i} \partial_v \sfX_{\sfn j}}{\widetilde\sfR_\sfm  ^2 \sfR_\sfn ^2|\sfR_\sfn - \widetilde\sfR_\sfm |^2} ~.
\nn
\end{align}
\end{subequations}
where
\begin{equation}
	\widetilde\sfR_\sfm \equiv \sfx-\sfF_\sfm (v) ~~,~~~~ \sfR_\sfn  \equiv \sfx-\sfX_\sfn (v)~.
\end{equation}
\begin{equation}
\label{CurlyA recap}
	\mathcal{A}_{ij} \equiv\widetilde\sfR_{\sfm i}\sfR_{\sfn j} -\sfR_{\sfn i}\widetilde\sfR_{\sfm j} -\epsilon_{ijkl}\widetilde\sfR_{\sfm k}\sfR_{\sfn l}~,
\end{equation}
and we recall that the $\Theta_i$ ($i=1,2$) encode the harmonic forms $a_i,\gamma_i$ appearing in the supergravity fields via~\eqref{Thetadefs}.

\section{Strings in the round NS5-P supertube}
\label{app:GWZW}

Perturbative string dynamics in the round NS5-P supertube is exactly solvable~\rcite{Martinec:2017ztd,Martinec:2018nco,Martinec:2020gkv,Martinec:2022okx,Martinec:2025xoy,Bena:2025uyg}, however the focus in these works is predominantly on strings in the T-dual NS5-F1 supertube.  The NS5-P background is of course related by T-duality along the $y$-circle, which is just a field redefinition of the worldsheet theory, nevertheless for completeness we summarize here the properties of BPS perturbative strings in the round NS5-P background.  More details may be found in the aforementioned references.
We then use the results to link our solution for round sources in section~\ref{sec:helices} to the effective superstratum analysis of~\rcite{Bena:2025uyg} summarized in section~\ref{sec:singlemode}.

\subsection{Setup}
\label{app:setup}

\begin{figure}[ht]
\centering
\includegraphics[scale=0.37]{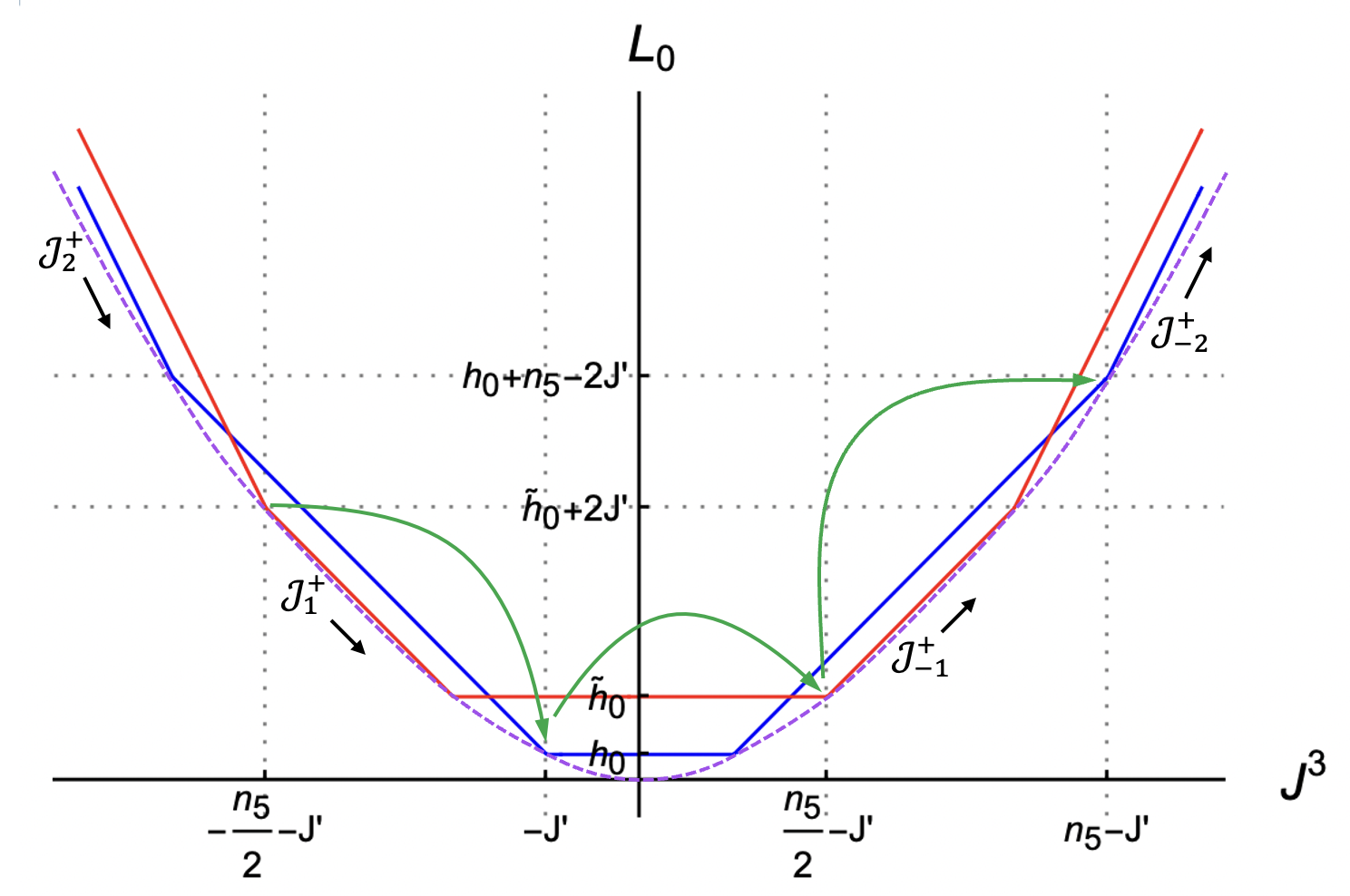}
\caption{\it Even amounts of spectral flow relate extremal weights in the affine weight diagram of $\sutwo$ spin $j'$ (the blue polygon).  Allowed weights lie on or within the polygonal bound.  Odd amounts of spectral flow map an extremal weight of spin $j'$ to an extremal weight of spin $\frac\nfive2-j'$ (the red polygon).  The conformal dimensions of the ground states are $h_0=\frac{j'(j'+1)}{\nfive}$ and $\tilde h_0 = \frac{\tilde j'(\tilde j'+1)}{\nfive}$, where $\tilde j'=\frac\nfive2-j'$.  The dashed purple curve is the trajectory of the extremal weights under continuous spectral flow.}
\label{fig:SU2specflow}
\end{figure}

The null-gauged Wess-Zumino-Witten (WZW) model
\be
\label{cosets}
\frac\cG\cH \,=\, \frac{\sltwo\times\sutwo\times \bR_t\times\bS^1_y}{U(1)_L\times U(1)_R} \times \bT^4 ~,
\ee
with $\cH$ a pair of null isometries of $\cG$ generated by the currents
\begin{align}
\begin{split}
\label{null-currents}
\cJ \,=\, J^3_\sl + J^3_\su -\frac{\sfp}{R_y}\big( \,i\partial \sft + i\partial \sfy\big)
~~,~~~~
\bar\cJ \,=\, \bar J^3_\sl + \bar J^3_\su -\frac{\sfp}{R_y}\big( \,i\bar\partial \sft + i\bar\partial \sfy\big)  ~,
\end{split}
\end{align}
is an exactly solvable worldsheet string model in the background of the round NS5-P supertube whose source profile is given by~\eqref{roundF}.
The gauging of these currents imposes constraints on the quantum numbers of physical string states
\begin{align}
&(2\Msl + n_5 \wsl) + (2\Msu+n_5
    \wsu)  -\frac{\sfp}{R_y}\big( E + P_{y}\big) \;=\;0 ~ ,  
\nn \\[.1cm]
&(2 \bMsl + n_5 \bwsl) + (2\bMsu+ n_5 \bwsu) -\frac{\sfp}{R_y}\big( E  +  \bar{P}_{y}\big) \;=\;0 ~ , 
\label{null-constr}
\end{align}
where $\Msu,\bMsu$ are the eigenvalues of $J^3_\su,\bar J^3_\su$, with $\wsu,\bwsu$ corresponding spectral flow quantum numbers; similarly for $\Msl,\bMsl,w$ in $\sltwo$; also, $E,P_y,\bar P_y$ are the charges under the translation currents $i\partial t,i\partial y,i\bar\partial y$.  The effect of $\sutwo$ spectral flow is depicted in figure~\ref{fig:SU2specflow}.

The zero-mode Virasoro constraints are~\rcite{Martinec:2018nco}
\begin{align}
\label{Vir-constr}
\begin{split}
L_0 - \frac12 
\,&=\,  -\frac{j(j-1)}{n_5} +
\frac{j'(j'+1)}{n_5}  - \Msl \wsl -
\frac{n_5}{4}\wsl^2 
+ \Msu \wsu + \frac{n_5}{4}\wsu^2
-\frac14 E^2 +\frac14 P_{y}^2 + h_L \,=\,0\;,  
\\[.3cm]
\bar L_0 - \frac12 \,&=\,  -\frac{\jsl(\jsl-1)}{n_5} +
\frac{\jsu(\jsu+1)}{n_5}  - \bMsl \bwsl -
\frac{n_5}{4}\bwsl^2 
+ \bMsu \bwsu+ \frac{n_5}{4}\bwsu^2
-\frac14 E^2 +\frac14 \bar{P}_{y}^2 + h_R \,=\,0 \;,  
\end{split}
\end{align}
where $\jsl,\jsu$ are the principal quantum numbers of the underlying bosonic WZW models, and $h_{L,R}$ are the contributions of nonzero modes, plus the NS sector ground state energy $-1/2$.

\subsection{BPS ground states}
\label{app:BPS gd st}

Right-BPS states have
\begin{align}
\label{F1bps}
\jsl=\jsu+1
~~,~~~~
\bMsu=-\bMsl\approx -j
~~,~~~~
\bwsu=-\wsl 
~~,~~~~
h_R=0
~,
\end{align}
where the approximation sign in the second relation results from order one shifts in the total spin relative to the center-of-mass spin coming from the string's polarization state, which we will ignore because we work in the WKB limit of large $j$. 

The right-moving Virasoro constraint then sets $E=-\bar P_y$; parametrizing these as
\be
\label{EPvals}
E=w_yR_y+\frac\vareps{R_y}
~~,~~~~
P_y = w_y R_y + \frac{n_y}{R_y}
~~,~~~~
\bar P_y = -w_y R_y + \frac{n_y}{R_y}
~,
\ee
one has
\be
\label{epsbps}
\vareps = - n_y ~,
\ee
and the right gauge constraint is also satisfied.  

There is a spectral flow symmetry under large gauge transformations of the axial $\uone$, which shifts the charges according to
\be
\label{gauge specflow}
\delta\wsl = q
~~,~~~~
\delta\wsu = -q
~~,~~~~
\delta\bwsu = -q
~~,~~~~
\delta \vareps = -\sfp\,  q
~~,~~~~
\delta n_y = \sfp\, q ~.
\ee
We use this symmetry to set $\wsl=0$ (and thus $\bwsu=0$ by the BPS condition~\eqref{F1bps}).  Level matching (\ie\ $L_0-\bar L_0\in\bZ$) then requires $\wsu\in2\bZ$.  The left constraints are then solved by
\begin{align}
\label{ny and M}
n_y = -\bigg(\frac{h_L+\Msu\wsu+\frac\nfive4\wsu^2}{w_y}\bigg)
~~,~~~~
\Msl = -\Msu-\frac\nfive2\wsu+\sfp\, w_y  ~;
\end{align}
momentum quantization imposes $n_y\in\bZ$.
Note that the T-dual of superstratum excitations has $h_L=\wsu=0$ in order to be straight strings carrying no momentum.  This makes sense, because $\sutwo$ spectral flow is an internal automorphism of the representations, and thus can be re-expressed in terms of current raising operators (and thus oscillator excitations). 

The physical conserved charges of the string are measured by a set of currents that commute with the gauge currents~\rcite{Martinec:2025xoy},%
\footnote{Which are in particular are invariant under gauge spectral flow transformations.} 
and corresponding gauge invariant charges%
\begin{align}
\vareps + n_y
~~,~~~~
\cM = \Msl + \frac\nfive2\Big(\wsl +\frac{\vareps-n_y}{2\sfp}\Big)
~~&,~~~~
\cM' = \Msu + \frac\nfive2\Big(\wsu -\frac{\vareps-n_y}{2\sfp}\Big)
\nn\\[.1cm]
w_y
~~,~~~~
\bar\cM = \bMsl + \frac\nfive2\Big(\wsl +\frac{\vareps-n_y}{2\sfp}\Big)
~~&,~~~~
\bar\cM' = \bMsu + \frac\nfive2\Big(\bwsu -\frac{\vareps-n_y}{2\sfp}\Big)
\end{align}
When $\sfp>1$, these charges are not all integer quantized, because they measure charge against the charge per unit moementum of the fivebrane background.

Linear combinations of these gauge invariant charges can be identified with the physical charges of strings in the background NS5-P spacetime:
\begin{align}
\cP_u &= \cE+\cN_p = \vareps+n_y
~~,~~~~
\cJ_R = \bar\cM+\bar\cM' = \bMsl+\bMsu+\frac{\nfive}2\big( \wsl+\bwsu\big)
\nn\\[.2cm]
\cN_1 &= w_y
\qquad~,\qquad\qquad\quad~~
\cJ_L = \cM'-\bar\cM' = \Msu-\bMsu+\frac\nfive2\big(\wsu-\bwsu\big)
\\[.2cm]
\cP_v &= \cE - \cN_p = \frac{\sfp}{\nfive}\big(\bar\cM-\bar\cM'\big)  = \frac{\sfp}{\nfive}\Big[\bMsl-\bMsu + \frac\nfive2 \Big( w - \bwsu + \frac{\vareps-n_y}{\sfp}\Big)\Big] ~.
\nn\end{align}
For BPS strings~\eqref{F1bps}, \eqref{epsbps}, one has 
\begin{align}
\label{GI charges}
\cP_u &=0
~~,~~~~
\cP_v = - n_y + \frac{2j\sfp}{\nfive}  
~~,~~~~
\cJ_R =0
~~,~~~~
\cJ_L = \big(\Msu+j\big)+\frac\nfive2 \wsu.
\end{align}
where we have used~\eqref{ny and M}.
Vanishing Killing momentum $\cP_u$ is a requirement for BPS-ness; $\cJ_R=0$ agrees with the absence of a string contribution to right-moving angular momentum in~\eqref{round chgs}.  
Below, we will match $\cP_v$ to~\eqref{round chgs} using~\eqref{ny and M}, after determining the string ring size $\mfb$ in terms of the worldsheet quantum numbers.  Note the additional fractional contribution $\frac{2j\sfp}{\nfive}$ as compared to the naively expected integer-valued first term.
As discussed in~\rcite{Martinec:2025xoy}, the gauge invariant charge~\eqref{GI charges} includes a contribution to the charges carried by the string which corresponds to the amount of that charge which was carried by the fivebrane in the absence of the string, so that one maintains a fixed overall charge budget.  Here this charge is momentum, and in~\rcite{Martinec:2025xoy} it is string winding charge.  This term is in general not quantized, but renders the expression for the charge invariant under large gauge transformations~\eqref{gauge specflow}, which trade charge carried on the string for charge carried as flux in the fivebrane background.  
It arises as part of the term $\frac{|\partial_v\sfF|^2}{|\sfX-\sfF|^2}$ in the fivebrane worldvolume effective action~\eqref{Joint}.%
\footnote{This same fractional charge contribution was noted in~\cite{Giveon:2001up} (see Eq.~3.36 there), and interpreted in~\cite{Kim:2015gak} as the worldsheet formalism for global $AdS_3\times \bS^3$ working at fixed chemical potential rather than fixed charge.  In the context of the gauged WZW model, which describes perturbations around the fivebrane supertube, we have a different interpretation~-- the worldsheet formalism conserves charge, and the fractional contributions to gauge invariant charge in correlators account for the fractional amounts of charge being carried as background flux, that are being manipulated by the vertex operator making a transition in the state of the holographic dual theory.}

The group coordinates correspond to the spheroidal coordinates $r,\theta,\phi,\psi$ of~\eqref{GH-bipolar}.
The peak of the string center-of-mass wavefunction is located at
\be
\label{F1location}
\Big(\frac{r_1}{\mfa}\Big)^2 = \frac{\Msl-j}{2j} \equiv \frac{n}{2j}
~~,~~~~
\cos^2\theta_1 = \frac{j+\Msu}{2j} \equiv \frac{m}{2j}
\ee
up to shifts of order $j^{-1}$ stemming from the effects of the string polarization state,
with the width of the peak scaling as $j^{-1/2}$.  

The classical solution~\eqref{roundX} for a circular string corresponds to the WKB limit of large $j$, where the relative width of the wavefunction is tiny.  The location of the string is 
\be
\label{stringrad}
|\sfX|^2 = \mfb^2 = r_1^2 + \mfa^2 \sin^2\theta_1 = \Big(\frac{\Msl-\Msu}{2j}\Big)  \mfa^2 = \Big(\frac{-2\Msu-\frac\nfive2\wsu+\sfp w_y}{2j}\Big) \mfa^2 ~,
\ee
and it wraps a particular cycle in the three-torus parametrized by $y,\phi,\psi$ (choosing say $\sigma$ to be fixed as a gauge choice).
The separation of the circular string from the circular fivebrane is given by
\be
\label{NS5-F1 distance}
\mfb^2-\mfa^2 = \Big(\frac{-2j-2\Msu-\frac\nfive2\wsu+\sfp w_y}{2j}\Big) \mfa^2    ~.
\ee
For strings lying entirely in the $\sfx^1$-\,$\sfx^2$ plane, one has 
\begin{align}
\label{rtheta cases}
\begin{split}
r_1^2 &= \mfb^2-\mfa^2
~~,~~~~
\theta_1=\frac\pi2
~~\Longrightarrow~~
\Msu = -j
~~,~~~~
\mfb>\mfa
\\[.2cm]
r_1^2 &= 0 
~~\Longrightarrow~~
\Msl=j
~~~~,~~~~~~\,
\sin^2\theta_1 = \frac{\mfb^2}{\mfa^2}
~~,~~~~
\mfb<\mfa
~.
\end{split}
\end{align}

\subsection{Classical strings}
\label{app:classical}

One can also solve directly the classical WZW model~\rcite{Martinec:2020gkv,Martinec:2025xoy,Bena:2025uyg}.  Classical pointlike string trajectories are geodesics; those related to the above BPS string states are given by
\be
\label{geodesic gsl}
g_\sl(\xi_0) = 
\left(\begin{matrix} e^{i\tau}\cosh\rho ~&~  e^{i\sigma}\sinh\rho 
\\[.1cm] 
e^{-i\sigma}\sinh\rho ~&~ e^{-
i\tau} \cosh\rho \end{matrix}\right)
=
\left(\begin{matrix} 
e^{+i\nu\xi_0}\cosh\frac{\alpha}2 
~~&~  
e^{-i\nu\xi_0}\sinh\frac{\alpha}2 
\\[.1cm] 
e^{+i\nu\xi_0}\sinh\frac{\alpha}2 
~~&~~ 
e^{-i\nu\xi_0}\cosh\frac{\alpha}2 
\end{matrix}\right) ~,
\ee
where $\rho,\tau,\sigma$ are standard global coordinates on $\sltwo$ with $r=a\sinh\rho$, and $\sigma_\pm =\xi_0\pm\xi_1$, $\nu=2j/\nfive$, and $\cosh\alpha=M/j$.
One finds that the geodesic sits at the fixed radial position
\be
\label{rhopm}
\cosh \rho_1 = \cosh\big(\hf\alpha\big)
~~\Longrightarrow~~
r_1 = a\sinh\big(\hf\alpha\big)
~.
\ee
Similarly, for $\sutwo$ one finds spectrally flowed geodesics
\be
\label{flowed gsu}
g_\su = 
\left(\begin{matrix} e^{-i\phi}\sin\theta ~&~  e^{i\psi}\cos\theta 
\\[.1cm] 
- e^{-i\psi}\cos\theta ~&~ e^{i\phi} \sin\theta \end{matrix}\right)
=
\left(\begin{matrix} 
e^{-i[(2 \nu'+w' )\xi_0 + w' \xi_1]/2}\cos\frac{\alpha'}2 ~&~  
e^{+i[ (2 \nu'-w' )\xi_0 - w' \xi_1]/2}\sin\frac{\alpha'}2 
\\[.1cm] 
- e^{-i[ (2 \nu'-w' )\xi_0 - w' \xi_1]/2}\sin\frac{\alpha'}2 ~&~ 
e^{+i[(2 \nu'+w')\xi_0 + w' \xi_1]/2} \cos\frac{\alpha'}2 
\end{matrix}\right) ~,
\ee
where $\nu'=2j'/\nfive$, and $\cos\alpha'=-M'/j$.  The geodesic sits at the fixed polar angle
\be
\sin\theta_1 = \cos\big(\hf\alpha'\big)
~~\Longrightarrow~~
\theta_1 = \half\big(\pi-\alpha'\big) ~.
\ee
In addition, one has a classical string configuration along $\bR_t\times\bS^1_y$
\be
\sft = -\Big(\frac{n_y}{R_y}-w_y R_y\Big)\xi_0
~~,~~~~
\sfy = \frac{n_y}{R_y} \xi_0 + w_y R_y \,\xi_1 ~.
\ee
One should then impose the reparametrization and gauge constraints, for which a solutions is~\eqref{F1bps},~\eqref{ny and M}.  

We wish to choose the gauge condition $\tau=\sigma=0$, so that the physical coordinates in the target space are $t,y,r,\theta,\phi,\psi$, however the above geodesic motion is not in this gauge.  Therefore we make a gauge transformation
\begin{align}
\label{gaugetransfs}
\delta\tau & = l_1\alpha\tight+r_1\beta = (\alpha\tight+\beta) \;,\qquad\quad~~~
\delta\phi = -l_2\alpha\tight-r_2\beta = -(\alpha\tight+\beta)  \;,
\nn\\[1.5mm]
\delta\sigma &= l_1\alpha\tight-r_1\beta = (\alpha\tight-\beta)\;, \qquad\quad~~~
\delta\psi =  +l_2\alpha \tight-r_2\beta = (\alpha\tight-\beta) \;,
\\[1mm]
\delta \sft &= l_3\alpha\tight+r_3\beta = -\frac\sfp{R_y}(\alpha\tight+\beta)
\;,\qquad
\delta \sfy = { -l_4\alpha } 
\tight-r_4\beta = \frac{\sfp}{R_y}(\alpha\tight+\beta)\;.
\nn
\end{align}
with gauge parameters
\be
\alpha = 0
~~,~~~~
\beta = -\frac{2j\xi_0}{\nfive} ~.
\ee

We have already imposed the BPS conditions~\eqref{F1bps} as well as $\vareps=-n_y$.  The constraints then fix the location~\eqref{F1location}, which not surprisingly agrees with the peak of the quantized string wavefunction.  Assembling the components, one finds the classical string trajectory in Cartesian coordinates
\begin{align}
\label{Xgwzw}
\sfX^1+i\sfX^2 &= \sqrt{r_1^2+\mfa^2}\, \sin\theta_1\,\exp\Big[+\frac{i\sqrt{2}w'v}{2w_y R_y}\Big]
~~,~~~~
\sfX^3+i\sfX^4 = r_1\, \cos\theta_1\,\exp\Big[-\frac{i\sqrt{2}w'v}{2w_y R_y}\Big] ~.
\end{align}
We thus find a more general class of rotating strings than~\eqref{roundX}, not necessarily oriented in the $\sfx^1$-\,$\sfx^2$ plane.  Motion restricted to the $\sfx^1$-\,$\sfx^2$ plane imposes one or the other of the two options~\eqref{rtheta cases}.

\subsection{Comparing to superstrata}
\label{app:matching}

Comparing~\eqref{Xgwzw} to~\eqref{roundX}, we identify 
\be
k= \frac{\wsu}2 
\ee
so that the null constraint~\eqref{ny and M} reads
\be
\label{nullcon}
m+n=\sfp w_y - k\nfive ~.
\ee
In anticipation of the relation to the effective single-mode superstrata of section~\ref{sec:singlemode}, we set
\be
\ell = 2j ~.
\ee
Although for the analysis here this relation can be regarded as a definition, one can verify that the string's position~\eqref{F1location} and its separation from the origin~\eqref{stringrad} and the fivebrane~\eqref{NS5-F1 distance} match the corresponding features of the T-dual momentum source~\eqref{locations}, with $\ell,m,n$ having identical values.

Two rather useful relations for strings lying in the $\sfx^1$-\,$\sfx^2$ plane follow from applying~\eqref{rtheta cases} to~\eqref{stringrad}:
\be\label{inside out} 
\big|\mfb^2-\mfa^2\big| = \mfa^2\Big(\frac{\Msl+\Msu}{2j}\Big)
~~,~~~~
{\rm min}(\mfa^2,\mfb^2) = \mfa^2\Big(\frac{j-\Msu}{2j}\Big)  ~.
\ee
Inserting them in the expression for the smeared source angular momentum $J_L$,~\eqref{round chgs}, 
and using the null constraint~\eqref{nullcon} as well as~\eqref{stringrad},
the string contribution to the angular momentum charge $J_L$ in~\eqref{round chgs} evaluates to
\be
\label{JLstr}
J_{L,1} = \frac{n_1}{w_y} \, \frac{\sfp w_y(j+\Msu)+kn_5(\Msl-j)}{\Msl+\Msu}
= \frac{n_1}{w_y} \, \big( m+\nfive k\big) ~.
\ee 
We find agreement with the angular momentum~$\cJ_L$ per string~\eqref{GI charges}.  Note also that there is no pathology when $m=n=0$, and the string is sitting on top of the supertube.

While the expression~\eqref{round chgs} was derived using strings restricted to the $\sfx^1$-\,$\sfx^2$ plane, when expressed in terms of the quantum numbers of the general string solutions of section~\ref{app:BPS gd st}, it holds for strings in general position, having general $m$ and $n$.
We find agreement between the angular momentum $J_L$ in the smeared supergravity solution~\eqref{round chgs} and the exactly solvable worldsheet dynamics~\eqref{GI charges}.  
These examples involve spiraling string sources that carry momentum as well as winding along $\bS^1_y$; when specialized to strings having winding but carrying no momentum, we find agreement with the angular momentum of the supergravity wave of single-mode superstrata.

The relations~\eqref{stringrad}, \eqref{inside out} also allow us to write the momentum charge~\eqref{round chgs} as
\begin{align}
\label{Qp F1}
n_p &= \bigg(1+\frac{\none \mu^2 }{ |\mfb^2-\mfa^2|}\bigg)\frac{\sfp^2\mfa^2}{n_5\mu^2}
+ \frac{\none}{w_y}\bigg[\frac{\nfive k^2(\ell+n-m)}{w_y(m+n)} - \frac{2\sfp k (\ell-m)}{m+n} \bigg] 
\nn\\[.2cm]
&= \frac{\sfp\, 2J_R}{\nfive} + 
\frac{\none}{w_y}\bigg[
\frac{\sfp\ell}{\nfive}  + \frac{(2m-\ell) k +\nfive k^2}{w_y}  \bigg] 
\\[.2cm]
&\equiv n_{p,5} + n_{p,1} ~.
\nn
\end{align}
The term in square brackets on the second line is just $\cP_v$, the momentum~\eqref{GI charges}  of each string source.  Note that again, there is no pathology when $m=n=0$, and the string sits on top of the supertube.

To compare to (orbifolded) superstrata~\cite{Shigemori:2022gxf}, we set $k=0$, and then the first term on the R.H.S. of the first line of~\eqref{Qp F1} reproduces~\eqref{dualbubble}.
We find the T-dual of the ``strand budget'' relation~\eqref{strandbudget} of (orbifolded) superstrata
\be
\label{dual budget}
n_pn_5 = \sfp\Big( 2J_R + \frac{n_1\ell}{w_y} \Big) ~.
\ee
There are $N_0=\frac{n_1}{w_y}\in\bZ$ strings.  In the original superstratum frame, the third charge is momentum, 
\be
n_1=w_y N_0 = \frac{m+n}{\sfp}N_0 = \tilde n_{ p} ~,
\ee
where $N_0$ is the number of internal R-R scalar modes excited on the fivebrane, Eq.~\eqref{Qp 1mode}; the background strands have length $\sfp$, and their contribution to the string winding budget in that frame is $\sfp\ell N_0$, accounting for the second term on the R.H.S. of~\eqref{dual budget}.  
The first term on the R.H.S. is the contribution to the strand budget of the underlying fivebrane supertube, from modes which T-dualize to the part of the string winding budget coming from the number $N_+=2J_R$ of transversely polarized excitations on the fivebrane, each of strand length $\sfp$ and carrying a half unit of angular momentum on both left and right.

We thus have a complete match between (1) the string profiles $\sfX(v)$ in our classical solutions for round sources elaborated in section~\ref{sec:helices}; and (2) BPS ground states in the exact worldsheet theory of the round NS5-P supertube.  When the strings carry no momentum, we also have a match to single-mode superstrata via T-duality, and a stringy generalization when they do carry both winding and momentum, for $k\ne0$.

Since T-duality along $\bS^1_y$ does not affect the radial structure, the estimate~\eqref{Lstratum} of the length of the throat holds.  In particular, we see that the deepest throats indeed correspond to the cap descending to the non-abelian scale, where D-brane excitations become as light as string excitations.

We can estimate the $AdS_2$ throat depth of the more stringy solutions as well.  The momentum quantum is~\eqref{Qp F1};
for simplicity, consider the limit of large $k$, so that the momentum charge is predominantly carried on the F1 strings.  One has
\be
n_p \sim n_1 n_5 \frac{k^2}{w_y^2}  
~~,~~~~
\mfb^2 \sim 2J_R\frac{n_5 k}{\sfp \ell}
~.
\ee
The length of the $AdS_2$ throat comes entirely from the region between the strings at radius $\mfb$ and the top of the throat at the scale $Q_p$, because the (T-dual) $AdS$ scale below $r\sim \mfb$ is much smaller than beyond $r\sim\mfb$.  The throat length is thus
\begin{align}
\frac{L_{\rm throat}}{R_{AdS_2}} &\approx  \log\Big[\frac{Q_p}{\mfb^2} \Big] \approx  \log \Big[ \frac{n_1\sfp\ell k}{2J_R w_y^2} \Big] 
\nn\\[.2cm]
& \approx  \log\bigg[ \sqrt{n_1 n_5{\textstyle \big(\frac{n_1 n_5 k^2}{w_y^2}\big)}} \cdot \frac{\sfp\ell}{2J_R w_y n_5} \bigg] ~.
\end{align}
The argument of the square root $n_1n_5n_p$ is of order the extremal entropy, while the remaining factor in the argument of the log is smaller than one~-- affine $\sutwo$ highest weights are restricted to $\ell<\nfive$; and $\mfa^2=\frac{2J_R}{\sfp} \mu^2$, so for the fivebrane ring to be larger than the quantum scale one must have $2J_R>\sfp$.  Thus, once again the length of the throat is at most that of the extremal black hole, Eq.~\eqref{depth}.

\section{Contributions to $\omega$ and $\mathcal{F}$ for concentric supertubes}
\label{app:om-F aves}

The purpose of this appendix is to calculate the smeared $\omega_{2r}$ and $\cF_{2r}$ in the example of the circular profiles for the strings and fivebranes, 
\be
\label{roundF2}
\sfF^1 _{\sfm}+i\sfF^2 _{\sfm}= \mfa\, \exp\Big[\frac{i\sfp v\sqrt{2}}{n_5 R_y}+\frac{2\pi i \sfm}{\nfive}\Big]
~~,~~~~
\sfF^3+i\sfF^4 =0 ~.
\ee
\be
\label{roundX2}
\sfX^1 _{\sfn}+i\sfX^2 _{\sfn} = \mfb\, \exp\Big[ \frac{ik v\sqrt{2}}{w_y R_y} +\frac{2\pi i \sfn}{w_y}\Big]
~~,~~~~
\sfX^3+i\sfX^4 = 0~. 
\ee
The one-form and scalar $\omega_{2r}$ and $\cF_{2r}$ are defined through
\begin{align}
\label{omega2rApp}
 \omega_{2r}\equiv  \sum_{\sfm =1} ^{\nfive} \sum_{\sfn =1} ^{\none} 
 \frac{\mathcal{A}_{ij,\sfm\sfn} (\partial_v \sfF_{\sfm j} - \partial_v \sfX_{\sfn j})}{\sfR_\sfn  ^2 \widetilde\sfR_\sfm ^2 |\sfR_\sfn  - \widetilde\sfR_\sfm |^2}~,
\end{align}
\begin{align}
 \mathcal{F}_{2r}=- 2\sum_{\sfm =1} ^{\nfive} \sum_{\sfn =1} ^{\none}\frac{\mathcal{A}_{ij,\sfm \sfn} \partial_v \sfF_{\sfm i} \partial_v \sfX_{\sfn j}}{\widetilde\sfR_\sfm  ^2 \sfR_\sfn ^2|\sfR_\sfn - \widetilde\sfR_\sfm |^2}~,
\end{align}
where 
\begin{align}
     		\mathcal{A}_{ij,\sfm \sfn} \equiv (\sfx_i - \sfF_{i\sfm})(\sfx_j - \sfX_{j\sfn}) - (\sfx_{j\sfn} - \sfF_{j\sfm})(\sfx_i - \sfX_{i\sfn})-\epsilon_{ijkl} (\sfx_k - \sfF_{k\sfm}) (\sfx_l -\sfX_{l\sfn})~.
     	\end{align}
We will calculate all the contributions to
\begin{align}
  \bar{\bar{\omega}}_{2r} \equiv  \frac{1}{\sqrt{2} \pi R_y} \int_0 ^{\sqrt{2} \pi R_y} dv \frac{1}{2\pi}\int_0 ^{2\pi} d\phi~\omega_{2r}~.
\end{align} 
\begin{align}
  \bar{\bar{\cF}}_{2r} \equiv  \frac{1}{\sqrt{2} \pi R_y} \int_0 ^{\sqrt{2} \pi R_y} dv\frac{1}{2\pi}\int_0 ^{2\pi} d\phi ~\cF_{2r}~.
\end{align} 
Also, define
\begin{align}
\phi_5 = \frac{p\tilde{v}}{n_5} +\frac{2\pi m}{n_5}
~~,~~~~
\phi_1 = \frac{k \tilde{v}}{w_y} + \frac{2\pi n}{w_y}~.
\end{align}
We compute the components of $\mathcal{A}_{ij}$:
\begin{align}
     		\mathcal{A}_{12} &= 
            \mfa \sqrt{x_1 ^2 +x_2^2} \sin(\phi_5-\phi) + \mfb \sqrt{\sfx_1^2 +\sfx_2^2} \sin(\phi-\phi_1) +\mfa \mfb \sin(\phi_1-\phi_5)~. 
\nn\\
     		\mathcal{A}_{13} &= 
            \sfx_3 (\mfb \cos(\phi_1)-\mfa \cos(\phi_5))+\sfx_4 (\mfb\sin(\phi_1)-\mfa\sin(\phi_5))~.
\nn\\
     		\mathcal{A}_{14} &= \sfx_4 (\mfb\cos(\phi_1) - \mfa \cos(\phi_5)) +\sfx_3 (\mfa \sin(\phi_5) - \mfb \sin(\phi_1))~.
\nn\\
     		\mathcal{A}_{23} &=-\mathcal{A}_{14}~~,~~~~
     		\mathcal{A}_{24} =\mathcal{A}_{13}~~,~~~~
     		\mathcal{A}_{34}=\mathcal{A}_{12}~.
     	\end{align}
The $\phi$ component of $\omega_{2r}$ is related to the expressions
\begin{align}
     		\mathcal{A}_{\phi j} \partial_v F_j &=\frac{\sfp \mfa^2 (\sfx_1^2+\sfx_2^2)}{\nfive R_y \sqrt{2}} \big(1-\cos(2(\phi-\phi_5))\big)+\frac{\mfa^2 \mfb \sfp \sqrt{2}\sqrt{\sfx_1^2+\sfx_2^2}}{\nfive R_y } \sin(\phi_1 -\phi_5)\sin(\phi_5-\phi)\nonumber\\
     		&\quad -\frac{\sfp \mfa \mfb(\sfx_1^2+\sfx_2^2)}{\sqrt{2}\nfive R_y }\big(\cos(\phi_1-\phi_5)-\cos(\phi_1+\phi_5-2\phi)\big)~, 
\nn\\
     		-\mathcal{A}_{\phi j} \partial_v X_j &=\frac{k \mfb^2 (\sfx_1^2+\sfx_2^2)}{\sqrt{2} R_y w_y}(1-\cos(2(\phi-\phi_1)))+\frac{\mfa \mfb^2 k \sqrt{2}\sqrt{\sfx_1^2+\sfx_2^2} }{R_y w_y} \sin(\phi_1-\phi_5)\sin(\phi-\phi_1) \nonumber\\
     		& \quad -\frac{k \mfa \mfb (\sfx_1^2+\sfx_2^2)}{\sqrt{2}R_y w_y} \big(\cos(\phi_1-\phi_5)-\cos(\phi_1+\phi_5-2\phi)\big)~.
\end{align}
The denominators of $\omega_{2r}$ and $\cF_{2r}$ are determined by the following expressions
\begin{align}
           |\sfx-\sfF_{\sfm}|^2 &= |\sfx|^2 + \mfa^2 -2 \mfa \sqrt{\sfx_1^2+\sfx_2 ^2} \cos(\phi_5-\phi)~,
\nn\\
           |\sfx-\sfX_{\sfn}|^2 &= |\sfx|^2 + \mfb^2 -2 \mfb \sqrt{\sfx_1^2+\sfx_2 ^2} \cos(\phi_1-\phi)~,
\nn\\
           |\sfF_{\sfm} - \sfX_{\sfn}|^2 &= \mfa^2 +\mfb^2 -2\mfa \mfb \cos(\phi_1-\phi_5)~.
\end{align}
Define the functions
\begin{equation}
     		c_a (\mfa,\sfx) \equiv \frac{2\mfa\sqrt{\sfx_1 ^2 + \sfx_2 ^2}}{\mfa^2 + |\sfx|^2} ~,~ c_b(\mfb,\sfx) \equiv \frac{2 \mfb \sqrt{\sfx_1 ^2 +\sfx_2 ^2}}{\mfb^2 + |\sfx|^2}~,
\end{equation}
and
\begin{equation}
     		A \equiv \frac{2\mfa \mfb}{\mfa^2+\mfb^2}~.
\end{equation}
We start by defining the following contributions to $\omega_{2r,\phi}$ using Eq.~(\ref{omega2rApp}) and the equations above:
\begin{align}
     		\omega_{2r,\phi,1} &=\frac{\none (\sfx_1^2+\sfx_2^2)\Big(\frac{\sfp \mfa^2 }{\nfive  }+\frac{k \mfb^2}{w_y}\Big)}{w_y R_y\sqrt{2}(\mfa^2+\mfb^2)(|\sfx|^2+\mfa^2)(|\sfx|^2+\mfb^2)}\sum_{m,n} \frac{1}{1-\frac{2\mfa \mfb}{\mfa^2+\mfb^2}\cos(\phi_1-\phi_5)}\times\nonumber\\
     		&\qquad \qquad \qquad \qquad \qquad \qquad  \frac{1}{(1-c_a \cos(\phi-\phi_5))(1-c_b \cos(\phi-\phi_1))}~.
\nn\\
     	\omega_{2r,\phi,2} &=-\frac{\sfp \none}{w_y}\frac{\mfa^2 (\sfx_1^2+\sfx_2^2)}{\nfive R_y \sqrt{2}}\sum_{m,n}\frac{\cos(2(\phi-\phi_5))}{(\mfa^2+\mfb^2)(|\sfx|^2+\mfa^2)(|\sfx|^2+\mfb^2)}\times \nonumber\\
     	& \qquad \qquad \qquad \qquad  \frac{1}{1-\frac{2\mfa \mfb}{\mfa^2+\mfb^2}\cos(\phi_1-\phi_5)}\frac{1}{(1-c_a \cos(\phi-\phi_5))(1-c_b \cos(\phi-\phi_1))}~.
\nn\\[.2cm]
     		\omega_{2r,\phi,3}&=\frac{\sfp \none}{w_y}\frac{\mfa^2 \mfb\sqrt{2}\sqrt{\sfx_1^2+\sfx_2^2}}{\nfive R_y }\sum_{m,n} \frac{\sin(\phi_1 -\phi_5)\sin(\phi_5-\phi)}{(\mfa^2+\mfb^2)(|\sfx|^2+\mfa^2)(|\sfx|^2+\mfb^2)} \frac{1}{1-\frac{2\mfa \mfb}{\mfa^2+\mfb^2}\cos(\phi_1-\phi_5)}\times \nonumber\\
     		&\qquad \qquad \qquad \qquad  \qquad \qquad \frac{1}{(1-c_a \cos(\phi-\phi_5))(1-c_b \cos(\phi-\phi_1))}~.
\nn\\[.2cm]
     		\omega_{2r,\phi,4}&=-\frac{ \none (\sfp w_y +k \nfive)}{w_y}\frac{\mfa \mfb(\sfx_1^2+\sfx_2^2)}{\nfive w_y R_y \sqrt{2}}\sum_{m,n} \frac{\cos(\phi_1-\phi_5)}{(\mfa^2+\mfb^2)(|\sfx|^2+\mfa^2)(|\sfx|^2+\mfb^2)}\times \nonumber\\
     		& \qquad \qquad \frac{1}{1-\frac{2\mfa \mfb}{\mfa^2+\mfb^2}\cos(\phi_1-\phi_5)}\frac{1}{(1-c_a \cos(\phi-\phi_5))(1-c_b \cos(\phi-\phi_1))}~.
\nn\\[.2cm]
     		\omega_{2r,\phi,5} &= \frac{ \none}{w_y}\frac{\mfa \mfb(\sfx_1^2+\sfx_2^2)(w_y \sfp+\nfive k)}{\nfive w_yR_y \sqrt{2}}\sum_{m,n}\frac{\cos(\phi_1+\phi_5-2\phi)}{{(\mfa^2+\mfb^2)(|\sfx|^2+\mfa^2)(|\sfx|^2+\mfb^2)}}\frac{1}{1-\frac{2\mfa \mfb}{\mfa^2+\mfb^2}\cos(\phi_1-\phi_5)}\times\nonumber\\
     		&\qquad \qquad \qquad \qquad \qquad \qquad \qquad \frac{1}{(1-c_a \cos(\phi-\phi_5))(1-c_b \cos(\phi-\phi_1))}~.
\nn\\[.2cm]
         	\omega_{2r,\phi,6} &=-\frac{\none k}{w_y}\frac{\mfb^2 (\sfx_1^2+\sfx_2^2)}{w_y R_y \sqrt{2}}\sum_{m,n}\frac{\cos(2(\phi-\phi_1))}{(\mfa^2+\mfb^2)(|\sfx|^2+\mfa^2)(|\sfx|^2+\mfb^2)}\times \nonumber\\
         	& \qquad \qquad \qquad \qquad \qquad \frac{1}{1-\frac{2\mfa \mfb}{\mfa^2+\mfb^2}\cos(\phi_1-\phi_5)}\frac{1}{(1-c_a \cos(\phi-\phi_5))(1-c_b \cos(\phi-\phi_1))}~.
\nn\\[.2cm]
         	\omega_{2r,\phi,7}&=\frac{\none k}{w_y}\frac{\mfa \mfb^2\sqrt{\sfx_1^2+\sfx_2^2}\sqrt{2}}{R_y }\sum_{m,n} \frac{\sin(\phi_1 -\phi_5)\sin(\phi-\phi_1)}{(\mfa^2+\mfb^2)(|\sfx|^2+\mfa^2)(|\sfx|^2+\mfb^2)} \frac{1}{1-\frac{2\mfa \mfb}{\sfa^2+\sfb^2}\cos(\phi_1-\phi_5)}\times \nonumber\\
         	&\qquad \qquad \qquad \qquad \qquad \qquad  \frac{1}{(1-c_a \cos(\phi-\phi_5))(1-c_b \cos(\phi-\phi_1))}~.
\end{align}
Averaging over $\phi$ and carrying out the sums in $\omega_{2r,\phi,1}$, one finds
\begin{align}
     		\bar{\bar{\omega}}_{2r,\phi,1} &=\frac{\none (\sfx_1^2+\sfx_2^2)}{w_y R_y\sqrt{2}}\big(\sfp w_y \mfa^2 +\nfive k \mfb^2\big)\frac{1}{(|\sfx|^2+\mfa^2)(|\sfx|^2+\mfb^2)(\mfa^2+\mfb^2)}\times 
\nonumber\\
     		& \Big(\frac{1}{\sqrt{1-c_b^2}} \mathcal{I}_1 \Big(\frac{2\mfa \mfb}{\mfa^2+\mfb^2},\frac{c_a}{c_b},-i\frac{c_a}{c_b}\sqrt{1-c_b^2}\Big)+\frac{1}{\sqrt{1-c_a^2}} \mathcal{I}_1 \Big(\frac{2\mfa \mfb}{\mfa^2+\mfb^2},\frac{c_b}{c_a},i\frac{c_b}{c_a}\sqrt{1-c_a^2}\Big)\Big)~.
\end{align}
Here,
     			\begin{equation}
     				\mathcal{I}_1 (a,b,c) = \frac{1}{2\pi} \int_0 ^{2\pi} \frac{d\alpha}{(1-a \cos(\alpha))(1-b \cos(\alpha) - c \sin(\alpha))}~,
     			\end{equation}
                whose result appears in the next subsection in Eq.~(\ref{I1}).
The next contribution is
\begin{align}
\bar{\bar{\omega}}_{2r,\phi,2}=&~\frac{\sfx_1 ^2 +\sfx_2^2}{R_y \sqrt{2} (|\sfx|^2+\mfa^2)(|\sfx|^2+\mfb^2)}\frac{\none \sfp \mfa}{\mfb}\Big(1-\frac{\mfa^2+\mfb^2}{|\mfa^2-\mfb^2|}\Big)\nonumber\\
        &-\frac{\sfp \mfa^2(\sfx_1 ^2 +\sfx_2^2)\none}{2R_y \sqrt{2} (|\sfx|^2+\mfa^2)(|\sfx|^2+\mfb^2)(\mfa^2+\mfb^2)}\frac{2-c_a^2}{2c_a^2 \sqrt{1-c_a^2}}\times \mathcal{I}_1 \Big(\frac{2\mfa \mfb}{\mfa^2+\mfb^2},\frac{c_b}{c_a},\frac{ic_b}{c_a}\sqrt{1-c_a^2}\Big)\nonumber\\
        &- \frac{\sfp \mfa^2(\sfx_1 ^2+\sfx_2^2)\none}{2R_y \sqrt{2} (|\sfx|^2+\mfa^2)(|\sfx|^2+\mfb^2)(\mfa^2+\mfb^2)}\frac{ (1+\sqrt{1-c_b^2})^2}{2c_b^2 \sqrt{1-c_b^2}}\times \mathcal{I}_{e^{-2i\alpha}} \Big(\frac{2\mfa \mfb}{\mfa^2+\mfb^2},\frac{c_a}{c_b},-\frac{ic_a}{c_b}\sqrt{1-c_b^2}\Big)\nonumber\\
      	&  -\frac{\sfp \mfa^2(\sfx_1 ^2+\sfx_2^2)\none}{2R_y \sqrt{2} (|\sfx|^2+\mfa^2)(|\sfx|^2+\mfb^2)(\mfa^2+\mfb^2)}\frac{ \big(1-\sqrt{1-c_b^2}\big)^2}{2c_b^2 \sqrt{1-c_b^2} }\mathcal{I}_{e^{2i\alpha}} \Big(\frac{2\mfa \mfb}{\mfa^2+\mfb^2},\frac{c_a}{c_b},-\frac{ic_a}{c_b}\sqrt{1-c_b^2}\Big)~.
\end{align}
The integrals are written in Eqs.~(\ref{I1}),(\ref{Iexp2I}) and its complex conjugate formula. The result for $\bar{\bar{\omega}}_{2r,\phi,3}$ is
\begin{align}
\bar{\bar{\omega}}_{2r,\phi,3}=&\frac{\sfp \none \mfa^2 \mfb\sqrt{\sfx_1^2+\sfx_2^2}}{ R_y \sqrt{2}(\mfa^2+\mfb^2)(|\sfx|^2+\mfa^2)(|\sfx|^2+\mfb^2)}\times \nonumber\\
         	&\Bigg[\Big(\frac{1}{4c_b\sqrt{1-c_b^2}} \Bigg((1-\sqrt{1-c_b^2})\mathcal{I}_{e^{2i \alpha}}\Big(\frac{2\mfa \mfb}{\mfa^2+\mfb^2},\frac{c_a}{c_b},-\frac{ic_a}{c_b}\sqrt{1-c_b^2}\Big)+\nonumber\\\
         	&(1+\sqrt{1-c_b^2})\mathcal{I}_{e^{-2i\alpha}}\Big(\frac{2\mfa \mfb}{\mfa^2+\mfb^2},\frac{c_a}{c_b},-\frac{ic_a}{c_b}\sqrt{1-c_b^2}\Big)-2\mathcal{I}_{1} \Big(\frac{2\mfa \mfb}{\mfa^2+\mfb^2},\frac{c_a}{c_b},-i\frac{c_a}{c_b}\sqrt{1-c_b^2}\Big)\Bigg)\nonumber\\
         	&-\frac{i}{c_a} \mathcal{I}_{\sin(\alpha)}\Big(\frac{2\mfa \mfb}{\mfa^2+\mfb^2},\frac{c_b}{c_a},\frac{ic_b}{c_a}\sqrt{1-c_a^2}\Big)\Bigg]~.
\end{align}
The result of the integral involving a $\sin(\alpha)$ in the numerator is written in Eq.~(\ref{Isin}).
\begin{align}
      	\bar{\bar{\omega}}_{2r,\phi,4} =& \frac{ \none}{w_y}\frac{(\sfx_1^2+\sfx_2^2)}{2\sqrt{2}}(\sfp w_y +k \nfive)\frac{(\text{conditional})}{ \sqrt{(|\sfx|^2+\mfa^2)^2-4\mfa^2 (\sfx_1^2 + \sfx_2^2)}\sqrt{(|\sfx|^2+\mfb^2)^2-4\mfb^2 (\sfx_1^2 + \sfx_2^2)}}\nonumber\\
      	&+\frac{ \none}{w_y}\frac{(\sfx_1^2+\sfx_2^2)}{2\sqrt{2}}(\sfp w_y +k \nfive)\frac{1}{ (|\sfx|^2+\mfa^2)(|\sfx|^2+\mfb^2)}\times \nonumber\\
      	&\Bigg(\frac{1}{\sqrt{1-c_b^2}}\mathcal{I}_1 \Big(\frac{2\mfa \mfb}{\mfa^2+\mfb^2},\frac{c_a}{c_b},-\frac{ic_a}{c_b}\sqrt{1-c_b^2}\Big)+\frac{1}{\sqrt{1-c_a^2}}\mathcal{I}_1 \Big(\frac{2\mfa \mfb}{\mfa^2+\mfb^2},\frac{c_b}{c_a},-\frac{ic_b}{c_a}\sqrt{1-c_a^2}\Big)\Bigg)~.
\end{align}
The word ``conditional'' means that the contributions is nonzero if 
\be
c_b \Big(1-\sqrt{1-c_a^2}\Big) <c_a \Big(1-\sqrt{1-c_b^2}\Big)< c_b \Big(1+\sqrt{1-c_a^2}\Big)
\ee
or 
\be
c_a \Big(1-\sqrt{1-c_b ^2}\Big)< c_b \Big(1+\sqrt{1-c_a^2}\Big)<c_a \Big(1+\sqrt{1-c_b^2}\Big) ~.
\ee
The remaining contributions are
\begin{align}
\bar{\bar{\omega}}_{2r,\phi,5} =& \frac{ \none}{w_y}\frac{\mfa \mfb(\sfx_1^2+\sfx_2^2)}{R_y \sqrt{2}}(w_y \sfp+\nfive k)\frac{1}{{(\mfa^2+\mfb^2)(|\sfx|^2+\mfa^2)(|\sfx|^2+\mfb^2)}}\times\nonumber\\ 
      		&\Bigg(\frac{2(\mfa^2+\mfb^2)}{c_a c_b |\mfa^2-\mfb^2|}+\frac{(1+\sqrt{1-c_a ^2})^2}{2c_a^2\sqrt{1-c_a ^2} }\mathcal{I}_{e^{i\alpha} } \Big(\frac{2\mfa \mfb}{\mfa^2+\mfb^2},\frac{c_b}{c_a},\frac{ic_b}{c_a} \sqrt{1-c_a^2}\Big)\nonumber\\
      		&+\frac{(1-\sqrt{1-c_a^2})^2}{2\sqrt{1-c_a ^2} c_a ^2} \mathcal{I}_{e^{-i\alpha} } \Big(\frac{2\mfa \mfb}{\mfa^2+\mfb^2},\frac{c_b}{c_a},\frac{ic_b}{c_a} \sqrt{1-c_a^2}\Big) \nonumber\\
      		&+\frac{(1+\sqrt{1-c_b ^2})^2}{2c_b^2\sqrt{1-c_b ^2} } \mathcal{I}_{e^{-i \alpha}}\Big(\frac{2\mfa \mfb}{\mfa^2+\mfb^2},\frac{c_a}{c_b},-i\frac{c_a}{c_b}\sqrt{1-c_b^2}\Big)\nonumber\\
      		&+\frac{(1-\sqrt{1-c_b^2})^2}{2c_b ^2\sqrt{1-c_b ^2} }\mathcal{I}_{e^{i \alpha}}\Big(\frac{2\mfa \mfb}{\mfa^2+\mfb^2},\frac{c_a}{c_b},-i\frac{c_a}{c_b}\sqrt{1-c_b^2}\Big)
      		\Bigg)~.
\end{align} 
\begin{align}					\bar{\bar{\omega}}_{2r,\phi,6}=&~ -\frac{k \none \nfive}{w_y}\frac{\mfb^2 (\sfx_1^2+\sfx_2^2)}{ R_y \sqrt{2}} \frac{1}{(\mfa^2+\mfb^2)(|\sfx|^2+\mfa^2)(|\sfx|^2+\mfb^2)}\Bigg(\frac{\mfa^2+\mfb^2}{2\mfa \mfb}\Big(1-\frac{\mfa^2+\mfb^2}{|\mfa^2-\mfb^2|}\Big)\nonumber\\
    &+\frac{(1+\sqrt{1-c_a^2})^2}{2c_a^2 \sqrt{1-c_a^2}}\mathcal{I}_{e^{2i\alpha}} \Big(\frac{2\mfa \mfb}{\mfa^2+\mfb^2},\frac{c_b}{c_a},\frac{ic_b}{c_a}\sqrt{1-c_a^2}\Big) \nonumber\\
    &+\frac{(1-\sqrt{1-c_a^2})^2}{2c_a^2 \sqrt{1-c_a^2}}\mathcal{I}_{e^{-2i\alpha}} \Big(\frac{2\mfa \mfb}{\mfa^2+\mfb^2},\frac{c_b}{c_a},\frac{ic_b}{c_a}\sqrt{1-c_a^2}\Big)\nonumber\\
    &+\frac{2-c_b^2}{c_b^2 \sqrt{1-c_b^2}}\mathcal{I}_1 \Big(\frac{2\mfa \mfb}{\mfa^2+\mfb^2},\frac{c_a}{c_b},-\frac{ic_a}{c_b}\sqrt{1-c_b^2}\Big)
      					\Bigg)~.
\end{align}
\begin{align}
	\bar{\bar{\omega}}_{2r,\phi,7} =&~ \frac{k \none \nfive \mfa \mfb^2 \sqrt{2(\sfx_1^2+\sfx_2^2)}}{R_y (\mfa^2+\mfb^2)(|\sfx|^2+\mfa^2)(|\sfx|^2+\mfb^2)}  \times \nonumber\\
	&\Bigg[ \frac{1}{c_a} \mathcal{I}_{i\sin \cos} \Big(\frac{2\mfa \mfb}{\mfa^2+\mfb^2},\frac{c_b}{c_a},\frac{ic_b}{c_a}\sqrt{1-c_a^2}\Big)-\frac{1}{c_a\sqrt{1-c_a^2}} \mathcal{I}_{\sin^2} \Big(\frac{2\mfa \mfb}{\mfa^2+\mfb^2},\frac{c_b}{c_a},i\frac{c_b}{c_a}\sqrt{1-c_a^2}\Big) \nonumber\\
	& -\frac{i}{c_b} \mathcal{I}_{\sin}\Big(\frac{2\mfa \mfb}{\mfa^2+\mfb^2},\frac{c_a}{c_b},-i\frac{c_a}{c_b}\sqrt{1-c_b^2}\Big)\Bigg]~.
\end{align}
The other nonzero component of the double-smeared $\omega_{2r}$ is the $\psi$ direction, where
\begin{align}
\label{omega2rResult3}
     		\bar{\bar{\omega}}_{2r,\psi} =&~ \frac{\none \sfp \mfa\sqrt{2}}{w_y R_y (|\sfx|^2+\mfa^2)(|\sfx|^2+\mfb^2)(\mfa^2+\mfb^2)} \Bigg(\frac{2\sfx_3 \sfx_4 \mfa}{\sqrt{1-c_b^2}}\mathcal{I}_1 \Big(\frac{2\mfa \mfb}{\mfa^2+\mfb^2},\frac{c_a}{c_b},-\frac{ic_a}{c_b}\sqrt{1-c_b^2}\Big)\nonumber\\
     		&\frac{2\sfx_3 \sfx_4 \mfa}{\sqrt{1-c_a^2}}\mathcal{I}_1 \Big(\frac{2\mfa \mfb}{\mfa^2+\mfb^2},\frac{c_b}{c_a},\frac{ic_b}{c_a}\sqrt{1-c_a^2}\Big)-\frac{2\sfx_3 \sfx_4 \mfb}{\sqrt{1-c_b^2}}\mathcal{I}_{\cos(\alpha)}\Big(\frac{2\mfa \mfb}{\mfa^2+\mfb^2},\frac{c_a}{c_b},-\frac{ic_a}{c_b}\sqrt{1-c_b^2}\Big)\nonumber\\
     		& -\frac{2\sfx_3 \sfx_4 \mfb}{\sqrt{1-c_a^2}}\mathcal{I}_{\cos(\alpha)}\Big(\frac{2\mfa \mfb}{\mfa^2+\mfb^2},\frac{c_b}{c_a},\frac{ic_b}{c_a}\sqrt{1-c_a^2}\Big)-\frac{\sfx_3 ^2 + \sfx_4^2}{\sqrt{1-c_b^2}}\mathcal{I}_{\sin(\alpha)}\Big(\frac{2\mfa \mfb}{\mfa^2+\mfb^2},\frac{c_a}{c_b},-\frac{ic_a}{c_b}\sqrt{1-c_b^2}\Big)\nonumber\\
     		& -\frac{\sfx_3 ^2 + \sfx_4^2}{\sqrt{1-c_a^2}}\mathcal{I}_{\sin(\alpha)}\Big(\frac{2\mfa \mfb}{\mfa^2+\mfb^2},\frac{c_b}{c_a},\frac{ic_b}{c_a}\sqrt{1-c_a^2}\Big) \Bigg)\nonumber\\
     		&+\frac{\none \nfive k \mfb\sqrt{2}}{w_y R_y (|\sfx|^2+\mfa^2)(|\sfx|^2+\mfb^2)(\mfa^2+\mfb^2)} \Bigg(\frac{2\sfx_3 \sfx_4 \mfb}{\sqrt{1-c_b^2}}\mathcal{I}_1 \Big(\frac{2\mfa \mfb}{\mfa^2+\mfb^2},\frac{c_a}{c_b},-\frac{ic_a}{c_b}\sqrt{1-c_b^2}\Big)\nonumber\\
     		&\frac{2\sfx_3 \sfx_4 \mfb}{\sqrt{1-c_a^2}}\mathcal{I}_1 \Big(\frac{2\mfa \mfb}{\mfa^2+\mfb^2},\frac{c_b}{c_a},\frac{ic_b}{c_a}\sqrt{1-c_a^2}\Big)-\frac{2\sfx_3 \sfx_4 \mfa}{\sqrt{1-c_b^2}}\mathcal{I}_{\cos(\alpha)}\Big(\frac{2\mfa \mfb}{\mfa^2+\mfb^2},\frac{c_a}{c_b},-\frac{ic_a}{c_b}\sqrt{1-c_b^2}\Big)\nonumber\\
     		& -\frac{2\sfx_3 \sfx_4 \mfa}{\sqrt{1-c_a^2}}\mathcal{I}_{\cos(\alpha)}\Big(\frac{2\mfa \mfb}{\mfa^2+\mfb^2},\frac{c_b}{c_a},\frac{ic_b}{c_a}\sqrt{1-c_a^2}\Big)-\frac{\sfx_3 ^2 + \sfx_4^2}{\sqrt{1-c_b^2}}\mathcal{I}_{\sin(\alpha)}\Big(\frac{2\mfa \mfb}{\mfa^2+\mfb^2},\frac{c_a}{c_b},-\frac{ic_a}{c_b}\sqrt{1-c_b^2}\Big)\nonumber\\
     		& -\frac{\sfx_3 ^2 + \sfx_4^2}{\sqrt{1-c_a^2}}\mathcal{I}_{\sin(\alpha)}\Big(\frac{2\mfa \mfb}{\mfa^2+\mfb^2},\frac{c_b}{c_a},\frac{ic_b}{c_a}\sqrt{1-c_a^2}\Big) \Bigg)~.
\end{align}
Finally, we write contributions to the doubled-smeared $\mathcal{F}_{2r}$:
\begin{align}
\label{F2r smeared}
     		\bar{\bar{\mathcal{F}}}_{2r} =&~ -\frac{4 \mfa \mfb \sfp k \none}{w_y R_y^2 (\mfa^2+\mfb^2)(|\sfx|^2+\mfa^2)(|\sfx|^2+\mfb^2)}\times \nonumber\\
     		&\Bigg( \frac{\mfa}{c_b \sqrt{1-c_b^2}}\big(\mfb c_b + \sqrt{\sfx_1^2+\sfx_2^2}\big)\mathcal{I}_{\sin^2}\Big(\frac{2\mfa \mfb}{\mfa^2+\mfb^2},\frac{c_a}{c_b},i\frac{c_a}{c_b}\sqrt{1-c_b^2}\Big)\nonumber\\
     		&+\frac{\mfb}{c_a \sqrt{1-c_a^2}}\big(\mfa c_a + \sqrt{\sfx_1^2+\sfx_2^2}\big)\mathcal{I}_{\sin^2}\Big(\frac{2\mfa \mfb}{\mfa^2+\mfb^2},\frac{c_b}{c_a},-i\frac{c_b}{c_a}\sqrt{1-c_a^2}\Big)\nonumber\\
     		& -\frac{\mfa \sqrt{\sfx_1^2+\sfx_2^2}}{c_b}\mathcal{I}_{i\sin \cos} \Big(\frac{2\mfa \mfb}{\mfa^2+\mfb^2},\frac{c_a}{c_b},i\frac{c_a}{c_b}\sqrt{1-c_b^2}\Big) \nonumber\\
     		&+\frac{\mfb \sqrt{\sfx_1^2+\sfx_2^2}}{c_a}\mathcal{I}_{i\sin \cos} \Big(\frac{2\mfa \mfb}{\mfa^2+\mfb^2},\frac{c_b}{c_a},i\frac{c_b}{c_a}\sqrt{1-c_a^2}\Big)\nonumber\\
     		&+\frac{\mfa \sqrt{\sfx_1^2+\sfx_2^2}}{c_a}\mathcal{I}_{i\sin}\Big(\frac{2\mfa \mfb}{\mfa^2+\mfb^2},\frac{c_b}{c_a},-\frac{ic_b}{c_a}\sqrt{1-c_a^2}\Big)\nonumber\\
     		&-\frac{\mfb \sqrt{\sfx_1^2+\sfx_2^2}}{c_b}\mathcal{I}_{i\sin}\Big(\frac{2\mfa \mfb}{\mfa^2+\mfb^2},\frac{c_a}{c_b},-\frac{ic_a}{c_b}\sqrt{1-c_b^2}\Big)
     		\Bigg)~.
     	\end{align}

\subsection{Table of integrals}
\label{app:ints}

Several integrals are used for the calculations of smeared contributions to $\omega$ and $\mathcal{F}$. The method of residues is used to derive all integral formulas in this subsection. 
We start with the following integrals
\begin{equation}
     		\int_0 ^{2\pi}\frac{\cos(\alpha) d\alpha}{1-a \cos(\alpha) - b\sin(\alpha)}=\frac{2\pi a}{\sqrt{1-a^2-b^2}(1+\sqrt{1-a^2-b^2})}~.
\end{equation}
\begin{equation}
     		\int_0 ^{2\pi}\frac{\sin(\alpha) d\alpha}{1-a \cos(\alpha) - b\sin(\alpha)}=\frac{2\pi b}{\sqrt{1-a^2-b^2}(1+\sqrt{1-a^2-b^2})}~.
\end{equation}
\begin{equation}
     		\frac{1}{2\pi} \int_0 ^{2\pi} \frac{\cos(\phi-\phi_0)d\phi}{A+B \cos(\phi-\phi_1)} = \frac{\cos(\phi_1-\phi_0)}{B}\Big(1-\frac{A}{\sqrt{A^2-B^2}}\Big)~.
\end{equation}
More complicated integrals have a product structure to their denominator
\begin{align}
\label{I1}
     		\mathcal{I}_1 (a,b,c,d)&\equiv\frac{1}{2\pi} \int_0 ^{2\pi} \frac{d\phi}{1-a \cos(\phi)-b\sin(\phi)}\frac{1}{1-c \cos(\phi)-d\sin(\phi)}\nonumber\\
     		&=\frac{a^2+b^2}{\sqrt{1-a^2-b^2}\Big(a^2 + b^2 - ac -bd + i(bc-da)\sqrt{1-a^2-b^2}\Big)}\nonumber\\
     		&+\frac{c^2+d^2}{\sqrt{1-c^2-d^2}\Big(c^2 +d^2 - ac -bd + i(da-bc)\sqrt{1-c^2-d^2}\Big)}~.
\end{align}
\begin{align}
\label{Isin}
     		\mathcal{I}_{\sin(\alpha)}\equiv&\frac{1}{2\pi}\int_0 ^{2\pi}\frac{\sin(\alpha)d\alpha}{(1-a \cos(\alpha))(1-b \cos(\alpha)-c\sin(\alpha))}  \nonumber\\
     		&=\frac{-ib -\frac{ c}{\sqrt{1 - b^2 - c^2}}}{a b - 
     		i a c \sqrt{1 - b^2 - c^2} - b^2 - c^2} - 
     		\frac{i}{\big(b - a + i c \sqrt{1 - a^2}\big)}~.
\end{align}
\begin{align}
     		\mathcal{I}_{\cos(\alpha)}&\equiv\frac{1}{2\pi}\int_0 ^{2\pi} 
            \frac{\cos(\phi)d\phi}{(1-a \cos(\phi))(1-b \cos(\phi)-c \sin(\phi))} =-\frac{1}{a\sqrt{1-a^2}\sqrt{1-b^2-c^2}}\nonumber \\
     		&+\frac{1}{a} \Big[\frac{a}{\sqrt{1-a^2}(a-b-ic \sqrt{1-a^2})}+\frac{b^2+c^2}{\sqrt{1-b^2-c^2}\big(b^2+c^2-ab+i ac\sqrt{1-b^2-c^2}\big)}\Big]~.
\end{align}
\begin{align}
      		\mathcal{I}_{e^{i\alpha}} (A,a,b) \equiv \frac{1}{2\pi}\int_0 ^{2\pi} \frac{e^{i\alpha}}{1-A \cos(\alpha)} \frac{1}{1-a \cos(\alpha) - b \sin(\alpha)}d\alpha=\mathcal{I}_{\cos(\alpha)}+ i \mathcal{I}_{\sin (\alpha)}~,
      		\end{align}
\begin{align}
      		\mathcal{I}_{e^{-i\alpha}} (A,a,b) \equiv \frac{1}{2\pi}\int_0 ^{2\pi} \frac{e^{-i\alpha}}{1-A \cos(\alpha)} \frac{1}{1-a \cos(\alpha) - b \sin(\alpha)}d\alpha=\mathcal{I}_{\cos(\alpha)}- i \mathcal{I}_{\sin (\alpha)}~~.
      				\end{align}
\begin{align}
\label{Iexp2I}
     		\mathcal{I}_{e^{2i\alpha}}(a,b,c) \equiv& \frac{1}{2\pi}\int_0 ^{2\pi} \frac{e^{2i\alpha}}{(1-a \cos(\alpha))(1-b \cos(\alpha)-c \sin(\alpha))}d\alpha\nonumber\\
     		&= \frac{4}{a(b-ic)} -\frac{ a(1+\sqrt{1-a^2})}{\sqrt{1-a^2}(1-\sqrt{1-a^2})(b-ic\sqrt{1-a^2}-a)}\nonumber\\
     		&-\frac{\epsilon (b+ic)^2 (1+\sqrt{1-b^2-c^2})}{\sqrt{1-b^2-c^2}(1-\sqrt{1-b^2-c^2})\Big(ab + i ca\sqrt{1-b^2-c^2}-b^2-c^2\Big)}~.
\end{align}
The parameter $\epsilon$ takes the values $0$ or $1$ depending on whether $\frac{1}{1-b \cos(\alpha) - c \sin(\alpha)}$ has a pole inside the unit circle when writing $z = e^{i\alpha}$. The complex conjugate of the last integral is 
\begin{align}
     		\mathcal{I}_{e^{-2i\alpha}}(a,b,c) \equiv& \frac{1}{2\pi}\int_0 ^{2\pi} \frac{e^{-2i\alpha}}{(1-a \cos(\alpha))(1-b \cos(\alpha)-c \sin(\alpha))}d\alpha\nonumber\\
     		&= \frac{4}{a(b+ic)} -\frac{ a(1+\sqrt{1-a^2})}{\sqrt{1-a^2}(1-\sqrt{1-a^2})(b+ic\sqrt{1-a^2}-a)}\nonumber\\
     		&-\frac{\epsilon (b-ic)^2 (1+\sqrt{1-b^2-c^2})}{\sqrt{1-b^2-c^2}(1-\sqrt{1-b^2-c^2})\Big(ab - i ca\sqrt{1-b^2-c^2}-b^2-c^2\Big)}~.
\end{align}
\begin{align}
     	&\mathcal{I}_{\sin^2} (a,b,c) \equiv\frac{1}{2\pi}\int_0 ^{2\pi} \frac{\sin^2(\alpha)}{(1-a \cos(\alpha))(1-b \cos(\alpha)-c \sin(\alpha))}d\alpha\nonumber\\
     	&= \frac{4}{a(b+ic)} -\frac{ a(1+\sqrt{1-a^2})}{\sqrt{1-a^2}(1-\sqrt{1-a^2})(b+ic\sqrt{1-a^2}-a)}\nonumber\\
     	&-\frac{\epsilon (b-ic)^2 (1+\sqrt{1-b^2-c^2})}{\sqrt{1-b^2-c^2}(1-\sqrt{1-b^2-c^2})\Big(ab - i ca\sqrt{1-b^2-c^2}-b^2-c^2\Big)}~.
\end{align}
\begin{align}
 &\mathcal{I}_{\sin^2} \Big(A,\frac{c_a}{c_b},\frac{c_a}{c_b}\sqrt{1-c_b^2} \Big)=\nonumber\\
 &\frac{2 (1 -\sqrt{1 - A^2}) c_a^2 c_b^2 + 
     		2 A c_a c_b (-4 + (1 + \sqrt{1 - A^2}) c_b^2) + 
     		A^2 (4 c_b^2 - c_a^2 (-4 + c_b^2 (4 + \sqrt{1 - c_b^2})))}{2 A c_a c_b (2 A c_a c_b - c_a^2 c_b^2 + A^2 (-c_b^2 + c_a^2 (-1 + c_b^2)))}\nonumber\\
     		&+ \frac{\epsilon c_a\Big(2 c_a c_b \big(1 - c_b^2 + \sqrt{1 - c_a^2}\big) + 
     		A c_b^2 \big( \sqrt{1 - c_a^2} - \sqrt{1 - c_b^2}-2 + 
     		\sqrt{1 - c_a^2} \sqrt{1 - c_b^2} + 
     	\mathcal{C}\big)}{2c_b (1 - c_a^2 - \sqrt{1 - c_a^2})  (2 A c_a c_b - c_a^2 c_b^2 + A^2 (-c_b^2 + c_a^2 (-1 + c_b^2)))}~.
\end{align}
\begin{align}
     	\mathcal{C}\equiv A c_a^2 \Big( c_b^2\Big(2 + \sqrt{1 - c_b^2}\Big)-2\Big)~.
\end{align} 
Next, we define
\begin{align}
     	\mathcal{I}_{i \sin \cos} \big(A,\frac{c_a}{c_b},-i\frac{c_a}{c_b}\sqrt{1-c_b^2}\big) &\equiv \frac{i}{2\pi} \int_0 ^{2\pi}\frac{1}{(1-A \cos(\alpha))} \frac{\sin(\alpha)\cos(\alpha) d\alpha}{\big(1-\frac{c_a}{c_b} \cos(\alpha)+\frac{ic_a}{c_b}\sqrt{1-c_b^2}\sin(\alpha)\big)} \nonumber\\
     			& =\frac{1}{2\pi\times 4} \int_0 ^{2\pi}\frac{1}{(1-A \cos(\alpha))} \frac{(e^{2i \alpha}-e^{-2i\alpha}) d\alpha}{\big(1-\frac{c_a}{c_b} \cos(\alpha)+\frac{ic_a}{c_b}\sqrt{1-c_b^2}\sin(\alpha)\big)}~.
\end{align}
One obtains
\begin{align}
     			&\mathcal{I}_{i\sin \cos} (c_a,c_b,A)= -\frac{\sqrt{1-c_b^2} \Big(8 A \big( \sqrt{1 - A^2}-1\big) c_a c_b + 4 (1 - \sqrt{1-A^2}) c_a^2 c_b^2 + 
     			A^2 \mathcal{D}\Big)}{2 c_a c_b A (  1-\sqrt{1 - A^2})  \big(2 A c_a c_b - c_a^2 c_b^2 - 
    			A^2 \big(c_b^2 + c_a^2 (1 - c_b^2)\big)\big)}\nonumber\\
    			& +\epsilon \frac{c_a \big(1 + \sqrt{1 - c_a^2}\big) \sqrt{ 
    			1 - c_b^2} \Big(-2 c_a c_b + A (2 - 2 \sqrt{1 - c_a^2} + \sqrt{1 - c_a^2} c_b^2)\Big)}{2 c_b(1 - c_a^2 - \sqrt{1 - c_a^2})  \Big(-2 A c_a c_b + c_a^2 c_b^2 + 
    			A^2 (c_b^2 + c_a^2 (1 - c_b^2))\Big)}
    			~,
\end{align}
\begin{align}
     			\mathcal{D}\equiv 4 (1 - \sqrt{1 - A^2}) c_b^2 + 
     			c_a^2 \Big(4 - 4 \sqrt{1-A^2} - c_b^2(5 - 3 \sqrt{1 - A^2})\Big)~.
\end{align}
\begin{align}
     		&\int_0 ^{2\pi} \frac{\cos(2\phi'-2\phi)}{(1-c_a \cos(\phi-\phi_5))(1-c_b \cos(\phi-\phi_1))}d\phi=\nonumber\\
     		&\frac{4\pi}{c_a c_b} e^{2i\phi'-i(\phi_1+\phi_5)}-\frac{\pi e^{-2i \phi_5}c_a (1+\sqrt{1-c_a^2})\Big(e^{2i \phi'}+\frac{1}{c_a^4}e^{4i \phi_5-2i \phi'} \big(1-\sqrt{1-c_a^2}\big)^4\Big)}{\sqrt{1-c_a^2}(1-\sqrt{1-c_a^2})\big(c_b \cos(\phi_1-\phi_5)+ic_b\sqrt{1-c_a^2}\sin(\phi_1-\phi_5)-c_a\big)}
     		\nonumber\\
     		&-\frac{\pi e^{-2i \phi_1}c_b (1+\sqrt{1-c_b^2})\Big(e^{2i \phi'}+\frac{1}{c_b^4}e^{4i \phi_1-2i \phi'} \big(1-\sqrt{1-c_b^2}\big)^4\Big)}{\sqrt{1-c_b^2}(1-\sqrt{1-c_b^2})\big(c_a \cos(\phi_1-\phi_5)+ic_a\sqrt{1-c_b^2}\sin(\phi_5-\phi_1)-c_b\big)}~.
\end{align}


\vskip 3cm


\bibliographystyle{JHEP}      

\bibliography{fivebranes}


\end{document}
